\documentclass[12pt,a4paper,twoside,headsepline,titlepage,enabledeprecatedfontcommands]{scrreprt}
\usepackage[english]{babel}
\usepackage{latexsym}
\usepackage{graphics}
\usepackage{epsfig}
\usepackage{longtable}
\usepackage{color}
\usepackage[active]{Srcltx}

\newcommand{\bra}[1]{\langle{#1}|}
\newcommand{\ket}[1]{|{#1}\rangle}

\newcommand{\su}{\uparrow}
\newcommand{\sd}{\downarrow}

\begin{document}

\title{Quantum optical experiments towards atom-photon entanglement} 
\author{\\ \\ 
Dissertation at the Department of Physics  \\
of the\\
Ludwig-Maximilians-Universit\"at M\"unchen \\ \\ \\ \\ \\
Markus Weber \\ \\ \\ \\ \\ \\ }
\date{Erratum: This revised version (July 29th 2025) includes corrections of typos and a corrected model of the 4-level optical Bloch-equations as published in \ https://doi.org/10.1103/PhysRevA.73.043406}
\maketitle

\pagestyle{empty}

\begin{abstract}

\subsubsection{{\Large Abstract}}

In 1935 {\it Einstein, Podolsky and Rosen (EPR)} used the assumption of local
realism to conclude in a Gedankenexperiment with two entangled particles that
quantum mechanics is not complete. For this reason {\it EPR} motivated an
extension of quantum mechanics by so-called local hidden variables. Based on
this idea in 1964 {\it Bell} constructed a mathe\-matical inequality whereby
experimental tests could distinguish between quantum mechanics and
local-realistic theories. Many experiments have since been done that are
consistent with quantum mechanics, disproving the concept of local
realism. But all these tests suffered from loopholes allowing a
local-realistic explanation of the experimental observations by exploiting
either the low detector efficiency or the fact that the detected particles
were not observed space-like separated. In this context, of special interest
is entanglement between different quantum objects like atoms and photons,
because it allows one to entangle distant atoms by the interference of
photons. The resulting space-like separation together with the almost perfect
detection efficiency of the atoms allows a first loophole-free test of Bell's
inequality.

The primary goal of the present thesis is the experimental realization of
entanglement between a single localized atom and a single spontaneously
emitted photon at a wavelength suitable for the transport over long
distances. In the experiment a single optically trapped $^{87}$Rb atom is
excited to a state which has two selected decay channels. In the following
spontaneous decay a photon is emitted coherently with equal probability into
both decay channels. This accounts for perfect correlations between the
polari\-zation state of the emitted photon and the Zeeman state of the atom
after spontaneous decay. Because these decay channels are spectrally and in
all other degrees of freedom indistinguishable, the spin state of the atom is
entangled with the polarization state of the photon. To verify entanglement,
appropriate correlation measurements in complementary bases of the photon
polarization and the internal quantum state of the atom are performed. It is
shown, that the generated atom-photon state yields an entanglement fidelity of
$0.82$.

The experimental results of this work mark an important step towards the
generation of entanglement between space-like separated atoms for a first
loophole-free test of Bell's inequality. Furthermore entanglement between a
single atom and a single photon is an important tool for new quantum
communication and information applications, e.g. the remote state preparation
of a single atom over large distances.

\includegraphics[width=2.5cm]{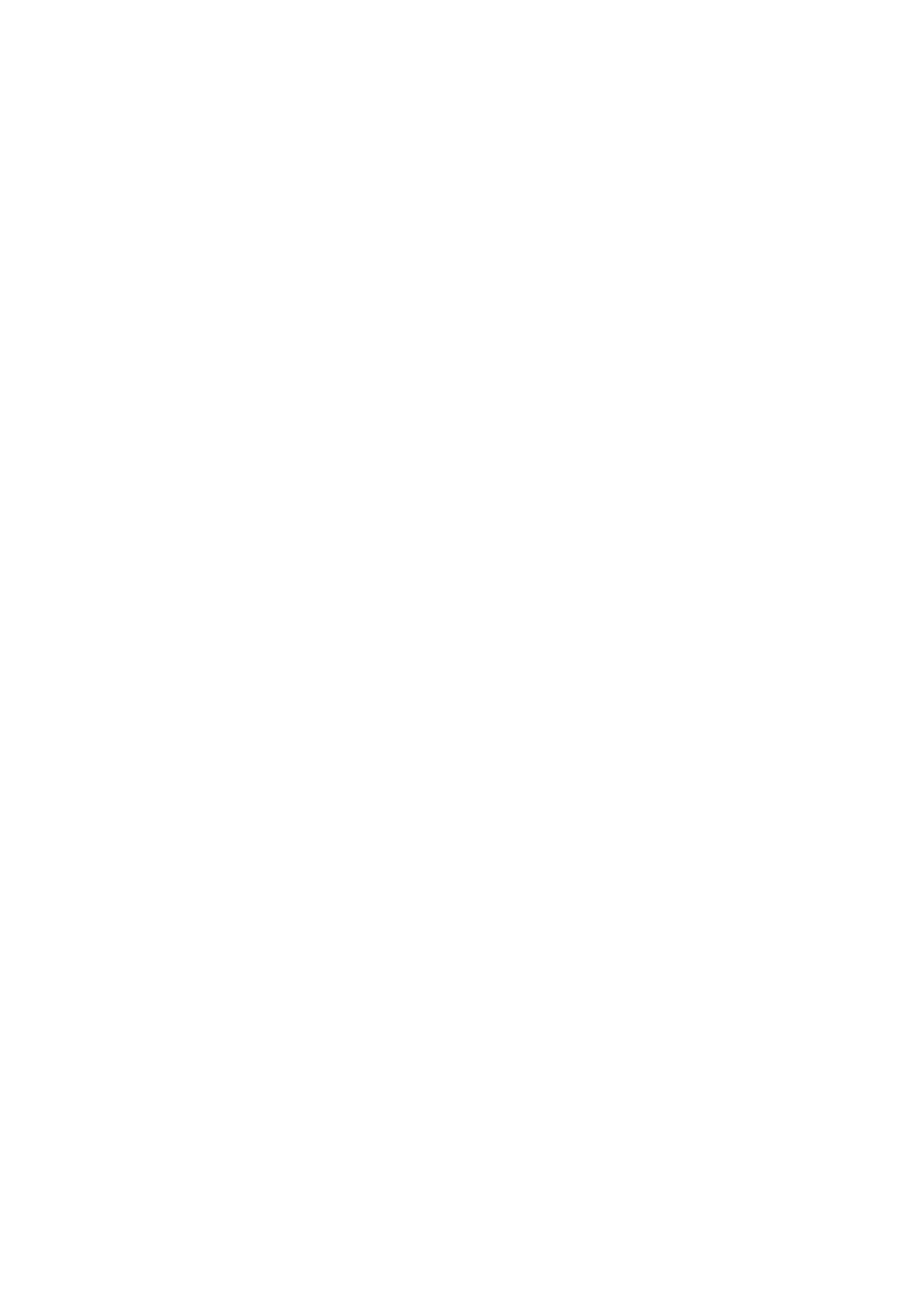}

\includegraphics[]{empty.eps}
\subsubsection{{\Large Zusammenfassung}}

Im Jahr 1935 ver\"offentlichten {\it Einstein, Podolsky und Rosen (EPR)} ein
Gedankenexperiment, in dem mit Hilfe zweier verschr\"ankter Teilchen und der
Annahme, dass jede physikalische Theorie lokal sein muss, gezeigt wurde, dass
die Quantenmechanik eine unvollst\"andige Theorie ist. {\it EPR} motivierten
damit die Erweiterung der Quantenmechanik durch sogenannte lokale verborgene
Parameter. Basierend auf dieser Idee konstruierte {\it Bell} 1964 eine
mathematische Ungleichung, anhand derer erstmals mit Hilfe von experimentellen
Tests zwischen der Quantentheorie und lokalen realistischen Theorien
unterschieden werden konnte. Seither wurden viele Experimente durchgef\"uhrt,
die die Quantentheorie best\"atigten und das Konzept der lokalen verborgenen
Parameter widerlegten. Aber all diese experimentellen Tests litten unter
sogenannten Schlupfl\"ochern, die eine lokal-realistische Erkl\"arung der
experimentellen Beobachtungen zulie{\ss}en. Entweder die verwendeten
Detektoren hatten eine zu niedrige Detektionseffizienz, oder die detektierten
Teilchen wurden nicht raumartig getrennt beobachtet. In diesem Zusammenhang
ist die Verschr\"ankung zwischen unterschiedlichen Quantenobjekten wie Atomen
und Photonen von besonderem Interesse, da hiermit zwei weit voneinander
entfernte Atome durch Interferenz von Photonen robust verschr\"ankt werden
k\"onnen. Die daraus resultierende raumartige Trennung erm\"oglicht zusammen
mit der beinahe perfekten Detektionseffizienz der Atome einen ersten
schlupflochfreien Test der Bell'schen Ungleichung.

Das vorrangige Ziel dieser Arbeit ist die experimentelle Realisierung von
Ver\-schr\"ankung zwischen einem einzelnen lokalisierten Atom und einem
einzelnen spontan emittierten Photon, mit einer Wellenl\"ange die sich gut zum
Transport \"uber gro{\ss}e Entfernungen eignet. In dem vorliegenden Experiment
wird ein einzelnes, optisch gefangenes, $^{87}$Rb Atom in einen Zustand
angeregt, der zwei ausgezeichnete Zerfallskan\"ale hat. Beim nachfolgenden
Spontanzerfall wird ein Photon mit gleicher Wahrscheinlichkeit koh\"arent in
beide Kan\"ale emittiert. Dies bedingt eine perfekte Korrelation zwischen der
Polarisation des emittierten Photons und dem Zeemanzustand des Atoms nach dem
Spontan\-zerfall. Da diese Kan\"ale spektral und in allen anderen
Freiheitsgraden ununterscheidbar sind, kommt es zur Verschr\"ankung des
Polarisationsfreiheitsgrads des Photons mit dem Spinfreiheitsgrad des
Atoms. Zum Nachweis der Verschr\"ankung werden geeignete
Kor\-relationsmessungen zwischen dem internen Zustand des Atoms und dem
Polari\-sationszustand des Photons in komplement\"aren Messbasen
vorgenommen. Es wird gezeigt, dass der generierte Atom-Photon Zustand mit
einer G\"ute von 82 Prozent verschr\"ankt ist.

Die in dieser Arbeit gewonnenen experimentellen Ergebnisse markieren einen
wich\-tigen Schritt in Richtung Verschr\"ankung zweier raumartig getrennter
Atome f\"ur einen ersten schlupflochfreien Test der Bell'schen
Ungleichung. Dar\"uberhinaus ist die Verschr\"ankung zwischen einem einzelnen
Atom und einem einzelnen Photon ein wichtiges Werkzeug zur Realisierung von
neuen Anwendungen auf dem Gebiet der Quantenkommunikation und
Quanteninformationsverarbeitung, wie zum Beispiel der Zustandspr\"aparation
eines einzelnen Atoms \"uber gro{\ss}e Entfernungen.

\newpage

\end{abstract}

\pagestyle{headings}
\setcounter{page}{1} 
\tableofcontents

%---------------------------------------------------------------------------
%---------------------------------------------------------------------------

\chapter{Introduction}

Since the early days of quantum theory {\it entanglement} - first introduced
by {Schr\"odinger} in his famous paper on {\it ``Die gegenw\"artige Situtation
in der Quantenmechanik} \cite{Schroedinger} - has been a considerable subject
of debate because it highlighted the counter-intuitive nonlocal aspect of
quantum mechanics. In particular, Einstein, Podolsky and Rosen (EPR)
\cite{EPR} presented an argument to show that there are situations in which
the general pro\-babilistic scheme of quantum theory seems not to describe the
physical reality. In this famous {\it Gedankenexperiment} EPR used the
assumption of local realism to conclude by means of two entangled particles
that quantum mechanics is incomplete. For this reason {\it EPR} motivated an
extension of quantum mechanics by so-called local hidden variables. Based on
this idea in 1964 John Bell constructed mathematical inequalities which allow
to distinguish between quantum mechanics and local-realistic theories
\cite{Bell64}. Many ex\-periments have since been done
\cite{Freedman72,Aspect81,Aspect82,Weihs98,Rowe01} that are consistent with
quantum mechanics, disproving the concept of local realism.

But all these tests suffered from at least one of two primary loopholes. The
first is called the locality loophole \cite{Bell88,Santos92}, in which the
correlations of apparently separate events could result from unknown
subluminal signals propagating between two different regions of the
measurement apparatus. An experiment was performed with entangled photons
\cite{Weihs98} enforcing strict relativistic separation between the
measurements. But this experiment suffered from low detection efficiencies
allowing the possibility that the subensemble of detected events agrees with
quantum mechanics even though the entire ensemble satisfies the predictions of
Bell's inequalities for local-realistic theories. This loophole is referred to
as detection loophole \cite{Pearle70,Santos95} and was addressed in an
experiment with two trapped ions \cite{Rowe01}, where the quantum state
detection of the atoms was performed with almost perfect efficiency. Because
the ion separation was only a few $\mu$m this experiment could not eliminate
the locality loophole. A possibility to close both loopholes in the same
experiment is the preparation of space-like separated entangled atoms
\cite{Saucke02,Simon03}. The key-element of this proposal is the faithful
generation of two highly entangled states between a single localized atom and
a single spontaneously emitted photon (at a wavelength suitable for low-loss
transport over large distances). The photons coming from each of the atoms
travel then to an intermediate location where a partial Bell-state measurement
is performed leaving the two distant atoms in an entangled state
\cite{Zukowski93,Simon03}. Because the atoms can be detected with an
efficiency up to 100 percent \cite{Rowe01} this finally should allow a first
loophole-free test of Bell's inequality \cite{Saucke02,Simon03}.

Nowadays there is also a large interest in the generation and engineering of
quantum entanglement for the implementation of quantum communication and
information \cite{Zeilinger00,Chuang00}. Until now entanglement was observed
mainly between quantum objects of similar type like single photons
\cite{Aspect82,Ou88,Weihs98}, single atoms
\cite{Haroche97a,Sackett00,Barrett04,Riebe04} and recently between optically
thick atomic ensembles \cite{Kuzmich03,Lukin03}. But all distributed quantum
computation and scalable quantum communication protocols \cite{Chuang00}
require to coherently transfer quantum information between photonic- and
matter- based quantum systems. The importance of this process is due to the
fact that matter-based quantum systems provide excellent long-term quantum
memory storage, whereas long-distance communication of quantum information
will be accomplished by coherent propagation of light, e.g. in the form of
single photons. The faithful mapping of quantum information between a stable
quantum memory and a reliable quantum information channel would allow, for
example, quantum communication over long distances and quantum teleportation
of matter. But, because quantum states cannot in general be copied, quantum
information could be distributed in these applications by entangling the
quantum memory with the communication channel. In this sense entanglement
between atoms and photons is necessary because it combines the ability to
store quantum information with an effective communication channel
\cite{Duan01,Saucke02,Duan03,Feng03,Simon03}. 

Atom-photon entanglement has been implicit in many previous experimental
systems, from early measurements of Bell inequality violations in atomic
cascade systems \cite{Freedman72,Aspect82} to fluorescence studies in trapped
atomic ions \cite{Eichmann93,DeVoe96} and atomic beam experiments
\cite{Kurtsiefer97}. Furthermore, on the basis of entanglement between matter
and light, different experimental groups combined in the last few years the
data storage properties of atoms with coherence properties of light. E.g., in
the microwave domain, coherent quantum control has been obtained between
single Rydberg atoms and single photons
\cite{Haroche97b,Haroche02,Haroche03}. Recently, Matsukevich {\it et al.}
\cite{Matsukevich04} reported the experimental realization of coherent quantum
state transfer from a matter qubit onto a photonic qubit using entanglement
between a single photon and a single collective excitation distributed over
many atoms in two distinct optically thick atomic samples. However,
atom-photon entanglement has not been directly observed until quite recently
\cite{Blinov04}, as the individual atoms and photons have not been under
sufficient control.

The primary goal of the present work is the experimental realization of
entanglement between a single localized atom and a single spontaneously
emitted photon (at a wavelength suitable for long-distance transport in
optical fibers and air) for a future loophole-free test of Bell's
inequality. This task can be managed using optically trapped neutral alkali
atoms like Rubidium or Cesium which radiate photons in the NIR of the
electromagnetic spectrum. 

\begin{figure}[h]
\centerline{\scalebox{1}{\includegraphics[]{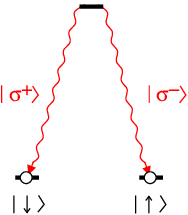}}}
\caption{Atomic dipole transition to generate atom-photon entanglement.}
\label{pict:lambda_general}
\end{figure}

To generate atom-photon entanglement in this experiment, a single $^{87}$Rb
atom (stored in an optical dipole trap) is excited to a state which has two
decay channels. In the following spontaneous emission the atom decays either
to the $\ket{\downarrow}$ ground state while emitting a
$\ket{\sigma^+}$-polarized photon or to the $\ket{\uparrow}$ state while
emitting a $\ket{\sigma^-}$-polarized photon. Provided these decay channels
are indistinguishable a coherent superposition of the two possibilities is
formed and the spin state of the atom is entangled with the polarization state
of the emitted photon. Thus, the resulting atom-photon pair is in the
maximally entangled quantum state
\begin{equation}
\ket{\Psi^+}=\frac{1}{\sqrt{2}}(\ket{\downarrow}\ket{\sigma^+}
+\ket{\uparrow}\ket{\sigma^-}).
\end{equation}

\markright{1 Introduction}
To verify entanglement of the generated atom-photon state one has to disprove
the possibility that the two-particle quantum system can be a statistical
mixture of separable states. This task is closely connected to a violation of
Bell's inequality and requires correlated local state measurements of the atom
and the photon in complementary bases. The polarization state of the photon
can be measured simply by a combination of a polarization filter and a single
photon detector. However, the spin state of a single atom is not trivial to
measure and therefore one of the challenges of this experiment.

In the context of the present work we set up an optical dipole trap which
allows one to localize and manipulate a single $^{87}$Rb atom. The internal
quantum state of the atom is analyzed by means of a
Stimulated-Raman-Adiabatic-Passage (STIRAP) technique, where the polarization
of the STIRAP light field defines the atomic measurement basis. To proof
atom-photon entanglement the internal quantum state of the atom is measured
conditioned on the detection of the polarization state of the photon. We
observe strong atom-photon correlations in complementary measurement bases
verifying entanglement between the atom and the photon.

\subsubsection{Overview}

In the second chapter I will introduce in general the property of entanglement
between two spin-1/2 particles and in particular spin-entanglement between a
single atom and a single photon. The third chapter deals with trapping single
atoms in a far-off-resonance optical dipole potential. After the theory of
optical dipole potentials is discussed, the trap setup is presented and the
observation of single atom resonance fluorescence is reported. A detailed
investigation of the resonance fluorescence spectrum is performed which allows
to determine the mean kinetic energy of the stored atom. In the fourth
chapter, the photon statistics of the emission from a single four-level atom
is analyzed. The measured photon-pair correlation functions are discussed and
compared with theoretical models. The fifth chapter describes in detail the
atomic state detection scheme. The theory of coherent-population-trapping
(CPT) and Stimulated-Raman-Adiabatic-Passage (STIRAP) is discussed and
experiments are presented which show the coherent analysis of Zeeman
superposition states of a single atom. In the sixth chapter I will report on
the observation of entanglement between a single optically trapped $^{87}$Rb
atom and a spontaneously emitted single photon. Finally in the seventh chapter
the experimental achievements are discussed and future applications of
atom-photon entanglement are highlighted.

%----------------------------------------------------------------------------
%----------------------------------------------------------------------------

\chapter{Theory of atom-photon entanglement} \label{chapt:AtomPhoton}

\section{Introduction}

In the context of this chapter I will introduce the concept of entanglement
between the spin state of a single atom and the polarization state of a single
photon. I will begin by very briefly recapitulating some basic features of
quantum mechanics which later on will become relevant for the understanding of
the experimental investigation of atom-photon entanglement in chapter
\ref{chapt:ObservationAtomPhoton}. Then I will establish in general the
property of ``entanglement'' between two quantum systems and important
applications in the field of quantum communication. Finally I will present the
basic idea of my thesis how to generate and analyze a spin-entangled
atom-photon state.

\section{Basics of quantum mechanics}

\subsection{The superposition principle}

\begin{figure}[h]
\centerline{\scalebox{1}{\includegraphics[]{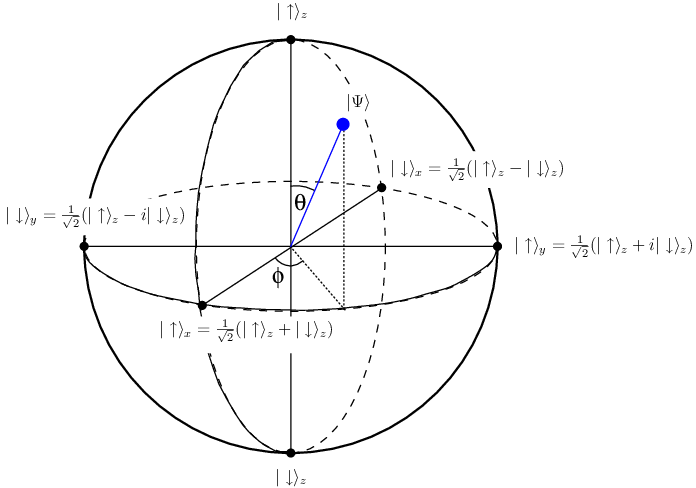}}}
\caption{An arbitrary spin-1/2 state $\ket{\Psi}$ and the three complementary
  bases {x,y,z} in the Bloch sphere representation.}
\label{pict:Blochsphere}
\end{figure}

One of the most important concepts in quantum mechanics is the {\it
  superposition} principle. Quantum states denoted by $\ket{\Psi}$ can exist
  in a superposition, i.e. a linear combination, of two possible orthogonal
  basis states $\ket{\su}$ and $\ket{\sd}$
\begin{equation}
\ket{\Psi}=a\ket{\su}+b\ket{\sd},
\end{equation}
where $a$ and $b$ are complex numbers and $|a|^2+|b|^2=1$. Examples of states
 in a two dimensional Hilbert space are the polarization states
 $\ket{\sigma^+}$ and $\ket{\sigma^-}$ of a single photon or the states
 $\ket{m_L =-1/2}$ and $\ket{m_L=+1/2}$ of the magnetic moment of a single
 atom with the angular momemtum $L=1/2$. The state $\ket{\Psi}$ can be written
 in a useful geometric representation as:
\begin{equation}
\ket{\Psi}=\cos\left(\frac{\theta}{2}\right) \ket{\su}
           + e^{i \phi} \sin\left(\frac{\theta}{2}\right) \ket{\sd},
\end{equation}
where $\theta$ and $\phi$ define a point on the three-dimensional Bloch
sphere (see Fig. \ref{pict:Blochsphere}).

\subsection{Quantum measurements}

A measurement in quantum mechanics is inherently {\it indeterministic}. If we
ask (measure) whether $\ket{\Psi}$ is in the state $\ket{\su}$, we obtain this
result with probability $|\bra{\su}\Psi\rangle|^2=|a|^2$. After the
measurement, the state is projected to $\ket{\su}$. In this ideal {\it Von
Neumann measurement}, it is not possible to measure an unknown quantum state
without disturbance.

\subsection{Complementary observables}

Two well known complementary observables are position and momentum of a single
particle. These two observables cannot be measured with arbitrary accuracy in
an experiment at the same time \cite{Bohr28}. In general, let $\hat{A}$ and
$\hat{B}$ denote two non-commuting Hermitian operators (observables) of a
quantum system of dimension $N$, $\hat{A}$ and $\hat{B}$ are said to be
complementary, or mutually unbiased, if their eigenvalues are non-degenerate
and the inner product between any two normalized eigenvectors $\ket{\Psi_A}$ of
$\hat{A}$ and $\ket{\Psi_B}$ of $\hat{B}$, always has the same magnitude. For
the case of a two-dimensional Hilbert space, there are three complementary
observables $\hat{\sigma}_x$, $\hat{\sigma}_y$ and $\hat{\sigma}_z$, which are
called Pauli spin-operators. These operators can be represented by $2\times2$
Hermitian matrices defined by
\begin{equation} \label{equ:Pauli}
\sigma_x=\left(\begin{array}{cc} 0 & 1 \\ 1 & 0 \end{array}\right) \quad
\sigma_y=\left(\begin{array}{cc} 0 & -i \\ i & 0 \end{array}\right) \quad
\sigma_z=\left(\begin{array}{cc} 1 & 0 \\ 0 & -1 \end{array}\right)
\end{equation}
The eigenvalues of $\sigma_x$, $\sigma_y$ and $\sigma_z$ are $\lambda=\pm 1$
and the corresponding eigenvectors are given by
\begin{eqnarray}
\ket{\su}_x & = & \frac{1}{\sqrt{2}} ( \ket{\su}_z + \ket{\sd}_z ) \\
\ket{\sd}_x & = & \frac{1}{\sqrt{2}} ( \ket{\su}_z - \ket{\sd}_z ) \\
\ket{\su}_y & = & \frac{1}{\sqrt{2}} ( \ket{\su}_z + i\ket{\sd}_z ) \\
\ket{\sd}_y & = & \frac{1}{\sqrt{2}} ( \ket{\su}_z - i \ket{\sd}_z ) 
\end{eqnarray}
and $\ket{\su}_z$ and $\ket{\sd}_z$, respectively. Fig. \ref{pict:Blochsphere}
shows these three complementary basis vectors on the Bloch sphere.

The inner product $\langle.|.\rangle$ between any two basis states belonging
to different bases is $1/\sqrt{2}$. This property guarantees that if a quantum
system is prepared in one basis, the outcome of a measurement in any
complementary basis is totally random.

\section{Entanglement}

Since the early days of quantum mechanics entanglement - first introduced by
{Schr\"odinger} in his famous paper on {\it ``Die gegenw\"artige Situtation in
der Quantenmechanik} \cite{Schroedinger} - has been a considerable subject of
debate because it highlighted the counter-intuitive nonlocal aspect of quantum
mechanics. In particular, Einstein, Podolsky and Rosen (EPR) \cite{EPR}
presented an argument to show that there are situations in which the general
probabilistic scheme of quantum theory seems not to describe the physical
reality properly. In this famous {\it Gedankenexperiment} EPR used the
assumption of local realism to conclude by means of two entangled particles
that quantum mechanics is incomplete. Here I will introduce the essential
features of entanglement between two spin-1/2 particles following Bohm
\cite{Bohm51}.
  
Let us consider a quantum state $\ket{\Psi_{AB}}$ of two spin-1/2
particles $A$ and $B$. Up to some time $t=0$, these particles are taken to be
in a bound state of zero angular momentum. Then we turn off the binding
potential (e.g., we disintegrate the bound system in two parts), but introduce
no angular momentum into the system and do not disturb the spins in any
way. The seperate parts of the system are now free but due do conservation of
angular momentum the two particles are entangled and the state is given by
\begin{equation} \label{equ:Bellstate}
\ket{\Psi_{AB}}=\frac{1}{\sqrt{2}}(\ket{\su_A} \ket{\sd_B} - \ket{\sd_A}
\ket{\su_B}).
\end{equation}
This state has four important properties:
\begin{enumerate}
\item it can not be expressed as a tensor product $\ket{\Psi_A} \otimes
\ket{\Psi_B}$ of two single particle states.
\item it is rotationally invariant.
\item the expectation value of the spin of a single particle is zero.
\item the spin states of both particles are anti-correlated in any analysis
  direction.
\end{enumerate}

The probability to simultaneously ``measure'' the spin of particle $A$ on the
equator of the Bloch-sphere at the angle $\phi_A$ and particle $B$ at the
angle $\phi_B$, respectively, is
\begin{equation}
P_{AB}(\phi_A,\phi_B)=\bra{\Psi_{AB}}\hat{\pi}_{\phi_A}^A \hat{\pi}_{\phi_B}^B
\ket{\Psi_{AB}},
\end{equation}
where the projection operators $\hat{\pi}_{\phi_A}^{A}$ and
$\hat{\pi}_{\phi_B}^{B}$ corresponding to particles $A$ and $B$ are given by
\begin{equation}
\hat{\pi}_{\phi_i}^{i}=\frac{1}{2}(\hat{I}^{i}+\hat{\sigma}_x^{i}
\cos\phi_i +  \hat{\sigma}_y^{i} \sin\phi_i), \quad i=A,B.
\end{equation}
After some calculations, we find
\begin{equation}
P_{AB}(\phi_A,\phi_B)=\frac{1}{4}[1-\cos{(\phi_A -\phi_B)}]=
\frac{1}{2}\sin^2\left(\frac{\phi_A - \phi_B}{2}\right).
\end{equation}

If we now ask for the conditional probability to find particle $A$ along
$\phi_A$ if particle $B$ was measured along $\phi_B$ we get
\begin{equation}
P(\phi_A|\phi_B)=\sin^2\left(\frac{\phi_A - \phi_B}{2}\right).
\end{equation}

\begin{figure}[t]
\centerline{\scalebox{1}{\includegraphics[]{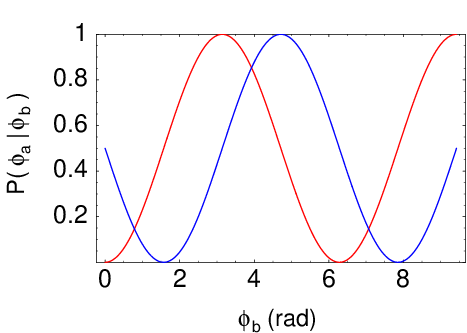}}}
\caption{Expected spin correlations of an entangled state $\ket{\Psi_{AB}}$
  for complementary measurement bases of particle $A$ as the
  analyzer-direction of particle $B$ is varied by an angle $\phi_B$ . (Red
  line) particle $A$ is measured along $\phi_A=0$ corresponding to
  $\ket{\su}_x$. (Blue line) particle $A$ is analyzed along $\phi_A=\pi/2$
  corresponding to $\ket{\su}_y$.}
\label{pict:EPR}
\end{figure}
This quantity is shown in Fig. \ref{pict:EPR} as a function of the analysis
direction $\phi_B$ if particle $A$ was measured as $\ket{\su}$ in $\sigma_x$
or $\sigma_y$. If the analysis direction of both particles is the same
($\phi_A=\phi_B$), the conditional probability of detection is zero, because
the spins are always anticorrelated. But for $\phi_A=\phi_B+\pi$ the
conditional probability is equal to one. This property holds true for any
initial choice of the analyzer direction $\phi_A$ and is invariant under the
exchange of the two particles $P(\phi_A|\phi_B)=P(\phi_B|\phi_A)$. The
described two-particle correlations are independent of the choice of the
measurement basis. This is the main signature of the entangled singlet state
$\ket{\Psi_{AB}}$.

For an arbitrary quantum system consisting of two spin-1/2 particles $A$ and
$B$ the four entangled states
\begin{eqnarray} \label{eqn:Bellstates}
\ket{\Psi^+}&=&\frac{1}{\sqrt{2}}(\ket{\su_A}\ket{\sd_B} +
\ket{\sd_A}\ket{\su_B}) \\
\ket{\Psi^-}&=&\frac{1}{\sqrt{2}}(\ket{\su_A}\ket{\sd_B} -
\ket{\sd_A}\ket{\su_B}) \\
\ket{\Phi^+}&=&\frac{1}{\sqrt{2}}(\ket{\su_A}\ket{\su_B} +
\ket{\sd_A}\ket{\sd_B}) \\
\ket{\Phi^-}&=&\frac{1}{\sqrt{2}}(\ket{\su_A}\ket{\su_B} -
\ket{\sd_A}\ket{\sd_B}) 
\end{eqnarray}
form an orthonormal basis of the $2\times2$ dimensional Hilbert space. These
states are called Bell states and violate a Bell inequality (see subsection
\ref{subsect:BellIn}) by the maximum value predicted by quantum
mechanics. Furthermore the Bell states have the important property that each
state can be transformed into any other of the four Bell states by unitary
single particle rotations.

\subsection{The EPR ``paradox''}

Every spin-entangled state has the property that none of the two spins has a
defined value. Therefore it is impossible to predict the outcome of a spin
measurement on one particle with certainty. But if we perform a measurement on
one spin, instantaneously the outcome of a spin measurement on the other
particle is known. This holds true even if the spins are separated by an
arbitrary distance. These counterintuitive features of entangled quantum
systems were used by Einstein, Podolsky and Rosen (EPR) \cite{EPR} to argue
that the quantum mechanical description of the physical reality can not be
considered complete. EPR understand a theory to be complete if every element
of the {\bf physical reality} is represented in the physical theory in the
sense that:
\newline

{\it ``If, without in any way disturbing a system, we can predict with
certainty (i.e. with probability equal to unity) the value of a physical
quantity, then there exists an element of physical reality corresponding to
this quantity.''}
\newline

The first part of the reality criterion requires that the prediction can be
made without disturbing the object in question. It is, for instance, possible
to predict the value of the mass of the next pion crossing a bubble chamber
without interacting with it. In every case in which the prediction can be made
in this way, the EPR reality criterion assigns to the object an {\it element
of reality}, i.e., something real that does not necessarily coincide with the
observed property but generates it deterministically when a measurement is
made.

Furthermore EPR require that any physical theory has to be local. In other
words the {\bf locality postulate} is based on the reasonable belief that the
strength of interaction between objects depends inversily on their
separation. If an electron is observed in a laboratory, another electron 10 or
$10^{10}$ m away acquires no new property. To summarize, the idea of locality
can be formulated as follows:
\newline

{\it ``Given two separated objects A and B, the modifications of A due to
  anything that may happen to B can be made arbitrarily small for any
  measurable physical quantity, by increasing their separation.''}
\newline

In 1951 David Bohm restated and simplified the EPR argument on the basis of
the spin-entangled state $\ket{\Psi_{AB}}$ from (\ref{equ:Bellstate}) as
follows
\cite{Bohm51}:
\begin{enumerate}
\item Pick an arbitrary analysis basis, e.g. the $z$-basis and measure the
  spin of particle $A$. Then the result will be either $\su$ or $\sd$, say
  $\su$.
\item Knowing that the spin of particle $A$ is ``up'', we know with certainty
  that the result of a spin-measurement of $B$ will be ``down''. But if we
  then measure particle $B$ in the $x$-basis we will find that particle $B$
  has a definite spin (either $\su_x$ or $\sd_x$), i.e., we know the value of
  $\sigma_x$.
\item Therefore, we know both the $z$ and $x$ components of the spin of
  particle $B$, which is a violation of complementarity.
\end{enumerate}
In the words of EPR this implies a very unsatisfactory state of affairs:
``Thus, it is possible to assign two different state vectors to the same
reality.'' One way out of this problem is to argue that in a single experiment
one can measure particle $B$ only in one basis and not in two bases at the
same time. The knowledge of a fictitious measurement result does not replace
the measurement process itself. From this point of view the EPR assumption
about elements of reality becomes useless: it can only lead to the conclusion
that {\it ``an element of reality is associated with a concretely performed
act of measurement''}. This argument was used by Bohr \cite{Bohr36} in a reply
to EPR, whereas EPR used the inability of quantum mechanics to make definite
predictions for the outcome of a certain measurement to postulate the
existence of ``hidden'' variables which are not known and perhaps not
measurable. It was hoped that an inclusion of these hidden variables (LHV)
would restore the completness and determinism to the quantum theory.

\subsection{Bell's inequality} \label{subsect:BellIn}

In 1964, John Bell \cite{Bell64} constructed an inequality for observables of
spin correlation experiments, for which predictions of hidden variable
theories based on Einstein's locality principle do not agree with the
statistical predictions of quantum mechanics. Here I will sketch the simplest
derivation of Bell's inequality as given by Bell in 1970 \cite{Selleri99}.
\begin{figure}[h]
\centerline{\scalebox{1}{\includegraphics[]{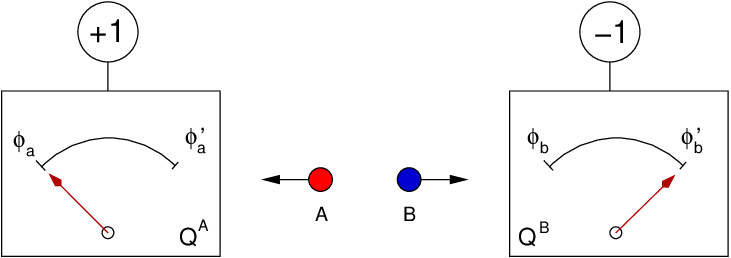}}}
\caption{Experimental apparatuses measuring one of the dichotomic observables
  $Q^A(\phi_A)$, $Q^A(\phi_A')$, $Q^B(\phi_B)$, $Q^B(\phi_B')$ on the incoming
  physical systems $A$ and $B$, respectively. The results (+1 and -1) of the
  last measurement are shown on the upper screens.}
\label{pict:Bell}
\end{figure}

It is assumed that in an EPR experiment dichotomic observables
$Q^A(\phi_A)=\pm 1$ and $Q^B(\phi_B)=\pm 1$ are measured on the two particles
$A$ and $B$, respectively, moving in opposite directions, as in
Fig. \ref{pict:Bell}. These observables depend on instrumental parameters
$\phi_A$ and $\phi_B$ (polarizers' axes, directions of magnetic fields, etc.)
that can be varied. In practice, only two observables [$Q^A(\phi_A)$ and
$Q^A(\phi_A')$] are of interest for particle $A$, and two [$Q^B(\phi_B)$ and
$Q^B(\phi_B')$] for particle $B$. In general, it is expected that
$Q^A(\phi_A)$ and $Q^A(\phi_A')$ are incompatible and hence cannot be measured
at the same time, and that the same holds for $Q^B(\phi_B)$ and
$Q^B(\phi_B')$.

It is assumed that hidden variables belonging to $A$ and $B$ fix the outcome
of all possible measurements. These hidden variables are collectively
represented by $\lambda$, assumed to vary in a set $\Lambda$ with a probability
density $\rho(\lambda)$. The normalization condition
\begin{equation} \label{equ:Bell_normal}
\int_{\Lambda} d\lambda \rho(\lambda)=1
\end{equation}
holds. Thus one can write
\begin{equation}
Q^A(\phi_A,\lambda)=\pm 1; \quad Q^A(\phi_A',\lambda)=\pm 1; \quad
Q^B(\phi_B,\lambda)=\pm 1; \quad Q^B(\phi_B',\lambda)=\pm 1
\end{equation}
meaning that, given $\lambda$, every one of the four dichotomic observables
assumes a well-defined value. The correlation function $E(\phi_A,\phi_B)$ is
defined as the average product of two dichotomic observables. In the
hidden-variable approach it can be written
\begin{equation}
E(\phi_A,\phi_B)=\int_{\Lambda} d\lambda \rho{(\lambda)} Q^A(\phi_A,\lambda)
Q^B(\phi_B,\lambda). 
\end{equation}  
This is a local expression, in the sense that neither $Q^A$ depends on
$\phi_B$ nor $Q^B$ on $\phi_A$.

It is easy to show that \cite{Selleri99}
\begin{eqnarray*}
|E(\phi_A,\phi_B)-E(\phi_A,\phi_B')+E(\phi_A',\phi_B)+E(\phi_A',\phi_B')| \\
 \le \int_{\Lambda} d\lambda
      \{|Q^A(\phi_A,\lambda)||Q^B(\phi_B,\lambda)-Q^B(\phi_B',\lambda)|+
      |Q^A(\phi_A',\lambda)||Q^B(\phi_B,\lambda)-Q^B(\phi_B',\lambda)|\} \\
= \int_{\Lambda} d\lambda
      \{|Q^B(\phi_B,\lambda)-Q^B(\phi_B',\lambda)|+
       |Q^B(\phi_B,\lambda)-Q^B(\phi_B',\lambda)|\}
\end{eqnarray*}
since $|Q^A(\phi_A,\lambda)|=|Q^A(\phi_A',\lambda)|=1$. But the moduli
$Q^B(\phi_B,\lambda)$ and $Q^B(\phi_B',\lambda)$ are also equal to 1, so
\begin{equation} \label{equ:Bell_Q}
|Q^B(\phi_B,\lambda)-Q^B(\phi_B',\lambda)|+|Q^B(\phi_B,\lambda)+Q^B(\phi_B',\lambda)|=2
\end{equation}
From (\ref{equ:Bell_normal}) and (\ref{equ:Bell_Q}) one obtains the inequality
\begin{equation} \label{equ:CHSH}
S(\phi_A,\phi_B,\phi_A',\phi_B')=|E(\phi_A,\phi_B)-E(\phi_A,\phi_B')|+
                                 |E(\phi_A',\phi_B)+E(\phi_A',\phi_B')|\le2.
\end{equation}
This is Bell's inequality in the CHSH form \cite{CHSH69}, which is more
general than its original form \cite{Bell64}.

The present proof is based on a general form of realism because the hidden
variable $\lambda$ is thought to belong objectively to the real physical
systems $A$ and $B$. It is also based on locality for three reasons: (1) the
dichotomic observables $Q^A$ and $Q^B$ do not depend on the parameters
$\phi_A$ and $\phi_B$ of the experimental apparatus; (2) the probability
density $\rho(\lambda)$ does not depend on $\phi_A$ and $\phi_B$; (3) the set
$\Lambda$ of possible $\lambda$ values does not depend on $\phi_A$ and
$\phi_B$. The time arrow assumption is also implicit in the dependance of
$\rho(\lambda)$ on $\phi$. In principle, the choices of the values of the
instrumental parameters could be made when the particles $A$ and $B$ are in
flight from the source to the analyzers.

But now, supposed two particles are in the entangled singlet state
$\ket{\Psi_{AB}}$. Quantum mechanically, the measurement of the dichotomic
variables $Q^A$ and $Q^B$ is represented by the spin operators
$\vec{\sigma}_A \vec{a}$ and $\vec{\sigma}_B \vec{b}$. The corresponding
quantum mechanical correlations entering Bell's inequality are given by the
expectation value for the product of spin-measurements on particle $A$ and $B$
along the directions $\vec{a}$ and $\vec{b}$:
\begin{eqnarray*} 
E_{QM}(\vec{a},\vec{b}) & = & \bra{\Psi_{AB}} \vec{\sigma}_A \vec{a}
            \otimes \vec{\sigma}_B \vec{b} \ket{\Psi_{AB}} \\
& = & \frac{1}{2}(\bra{\su}\vec{\sigma}_A \vec{a}
		\ket{\su} \bra{\sd}\vec{\sigma}_B \vec{b}\ket{\sd} -
		\bra{\su}\vec{\sigma}_A \vec{a}
		\ket{\sd} \bra{\sd}\vec{\sigma}_B \vec{b}\ket{\su} \\
&  &          - \bra{\sd}\vec{\sigma}_A \vec{a}
		\ket{\su} \bra{\su}\vec{\sigma}_B \vec{b}\ket{\sd} +
		\bra{\sd}\vec{\sigma}_A \vec{a}
		\ket{\sd} \bra{\su}\vec{\sigma}_B \vec{b}\ket{\su}) \\
& = & -\vec{a}\vec{b} = \cos{(\phi_A-\phi_B)}
\end{eqnarray*}
Choosing the directions $\phi_A=0$,
$\phi_A'=\pi/2$, $\phi_B=\pi/4$ and $\phi_B'=3\pi/4$ of the spin analyzers,
one finds a maximal violation of inequality (\ref{equ:CHSH}), namely
\begin{equation}
S_{QM}(0,\pi/2,\pi/4,3\pi/4)=2\sqrt{2} > 2.
\end{equation}
Thus, for the spin entangled singlet state, the quantum mechanical
correlations between the measurement results of two distant observers are
stronger than any possible correlation predicted by LHV theories.

The great achievement of John Bell was, that he derived a formal expression
which allows to distinguish experimentally between local-realistic theories
and quantum theory. In real experiments it is hard to violate Bell's
inequality without the additional assumption of {\it fair sampling}. Pearle
\cite{Pearle70}, for example, noted that a subensemble of detected events can
agree with quantum mechanics even though the entire ensemble satisfies the
predictions of Bell's inequality for local-realistic theories.

\subsubsection{Clauser-Horne inequality}
In 1974 Clauser and Horne derived an inequality, which can be tested in real
experiments with an essential weaker assumption \cite{CH74}. Supposed during a
period of time, while the adjustable parameters of the apparatus have the
values $\phi_A$ and $\phi_B$, the source emits, say, $N$ two-particle systems
of interest. For this period, denote by $N_A(\phi_A)$ and $N_B(\phi_B)$ the
number of counts at detectors $A$ and $B$, respectively, and by
$N_{AB}(\phi_A,\phi_B)$ the number of simultaneous counts from the two
detectors (coincident events). If $N$ is sufficiently large, then the ensemble
probabilities of these results are
\begin{eqnarray}
P_A(\phi_A)&=&N_A(\phi_A)/N, \nonumber \\
P_B(\phi_B)&=&N_B(\phi_B)/N, \\
P_{AB}(\phi_A,\phi_B)&=&N_{AB}(\phi_A)/N. \nonumber
\end{eqnarray}
After some algebra they obtain 
\begin{equation} \label{equ:CH}
S(\phi_A,\phi_B,\phi_A',\phi_B')=\frac{P_{AB}(\phi_A,\phi_B)-P_{AB}(\phi_A,\phi_B')+P_{AB}(\phi_A',\phi_B)+P_{AB}(\phi_A',\phi_B')}{P_A(\phi_A')+P_B(\phi_B)}
\le 1.
\end{equation}
In this inequality the probabilities can be replaced simply by count rates
because the normalization to the real number of emitted pairs $N$ cancels.

\subsubsection{Loopholes}

To exclude any local realistic theory in a Bell-type experiment one generally
assumes: (1) the detection probability of a pair of particles which has passed
the analyzers is independent of the analyzer settings \cite{CHSH69}; (2) the
detected subset is a sample of the whole emitted pairs; (3) a distant
apparatus does not influence a space-like and time-limited measurement due to
any relativistic effect. If any of these assumptions is dropped a hard proof
that local-realistic theories can not describe our physical reality is
impossible. Most prominently the {\it detection loophole} (2) and the {\it
locality loophole} (3) are used by critics to argue that no experiment
disproves the concept of local realism.
\newline

In 1972 Freedman and Clauser \cite{Freedman72} published an experiment which
was designed for a first test of Bell's inequality. A cascade transition in
$^{40}$Ca was used to generate polarization entangled photon pairs. Ten years
later Aspect {\it et al.} \cite{Aspect81,Aspect82} modified the Ca experiments
by Clauser in a way that they used a non-resonant two-photon process for the
excitation instead of a Deuterium arc lamp. Furthermore this experiment took
great care to use non-absorptive state analyzers and to switch fast between
different measurement bases in order to fulfill the requirements assumed by
John Bell in his original work. Aspect {\it et al.}  retrieved experimental
data violating the CHSH-inequality by $2.697\pm0.015$. A new era of Bell
experiments was opened by the application of the nonlinear optical effect of
parametric down-conversion to generate entangled photon pairs \cite{Ou88}. A
sequence of different experiments have been performed which culminated in two
Bell experiments \cite{Weihs98,Tittel98} highlighting the strict relativistic
separation between measurements. But both experiments suffered from low
detection efficiencies allowing the possibility that the subensemble of
detected events agrees with quantum mechanics even though the entire ensemble
satisfies the predictions of Bell's inequalities for local-realistic
theories. This loophole is referred to as detection loophole and was addressed
in an experiment with two trapped ions \cite{Rowe01}, whereby the quantum
state detection of the atoms was performed with almost perfect efficiency. But
the ion separation was not large enough to eliminate the locality loophole.

\subsection{Quantifying entanglement} \label{sect:QE}

The quantification of entanglement is a long standing problem in quantum
information theory. Any good measure of entanglement should satisfy certain
conditions. An important condition is that entanglement cannot increase by
local operations and classical communications (for more detail see
\cite{Vidal00}). The question which amount of entanglement is contained in an
arbitrary two-particle quantum state represented by a density matrix $\rho$ is
connected closely to the general question ``how close are two quantum
states''. One measure of distance between quantum states is the {\it
fidelity} $F$. 
\newline

For an unknown quantum system consisting of two spin-1/2 particles the state
is represented by a $4\times4$ density matrix $\rho$. The entanglement
fidelity with respect to a particular maximally entangled pure state
$\ket{\Psi_{AB}}$ is given by \cite{Bennett96}
\begin{equation} \label{equ:EntanglementFidelity}
F(\ket{\Psi_{AB}},\rho)=\bra{\Psi_{AB}}\rho\ket{\Psi_{AB}}=
\frac{1}{2}(\bra{\sd\su}\rho\ket{\sd\su}
+ \bra{\su\sd}\rho\ket{\su\sd} + \bra{\sd\su}\rho\ket{\su\sd} +
\bra{\su\sd}\rho\ket{\sd\su}),
\end{equation}
the overlap between $\ket{\Psi}$ and $\rho$. The first two terms in this
expression are the measured conditional probabilities of detecting
$\ket{\sd}_A\ket{\su}_B$ and $\ket{\su}_A\ket{\sd}_B$. The last two terms can
be determined by repeating the experiment while rotating each spin through a
polar angle of $\pi/2$ in the Bloch sphere before measurement. The rotated
quantum state is then given by $\tilde{\rho}$. Without a complete state
tomography of $\rho$ the entanglement fidelity $F$ can not be determined
accurately. But one can derive a lower bound of $F$, expressed only in terms
of diagonal density matrix elements in the original and rotated basis
\cite{Blinov04}:
\begin{eqnarray}
F \ge
\frac{1}{2}(\bra{\sd\su}\rho\ket{\sd\su}+\bra{\su\sd}\rho\ket{\su\sd}-
2 \sqrt{\bra{\sd\sd}\rho\ket{\sd\sd}\bra{\su\su}\rho\ket{\su\su}}+
\bra{\sd\su}\tilde{\rho}\ket{\sd\su}+\bra{\su\sd}\tilde{\rho}\ket{\su\sd}
\nonumber \\
-\bra{\sd\sd}\tilde{\rho}\ket{\sd\sd}-\bra{\su\su}\tilde{\rho}\ket{\su\su})
\end{eqnarray}

Another possibility to determine the entanglement fidelity of an unknown
quantum state $\rho$ is to do the following. Suppose, a source emits two
spin-1/2 particles in the entangled pure state $\ket{\Psi_{AB}}$. But due to
experimental imperfections in the state detection of spin $A$ and $B$, the
state $\ket{\Psi_{AB}}$ can be measured with the probability $0\le p \le
1$. For the rest of the unperfect state we expect white noise, corresponding
to the maximally mixed state $\frac{1}{4}\hat{I}$. Technically speaking, the
imperfect detection is modeled by an ideal preparation accompanied/followed by
a noisy channel, which transforms the maximally entangled initial state
$\rho=\ket{\Psi_{AB}}\bra{\Psi_{AB}}$ as
\begin{equation} \label{equ:rho_noisy}
\rho \rightarrow \rho=p\ket{\Psi_{AB}}\bra{\Psi_{AB}} + \frac{(1-p)}{4}\hat{I},
\end{equation}
where $\hat{I}$ is the identity operator. If we plug this expression into the
definition of the entanglement fidelity $F$ in
Eq. \ref{equ:EntanglementFidelity}, we obtain
\begin{equation}
F=p+\frac{(1-p)}{4}
\end{equation} 
For a mixed state denoted by a density matrix $\rho$, the correlations
entering a Bell inequality are given by
\begin{equation} \label{equ:trace}
E(\vec{a},\vec{b})=Tr\{\rho \vec{\sigma}_A \vec{a} \otimes \vec{\sigma}_B
\vec{b}\} = \sum_{i,k}^4 \bra{u_i}\rho\ket{u_k}\bra{u_k}\vec{\sigma}_A \vec{a} 
\otimes \vec{\sigma}_B
\vec{b}\ket{u_i},
\end{equation}
where $\ket{u_i}$ denotes the two particle spin-1/2 basis states
$\ket{\su\su}$, $\ket{\su\sd}$, $\ket{\sd\su}$ and $\ket{\sd\sd}$. On the
basis of the mixed state density matrix (\ref{equ:rho_noisy}) and
(\ref{equ:trace}) it is possible to derive the minimal amount of entanglement
necessary to violate a Bell inequality. One obtains $p\ge0.707$ corresponding
to a minimum entanglement fidelity of $F=0.78$. For fidelities $F>0.5$, the
underlying quantum state is entangled \cite{Bennett96}.

\subsection{Applications of entanglement}

Until the early 1990s the general physical interest concerning entanglement
was focused on fundamental tests of quantum mechanics. But in 1991 A. Ekert
\cite{Ekert91} realized that the specific correlations of entangled spin-1/2
particles and Bell's theorem can be used to distribute secret keys. Two years
later Bennet, Brassard, Crepeau, Josza, Peres and Wootters \cite{Bennett93}
proposed to transfer the quantum state of a particle to another particle at a
distant location employing entanglement as a resource, and it was realized by
Zukowski {\it et al.} \cite{Zukowski93} that two particles can be entangled by
a projection measurement on entangled Bell-states although they never
interacted in the past. In the last decade the physical interest about
entanglement focused also on applications for computational tasks. On the next
pages I will give a short review on quantum teleportation and entanglement
swapping because entanglement between a single atom and a single photon can be
used to map quantum information between light- and matter-based quantum
systems and for entangling space-like separated atoms for a loophole-free test
of Bell's inequality.

\subsubsection{Quantum Teleportation}

\begin{figure}[t]
\centerline{\scalebox{1}{\includegraphics[]{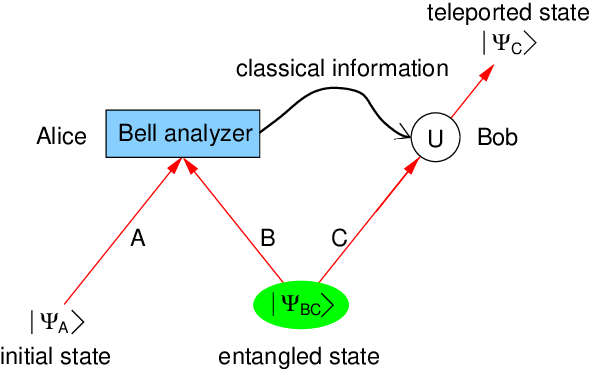}}}
\caption{Principle of quantum teleportation using two-particle
  entanglement. Alice performs a Bell state measurement on the initial
  particle and particle B. After she has sent the measurement result as
  classical information to Bob, he performes a unitary transformation (U) on
  particle C depending on Alice' measurement result to reconstruct the initial
  state.}
\label{pict:Teleportation}
\end{figure}
The idea of of quantum teleportation \cite{Bennett93} is that Alice has a
spin-1/2 particle in a certain quantum state
$\ket{\Psi_A}=a\ket{\su_A}+b\ket{\sd_A}$.  She wishes to transfer this quantum
state to Bob, but she cannot deliver the particle directly to him because
their only connection is via a classical channel. According to the projection
postulate of quantum mechanics we know that any quantum measurement performed
by Alice on her particle will destroy the quantum state at hand without
revealing all the necessary information for Bob to reconstruct the quantum
state. So how can she provide Bob with the quantum state? The answer is to use
an ancillary pair of entangled particles $B$ and $C$ in the singlet state
\begin{equation}
\ket{\Psi^-_{BC}}=\frac{1}{\sqrt{2}}(\ket{\su_B}\ket{\sd_C} 
-\ket{\sd_B}\ket{\su_C})
\end{equation}
shared by Alice and Bob.

Although initially particles $A$ and $B$ are not entangled, their joint state
can always be expressed as a superposition of four maximally entangled Bell
states, given by (\ref{eqn:Bellstates}), since these states form a complete
orthonormal basis. The total state of the three particles can be written as:
\begin{eqnarray}
\ket{\Psi_{ABC}}=\ket{\Psi_A}\otimes\ket{\Psi_{BC}}=\frac{1}{2}&[
&\ket{\Psi^-_{AB}} \otimes(-a \ket{\su_C}-b\ket{\sd_C}) \nonumber \\
&+&\ket{\Psi^+_{AB}} \otimes(-a \ket{\su_C}+b\ket{\sd_C}) \nonumber \\
&+&\ket{\Phi^-_{AB}} \otimes(+a \ket{\sd_C}+b\ket{\su_C}) \nonumber \\
&+&\ket{\Phi^+_{AB}} \otimes(+a\ket{\sd_C}-b\ket{\su_C})].
\end{eqnarray}
Alice now performs a Bell state measurement (BSM) on particles $A$ and $B$,
that is, she projects her two particles onto one of the four Bell states. As a
result of the measurement Bob's particle will be found in a state that is
directly related to the initial state. For example, if the result of Alice's
Bell state measurement is $\ket{\Phi^-_{AB}}$ then particle $C$ in the hands of
Bob is in the state $a\ket{\sd_C}+b\ket{\su_C}$. All that Alice has to do is
to inform Bob via a classical communication channel on her measurement result
and Bob can perform the appropriate unitary transformation (U) on particle $C$
in order to obtain the initial state of particle $A$.

\subsubsection{Entanglement Swapping} 

Entanglement can be realized by having two entangled particles emerge from a
common source, or by allowing two particles to interact with each other. Yet,
another possibility to obtain entanglement is to make use of a projection of
the state of two particles onto an entangled state. This projection
measurement does not necessarily require a direct interaction between the two
particles. When each of the two particles is entangled with another particle,
an appropriate measurement, for example, a Bell-state measurement, of the
partner particles will automatically collapse the state of the remaining two
particles into an entangled state. This striking application of the projection
postulate is referred to as {\it entanglement swapping} \cite{Zukowski93} or
teleportation of entanglement.
\begin{figure}[h]
\centerline{\scalebox{1}{\includegraphics[width=7cm]{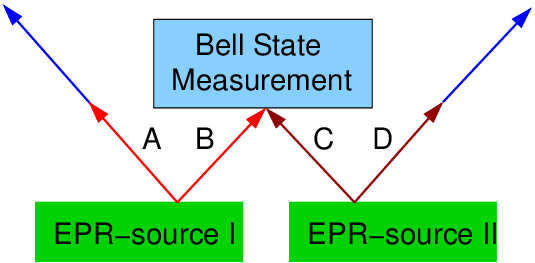}}}
\caption{Principle of entanglement swapping. Two sources produce two pairs of
  entangled particles, pair A-B and pair C-D. One particle from each pair
  (particles B and C) is subjected to a Bell-state measurement. This results
  in a projection of particles A and D onto an entangled state.}
\label{pict:EntSwapping}
\end{figure}

Consider two EPR sources, simultaneously emitting a pair of entangled
particles (Fig.\ref{pict:EntSwapping}) each. We assume that the entangled
particles are in the state
\begin{equation} \label{equ:EntSwapping1}
\ket{\Psi_{ABCD}}=\frac{1}{2}(\ket{\su_A}\ket{\sd_B} 
-\ket{\sd_A}\ket{\su_B})\otimes(\ket{\su_C}\ket{\sd_D} 
-\ket{\sd_C}\ket{\su_D}).
\end{equation}
If one now performs a joint Bell-state measurement on particles $B$ and $C$,
the particles $A$ and $D$ are projected onto a Bell state. This is a
consequence of the fact that the state of (\ref{equ:EntSwapping1}) can be
written as
\begin{eqnarray} \label{equ:EntSwapping2}
\ket{\Psi_{ABCD}}=\frac{1}{2}&(& \ket{\Psi^+_{AD}}\ket{\Psi^+_{BC}} -
                                  \ket{\Psi^-_{AD}}\ket{\Psi^-_{BC}} \nonumber
                                  \\
&-& \ket{\Phi^+_{AD}}\ket{\Phi^+_{BC}} + \ket{\Phi^-_{AD}}\ket{\Phi^-_{BC}}).
\end{eqnarray}
In all cases particles $A$ and $D$ emerge entangled, despite the fact that
they never interacted in the past.

\section{Atom-photon entanglement}

When a single atom is prepared in an excited state $\ket{e}$ it can
spontaneously decay to the ground level $\ket{g}$ and emit a single
photon. Due to conservation of angular momentum in spontaneous emission the
polarization of the emitted photon is correlated with the final quantum state
$\ket{g}$ of the atom. For a simple two-level atom, after spontaneous
emission, the system is in a tensor product state of the atom and the
photon. But for multiple decay channels to different ground states the
resuling state of atom and photon is entangled.

The physical process of spontaneous emission can not be explained by a
semiclassical treatment of the light field but only by a quantum field
approach (this can be found in \cite{Tannoudji98}). I do not intend to give a
sophisticated treatment of spontaneous emission, but I rather give a
phenomenological approach, which will be sufficient to understand atom-photon
entanglement in the context of the present work.

\subsection{Weisskopf-Wigner theory of spontaneous emission}

Let us consider a single atom at time $t=0$ in the excited state $\ket{e}$ and
the field modes in the vacuum state $\ket{0}$. In the {\it ``dressed state
picture''} the state of the system is then given by
$\ket{e,0}=\ket{e}\otimes\ket{0}$, the product of the atomic state $\ket{e}$
and the vacuum state $\ket{0}$ of the electromagnetic field. Due to coupling
of the atom to vacuum fluctuations of the electromagnetic field (this can be
found in \cite{Weisskopf30}) the atom will decay with a characteristic time
constant $\tau$, called the natural lifetime, to the ground state $\ket{g}$
and emit a single photon into the field mode $k$. For $t \rightarrow \infty$
the sytem atom+photon will be in the state $\ket{g,1}$.

\begin{figure}[h]
\centerline{\scalebox{1}{\includegraphics[]{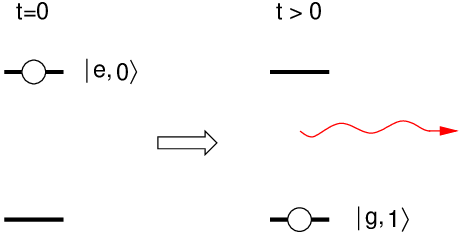}}}
\caption{Spontaneous emission of a photon. At time $t=0$ the atom is in the
  excited state $\ket{e}$. After spontaneous emission the atom passes to the
  ground state $\ket{g,1}$, and the electromagnetic field mode is occupied
  with one photon.}
\label{pict:spontemiss}
\end{figure}

The time evolution of the system is governed by the time-dependent
Schr\"odinger equation
\begin{equation}
\ket{\dot{\Psi}(t)} = -\frac{i}{\hbar} H \ket{\Psi(t)},
\end{equation}
where $H$ denotes the Hamiltonian describing the interaction of a single
two-level atom with a multi-mode radiation field. In the rotating-wave
approximation the simplified Hamiltonian $H$ is given by
\begin{equation} \label{equ:AtomPhoton}
H=\sum_{k} \hbar \omega_k \hat{a}_k^{\dagger} \hat{a}_k
+\frac{1}{2}\hbar\omega_{eg}\hat{\sigma}_z +\hbar \sum_{k} g_k (\hat{\sigma}_+
\hat{a}_k + \hat{a}_k^{\dagger} \hat{\sigma}_- ).
\end{equation}  
This Hamiltonian consists of three parts. The first term in
Eq. (\ref{equ:AtomPhoton}) describes the energy of the free radiation field in
terms of the creation and destruction operators $\hat{a}_k^{\dagger}$ and
$\hat{a}_k$, respectively. The second term $\hbar\omega_{eg}\hat{\sigma}_z/2$
desrcibes the energy of the free atom, whereby $\hat{\sigma}_z$ is given by
$\ket{e}\bra{e}-\ket{g}\bra{g}$. The third term finally characterizes the
interaction energy of the radiation field with the two-level atom. In detail
$\hat{\sigma}_+$ and $\hat{\sigma}_-$ are operators which take the atom from
the lower state to the upper state and vice versa. Hence,
$\hat{a}_k^{\dagger}\hat{\sigma}_-$ describes the process in which the atom
makes a transition from the upper to the lower level and a photon in the mode
$k$ is created, whereas $\hat{\sigma}_+ \hat{a}_k$ describes the opposite
process in which the atom is excited from the lower to the upper level and a
photon is annihilated.

In the Weisskopf-Wigner approximation the eigenstate vector is given
by (see \cite{Scully}, pp. 206)
\begin{equation} \label{equ:Weisskopf}
\ket{\Psi(t)}= e^{-\Gamma t/2} \ket{e,0} + \ket{g} \sum_k W_k
e^{-ik r_0} \left[ \frac{1-e^{i\Delta t-\Gamma
t/2}}{i\Gamma/2+\Delta} \right] \ket{1_k}.
\end{equation}
Here, the form of the probability amplitude of the state $\ket{e,0}$ signals
that an atom in the excited state $\ket{e}$ in vacuum decays exponentially
with the lifetime $\tau=1/\Gamma$ and emits a photon of angular frequency
$\omega_{k}$. The probability amplitude of the state $\ket{1_k}$ describes the
temporal occupation of the modes $k$ of the radiation field, where $W_k$
denotes the overlap between the atomic states $\ket{g}$ and $\ket{e}$ in the
field mode $k$, and $\Delta=\omega_k-\omega_{eg}$ the detuning in respect to
the atomic transition frequency $\omega_{eg}$. For times long compared to the
radiative decay the first term in (\ref{equ:Weisskopf}) is negligible and the
state of the system is given by a linear superposition of single-photon states
with different wave vectors.

\subsection{Properties of the emitted photon}

Because atomic states are eigenstates of the total angular momentum, the modes
of the electromagnetic field after spontaneous emission are also eigenstates
of angular momentum. Therefore the polarization state of a photon
spontaneously emitted from an atomic dipole depends upon the change in angular
momentum $\Delta m$ of the atom along the dipole axis, and the direction of
the emission. For $\Delta m=0,\pm 1$, the polarization state of a
spontaneously emitted photon is
\begin{eqnarray}
\ket{\Pi_0}=\sin{\theta} &\ket{\pi}& \quad \hbox{for} \quad \Delta m = 0 \\
\ket{\Pi_{\pm1}}=\sqrt{1+\cos^2{\theta}}/\sqrt{2} &\ket{\sigma^{\pm}}& \quad
\hbox{for} \quad \Delta m = \pm 1,
\end{eqnarray}
where $\theta$ is the spherical polar angle with respect to the dipole
(quantization) axis, and $\ket{\pi}$ and $\ket{\sigma^{\pm}}$ denote the
polarization state of the photon. Note that along a viewing axis parallel to
the dipole ($\theta=0$) only $\ket{\sigma^{\pm}}$-polarized radiation is
emitted.

\begin{figure}[t]
\centerline{\scalebox{1}{\includegraphics[width=10cm]{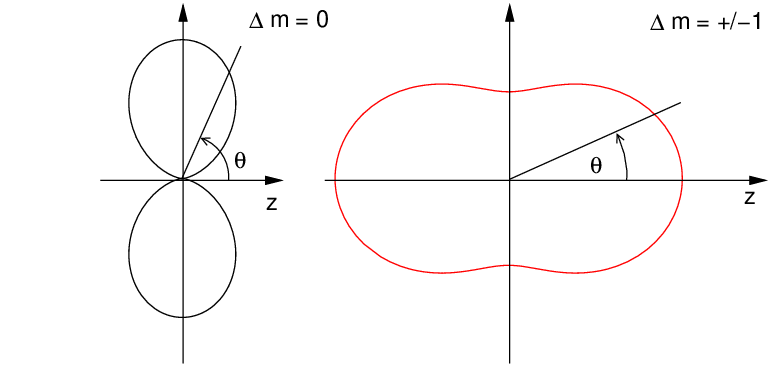}}}
\caption{Emission characteristics of light emitted from dipole transitions
  with $\Delta m=0,\pm1$.}
\label{pict:emissionchar}
\end{figure}

\subsection{Spontaneous emission as a source of atom-photon entanglement}

\begin{figure}[t]
\centerline{\scalebox{1}{\includegraphics[width=7cm]{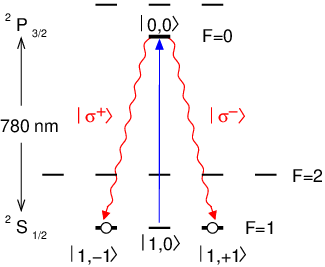}}}
\caption{Atomic level structure in $^{87}$Rb used to generate atom-photon
  entanglement. Provided the emission frequencies for $\sigma^+$, $\sigma^-$
  and $\pi$ polarized transitions are indistinguishable within the natural
  linewidth of the transition, the polarization of a spontaneously emitted
  photon will be entangled with the spin state of the atom.}
\label{pict:lambda}
\end{figure}

Until now we considered a single atom in free space which spontaneously emits
a photon from a two-level transition. According to Weisskopf and Wigner the
state of the system atom+photon is a simple tensor product state of the form
$\ket{g}\ket{\Pi_{\Delta m}}$, where the atom is in the ground state $\ket{g}$
and the photon in the state $\ket{\Pi_{\Delta m}}$. For multiple decay
channels of spontaneous emission to different ground states of atomic angular
momentum $F$, the resulting (unnormalized) state of photon and atom is
\begin{equation} \label{equ:AP}
\ket{\Psi}=\sum_{\Delta m} C_{\Delta m} \ket{F_{\Delta m}} \ket{\Pi_{\Delta
    m}},
\end{equation}
where $C_{\Delta m}$ are atomic Clebsch-Gordon coefficients for the possible
decay channels and $\ket{F_{\Delta m}}$ denote the respective atomic ground
states. The state in (\ref{equ:AP}) can not be represented as a tensor
product, only as a linear superposition of different product states. Therefore
the spin degree of freedom of the atom and the polarization of the photon are
entangled.

In the current experiment we excite a single $^{87}$Rb atom to the $^2P_{3/2},
\ket{F=0,m_F=0}$ state by a short optical $\pi$-pulse. In the following
spontaneous emission the atom decays either to the $\ket{1,-1}$ ground state
while emitting a $\ket{\sigma^+}$-polarized photon, or to the $\ket{1,0}$
state while emitting a $\ket{\pi}$-polarized photon, or to the $\ket{1,+1}$
ground state and emits a $\ket{\sigma^-}$-polarized photon. Provided these
decay channels are spectrally indistinguishable, a coherent superposition of
the three possibilities is formed and the magnetic quantum number $m_F$ of the
atom is entangled with the polarization of the emitted photon resulting in the
atom-photon state
\begin{equation} \label{equ:AtomPhotonGeneral}
\ket{\Psi}=\frac{1}{\sqrt{2}} \left[\sqrt{(1+\cos^2{\theta})/2}
  \left(\ket{1,-1}\ket{\sigma^+} + \ket{1,+1}\ket{\sigma^-}\right) +
  \sin{\theta}\ket{1,0}\ket{\pi}\right],
\end{equation}
where the first index in the atomic basis state $\ket{F,m_F}$ denotes the
total angular momentum $F$ and the second index indicates the respective
magnetic quantum number $m_F$. 

If we now put a single photon detector in the far-field region of the atom and
detect the spontaneously emitted photon along the quantization axis $z$ -
defined by the optical axis of the detection optics - then the
$\ket{\pi}$-polarized light is not detected (see
Fig. \ref{pict:emissionchar}). Thus the resulting atom-photon state is
maximally entangled:
\begin{equation}
\ket{\Psi^+}=\frac{1}{\sqrt{2}}(\ket{1,-1}\ket{\sigma^+}
+\ket{1,+1}\ket{\sigma^-}).
\end{equation}

\subsection{Experimental proof of atom-photon entanglement}

To verify atom-photon entanglement in an experiment, one has to disprove the
alternative description of the system being in a statistical mixture of
seperable states. Experimentally one has to determine the diagonal density
matrix elements in at least two complementary measurement bases (see
sect. \ref{sect:QE}). The choice of the measurement basis can be realized in
two ways. First, the atomic and photonic spin-state is rotated by an active
unitary transformation into the new measurement basis. Second, the atomic and
photonic spin-state stay unchanged, but the spin-analyzer is rotated by a
respective passive transformation. For photons, the state-measurement can be
realized relatively simple by a rotable birefringent waveplate followed by a
polarizer. For atomic spin-states $\ket{\su}_z=\ket{1,-1}$ and
$\ket{\sd}_z=\ket{1,+1}$ active rotations can be realized by suitable optical
Raman laser pulses that perform the transformation:
\begin{eqnarray} \label{equ:statetransform}
\ket{\su}_z+e^{i\phi}\ket{\sd}_z &\rightarrow& \ket{\su}_z \\
\ket{\su}_z-e^{i\phi}\ket{\sd}_z &\rightarrow& \ket{\sd}_z,
\end{eqnarray}
where $\phi$ is the relative geometric phase between the basis states
$\ket{\su}_z$ and $\ket{\sd}_z$.

In the present experiment the atomic state detection is realized by a suitable
Stimulated-Raman-Adiabatic-Passage (STIRAP) laser pulse which transfers,
e.g. the superposition state $\ket{\su}_z+e^{i\phi}\ket{\sd}_z$ to the
hyperfine ground state $F=2$, whereas the orthogonal superposition state
$\ket{\su}_z-e^{i\phi}\ket{\sd}_z$ does not couple to the laser field and
remains ``dark''. By scattering light from a preceding laser pulse which
couples only to $F=2$ both states can be identified with nearly perfect
efficiency (see chapter \ref{chapter:AtomicStateDetection}).

\section{Summary}

In this chapter I gave a short theoretical review about the properties of
entangled states. A discussion about fundamental tests of quantum
mechanics on the basis of Bell's inequalities is presented and applications of
entanglement in quantum information and communication is discussed. Finally
the spontaneous decay of single atoms is introduced as a source for the
generation of entangled atom-photon states.

%-----------------------------------------------------------------------
%-----------------------------------------------------------------------

\chapter{Single atom dipole trap} \label{chapter:diptrap}
\section{Introduction}

To investigate the nonclassical correlation properties of an entangled
atom-photon pair experimentally it is necessary to isolate a single atom and
to detect the atom and the spontaneously emitted photon with high
efficiency. For this reason it is convenient to localize the atom in a region
of a few optical wavelengths. This experimental requirement can be
accomplished with various kinds of traps, but due to the intrinsic properties
of the generation process of atom-photon entanglement there exist important
constraints on the trapping mechanism: (1) All magnetic substates of the
atomic ground state have to experience the same binding potential; (2) The
energy splitting of the atomic qubit states has to be less than the natural
linewidth of the transition in order to fulfill the requirement of spectral
indistinguishability; (3) The trap must preserve the internal state of the
atom.

\begin{figure}[h]
\centerline{\scalebox{1}{\includegraphics[width=10cm]{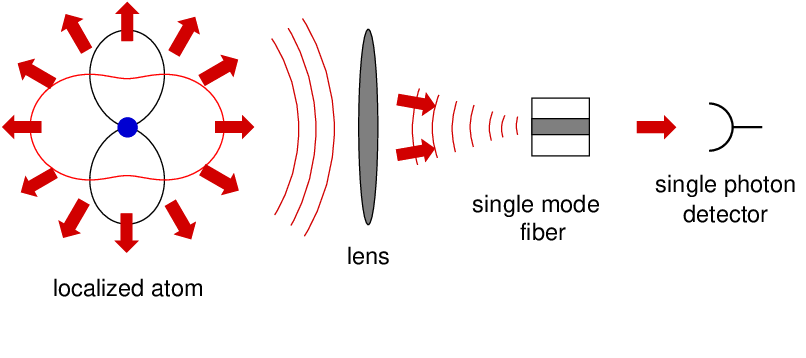}}}
\caption{Schematic setup for the detection of light emitted from a single
  atom.}
\label{pict:spont_emiss}
\end{figure}

Furthermore, the present experiment should be designed such, that it should
allow in a next step the faithful generation of entanglement between
space-like separated atoms by the interference of photons emitted from the
atoms. A possible way to do that is to use single ions stored in
electrodynamical traps \cite{Blinov04} that are separated by large
distances. But, because most ions radiate photons from atomic transitions in
the visible and ultraviolet range of the electromagnetic spectrum the
transport of emitted photons over large distances is complicated by high
transmission losses in optical fibers and air. This problem can be avoided by
the use of single trapped $^{87}$Rb atoms which radiate light in the
near-infrared (NIR). Different kinds of traps for neutral atoms have been
realized in the last 20 years, but not all trapping mechanisms are applicable
for the investigation of atom-photon and atom-atom entanglement. {\it Magnetic
traps} \cite{Migdall85,Bergeman87} are based on the state-dependent force of
the magnetic dipole moment in an inhomogenous field. Hence, this kind of trap
is unsuitable for trapping atomic spin states with different sign. In
addition, magnetic traps do not preserve the spectral indistinguishability of
the possible photonic emission paths due to different Zeeman-splitting of the
magnetic substates $m_F=-1$ and $m_F=+1$. The trapping mechanism of {\it
Magneto-optical traps} (MOT) \cite{Pritchard86,Raab87} relies on near resonant
scattering of light destroying any atomic coherence on a timescale of the
excited state lifetime.  Optical dipole traps - first proposed by Letokhov in
1968 \cite{Askaryan62,Letokhov68} - rely on the electric dipole interaction
with far-detuned light at which the optical excitation can be kept extremely
low. In optical dipole traps atomic coherence times up to several seconds are
possible \cite{Frese00}, and under appropriate conditions the trapping
mechanism is independent of the particular magnetic sub-level of the
electronic ground state. Therefore a localized single $^{87}$Rb atom in a
far-off-resonance optical dipole trap satisfies all necessary requirements for
the investigation of atom-photon and atom-atom entanglement.
\newline

In the context of this thesis our group has set up a microscopic optical
dipole trap, which allows to trap single Rubidium atoms one by one with a
typical trap life- and coherence time of several seconds. In order to
understand the operating mode of our trap I will first indroduce the basic
theoretical concepts of atom trapping in optical dipole potentials. Then I
will describe the experimental setup of our dipole trap and measurements which
prove the subpoissonian occupation statistics. Finally the mean kinetic energy
of a single atom was determined via the spectral analysis of the emitted
fluorescence light.

\section{Theory of optical dipole traps for neutral atoms} 
\label{section:DipoleTraps}

Following Grimm et al. \cite{Grimm00} I will introduce the basic concepts of
atom trapping in optical dipole potentials that result from the interaction
with {\it far-detuned} light. In this case of particular interest, the optical
excitation is very low and the radiation force due to photon scattering is
negligible as compared to the dipole force. In
Sec. \ref{subsect_dipole:potentials}, I consider the atom as a simple
classical or quantum-mechanical oscillator to derive the main equations for
the optical dipole interaction. Then the case of real multi-level atoms is
considered, which allows to calculate the dipole potential of the $^2S_{1/2}$
ground and $^2P_{3/2}$ excited state in $^{87}$Rb. Atom trapping in dipole
potentials requires cooling to load the trap and eventually also to counteract
heating in the dipole trap. I therefore briefly review two laser cooling
methods and their specific features with respect to our experiment. Then I
discuss sources of heating, and derive explicit expressions for the heating
rate in the case of thermal equilibrium in a dipole trap. Finally I will
present a simple model which allows to understand the atom number limitation
in a small volume dipole trap due to light-induced two-body collisions. This
effect opens the possibility to trap only single atoms and simplifies the
experimental investigation of an entangled atom-photon state.
   
\subsection{Optical dipole potentials} \label{subsect_dipole:potentials}
\subsubsection{Oscillator model}

The optical dipole force arises from the dispersive interaction of the induced
atomic dipole moment with the intensity gradient of the light field
\cite{Askaryan62}. Because of its conservative character, the force can be
derived from a potential, the minima of which can be used for atom
trapping. The absorptive part of the dipole interaction in far-detuned light
leads to residual photon scattering of the trapping light, which sets limits
to the performance of dipole traps. In the following I will derive the basic
equations for the dipole potential and the scattering rate by considering the
atom as a simple oscillator subject to the classical radiation field.

When an atom is irradiated by laser light, the electric field ${\bf E}$
induces an atomic dipole moment ${\bf p}$ that oscillates at the driving
frequency $\omega$. In the usual complex notation ${\bf{E}}({\bf{r}},t)=
\hat{{\bf{e}}}\hat{E}({\bf{r}})e^{i\omega t}+\hat{{\bf{e}}}\hat{E}^*
({\bf{r}})e^{-i\omega t}$ and
${\bf{p}}({\bf{r}},t)=\hat{{\bf{e}}}\hat{p}({\bf{r}})e^{i\omega
t}+\hat{{\bf{e}}}\hat{p}^* ({\bf{r}})e^{-i\omega t}$, where $\hat{{\bf{e}}}$
is the unit polarization vector, the amplitude $\hat{p}$ of the dipole moment
is simply related to the field amplitude $\hat{E}$ by
\begin{equation} \label{equ:polarizability1} 
\hat{p} = \alpha(\omega) \hat{E}.
\end{equation}
Here $\alpha$ is the {\it complex polarizability}, which depends on the
driving frequency $\omega$. 

The interaction potential of the induced dipole moment ${\bf p}$ in the
driving field ${\bf E}$ is given by
\begin{equation} \label{equ:dippot1}
U_{dip}= - \frac{1}{2} \langle{\bf{p}}{\bf{E}}\rangle = - \frac{1}{2 \epsilon_0
  c}{\rm{Re}}(\alpha)I
\end{equation} 
where $<>$ denote the time average over the rapid oscillating
terms. The laser field intensity is 
\begin{equation} \label{equ:intensity}
I = \frac{1}{2} \epsilon_0 c |\hat{E}|^2,
\end{equation}
and the factor $1/2$ takes into account that the dipole moment is an induced,
not a permanent one. The potential energy of the atom in the field is thus
proportional to the intensity $I$ and to the real part of the polarizability,
which describes the in-phase component of the dipole oscillation being
responsible for the dispersive properties of the interaction. The {\it dipole
force} results from the gradient of the interaction potential
\begin{equation} \label{equ:dipforce}
{\bf{F}}_{dip} ({\bf{r}}) = - \nabla V_{dip}({\bf{r}}) = \frac{1}{2 \epsilon_0
c} {\rm{Re}}(\alpha) \nabla I({\bf{r}}).
\end{equation} 
It is thus a conservative force, proportional to the intensity gradient of the
driving field.

The power absorbed by the oscillator from the driving field (and re-emitted as
dipole radiation) is given by
\begin{equation} \label{equ:powerabs}
P_{abs}= \langle\dot{\bf{p}} {\bf{E}} \rangle =\frac{\omega}{\epsilon_0
  c}{\rm{Im}}(\alpha)I({\bf{r}}). 
\end{equation} 
The absorption results from the imaginary part of the polarizability, which
describes the out-of-phase component of the dipole oscillation. Considering
the light as a stream of photons $\hbar \omega$, the absorption can be
interpreted in terms of photon scattering in cycles of absorption and
subsequent spontaneous reemission processes. The corresponding ${\it
scattering rate}$ is
\begin{equation} \label{equ:scatter1}
\Gamma_{Sc}({\bf{r}}) = \frac{P_{abs}}{\hbar \omega} = \frac{1}{\hbar
\epsilon_0 c} {\rm{Im}}(\alpha) I({\bf{r}}).
\end{equation} 

We have now expressed the two main quantities of interest for dipole traps,
the {\it interaction potential} and the {\it scattered radiation power}, in
terms of the position dependent intensity $I({\bf r})$ and the polarizibility
$\alpha(\omega)$. These expressions are valid for any polarizable neutral
particle in an oscillating electric field. This can be an atom in a
near-resonant or far off-resonant laser field, or even a molecule in an
optical or microwave field.

In order to calculate its polarizability $\alpha$, I first consider the atom
in Lorenz's model as a classical oscillator (see, e.g. \cite{Jackson62}). In
this simple and very useful picture, an electron (mass $m_e$, elementary
charge $e$) is considered to be bound elastically to the core with an
oscillation eigenfrequency $\omega_0$ corresponding to the optical transition
frequency. Damping results from the dipole radiation of the oscillating
electron according to Larmor's well-known formula for the power radiated by an
accelerated charge.

It is straightforward to calculate the polarizability by integration of the
equation of motion
\begin{equation}
\ddot{x}+ \Gamma_{\omega} \ \dot{x} + \omega_0^2x = - \frac{e E(t)}{m_e}
\end{equation} 
for the driven oscillation of the electron with the result
\begin{equation}  \label{equ:polarizability}
\alpha = \frac{e^2}{m_e}\frac{1}{\omega_0^2 - \omega^2 - i \omega
  \Gamma_{\omega}}.
\end{equation}
Here 
\begin{equation} \label{equ:gamma1}
\Gamma_{\omega} = \frac{e^2 \omega^2}{6 \pi \epsilon_0 m_e c^2}
\end{equation} 
is the classical damping rate due to the radiative energy loss. Introducing
the on-resonance damping rate $\Gamma \equiv \Gamma_{\omega_0} =
(\frac{\omega_0}{\omega})^2 \Gamma_{\omega}$, one can put
Eq. \ref{equ:polarizability} into the form
\begin{equation} \label{equ:polarizability2}
\alpha(\omega) = 6 \pi \epsilon_0 c^3 \frac{\Gamma/\omega_0^2}{\omega_0^2 -
  \omega^2 - i(\omega^3/\omega_0^2)\Gamma}.
\end{equation} 

In a {\it semiclassical approach} the atomic polarizability can be calculated
by considering the atom as a two-level quantum system interacting with a
classical radiation field. When saturation effects can be neglected, the
semiclassical calculation yields the same result as the classical calculation
with only one modification, the damping rate $\Gamma$ (corresponding the
spontaneous decay rate of the excited state) is determined by the dipole
matrix element $\langle e |\mu| g \rangle$ between the ground state
$|g\rangle$ and the excited state $|e\rangle$ by
\begin{equation} \label{equ:gamma2}
\Gamma = \frac{\omega_0^3}{3 \pi \epsilon_0 \hbar
c^2}|\langle{e}|\mu|g\rangle|^2.
\end{equation} 
For many atoms with a strong dipole-allowed transition starting from its
ground state, the classical formula Eq. \ref{equ:gamma1} nevertheless provides
a good approximation to the spontaneous decay rate. For the {\it D} lines of
the alkali atoms Na, K, Rb, and Cs, the classical result agrees with the true
decay rate to within a few percent.

An important difference between the quantum-mechanical and the classical
oscillator is the possible occurence of saturation. At very high intensities of
the driving field, the excited state gets strongly populated and the classical
result (Eq. \ref{equ:polarizability2}) is no longer valid. For dipole
trapping, however one is essentially interested in the far-detuned case with
very low saturation and thus very low scatttering rates ($\Gamma_{Sc} \ll
\Gamma$). Thus the expression in Eq. \ref{equ:polarizability2} for the atomic
polarizability can be used as an excellent approximation for the
quantum-mechanical oscillator and the explicit expressions for the dipole
potential $U_{dip}$ and the scattering rate $\Gamma_{Sc}$ can be derived.
\begin{equation} \label{equ:dipolepot}
U_{dip}({\bf{r}}) = - \frac{3 \pi c^2}{2
  \omega_0^3}\left(\frac{\Gamma}{\omega_0-\omega}+ \frac{\Gamma}{\omega_0 +
  \omega}\right)I({\bf{r}})
\end{equation} 
\begin{equation} \label{equ:scatterrate}
\Gamma_{Sc}({\bf{r}}) =  \frac{3 \pi c^2}{2 \hbar
  \omega_0^3}\left(\frac{\omega}{\omega_0}\right)^3\left(\frac{\Gamma}{\omega_0-\omega}+ \frac{\Gamma}{\omega_0 + \omega}\right)^2I({\bf{r}})
\end{equation} 
These general expressions are valid for any driving frequency $\omega$ and
show two resonant contributions: Besides the usually considered resonance at
$\omega = \omega_0$, there is also the so-called counter-rotating term
resonant at $\omega = -\omega_0$.

In most experiments, the laser is tuned relatively close to the resonance
$\omega_0$ such that the detuning $\Delta = \omega - \omega_{0}$ fulfills
$|\Delta| \ll \omega_0$. In this case, the counter-rotating term can be
neglected in the well-known {\it rotating wave approximation} and the general
expressions for the dipole potential and the scattering rate simplify to
\begin{equation} \label{equ:dipolepotnear}
U_{dip}({\bf{r}}) = \frac{3 \pi c^2}{2
\omega_0^3}\left(\frac{\Gamma}{\Delta}\right)I({\bf{r}}),
\end{equation}
\begin{equation} \label{equ:scatterratenear}
\Gamma_{Sc}({\bf{r}}) = \frac{3 \pi c^2}{2 \hbar
\omega_0^3}\left(\frac{\Gamma}{\Delta}
\right)^2I({\bf{r}}).
\end{equation}
The basic physics of dipole trapping in far-detuned laser fields can be
understood on the basis of these two equations. Obviously, a simple relation
exists between the scattering rate and the dipole potential,
\begin{equation}
\hbar \Gamma_{Sc} = \frac{\Gamma}{\Delta}U_{dip}.
\end{equation}
Equation \ref{equ:dipolepotnear} and \ref{equ:scatterratenear} show two
essential points for dipole trapping. First: below an atomic resonance
(``red'' detuned, $\Delta < 0$) the dipole potential is negative and the
interaction thus attracts atoms into the light field. Potential minima are
therefore found at positions with maximum intensity. Above resonance (``blue''
detuned, $\Delta > 0$) the dipole interaction repels atoms out of the field,
and potential minima correspond to minima of the intensity. Second: the dipole
potential scales as $I/\Delta$, whereas the scattering rate scales as
$I/\Delta^2$. Therefore, optical dipole traps usually use large detunings and
high intensities to keep the scattering rate as low as possible for a certain
potential depth.

\subsubsection{Dressed state picture}

In terms of the oscillator model, multi-level atoms can be described by
state-dependent atomic polarizabilies. An alternative description of dipole
potentials is given by the {\it dressed state picture} \cite{Tannoudji98}
where the atom is considered together with a quantized light field. In its
ground state the atom has zero internal energy and the field energy is $n\hbar
\omega$ depending on the number of photons. When the atom is put into an
excited state by absorbing a photon, the sum of its internal energy $\hbar
\omega_0$ and the field energy $(n-1)\hbar \omega$ becomes $\it{E_j}= \hbar
\omega_0 + (n-1)\hbar \omega = -\hbar \Delta_{ij} + n\hbar \omega$.

For an atom interacting with laser light, the interaction Hamiltonian is
$H_{int}=-\hat{\bf{\mu}} \bf{E}$ with $\hat{\bf{\mu}}= -e \bf{r}$ representing
the electric dipole operator. The effect of far-detuned laser light on the
atomic levels can be treated as a time-independent perturbation in the second
order of the electric field. Applying this perturbation theory for
nondegenerate states, the interaction Hamiltonian $H_{int}$ leads to an energy
shift of the $i$-th state with the unperturbed energy $\it{E_i}$, that is
given by
\begin{equation} \label{equ:lightshift1}
\Delta E_i = \sum_{j\neq i}
  \frac{|\bra{j}{\it{H_{int}}}\ket{i}|^2}{{\it{E_i}} - {\it{E_j}}}.
\end{equation}
For a {\it two-level atom}, this equation simplifies to 
\begin{equation}
\Delta E =\pm \frac{|\bra{e}\mu\ket{g}|^2}{\Delta} |E|^2= \pm \frac{3 \pi
  c^2}{2 \omega_0^3}\,\frac{\Gamma}{\Delta}I
\end{equation}
with the plus and the minus sign for the excited and ground state,
respectively. This optically induced energy shift (known as {\it Light Shift}
or {\it AC-Stark shift}) of the ground-state exactly corresponds to the dipole
potential for the two-level atom in Eq. \ref{equ:dipolepotnear}.
\begin{figure}[t]
\centerline{\scalebox{1}{\includegraphics[width=5cm]{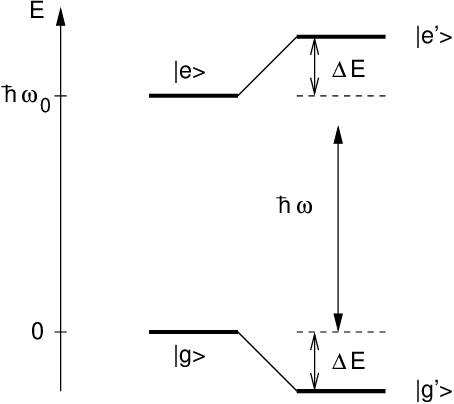}}}
\caption{Light shifts for a two-level atom. Red-detuned light ($\Delta < 0$)
  shifts the ground state $\ket{g}$ down and the excited state $\ket{e}$ up by
  the energy $\Delta$E. 
}\label{pict:twolevellightshift}
\end{figure}

\subsubsection{Semiclassical treatment of multi-level atoms}

When calculating the {\it light shift} on the basis of the semiclassical
treatement for a multi-level atom in a specific electronic ground state
$\ket{i}$ all dipole allowed transitions to the excited states $\ket{f}$ have
to be taken into account. Time-dependent perturbation theory yields an
expression for the energy shift of the atomic levels $\ket{i}$ - characterized
by their eigenenergies $E_i$ - due to interaction with a time-dependent
classical electric field with angular frequency $\omega$, which is given by
\cite{Shore}
\begin{equation} \label{equ:lightshiftgeneral}
\Delta E_{i}(\omega)=\frac{-|\hat{E}|^2}{4 \hbar} \sum_{f \neq i} |\bra{i}{e\bf
   r}\ket{f}|^2 \left( \frac{1}{\omega_{if}-\omega} +
  \frac{1}{\omega_{if}+\omega} \right), 
\end{equation}
where the sum covers all atomic states $\ket{f}$ except for the initial state
$\ket{i}$, and $\omega_{if} = \omega_f-\omega_i$ denotes the atomic transition
frequencies. 

The transition matrix elements $\bra{i}{e\bf r}\ket{f}$ in general depend on
the quantum numbers of the initial state $\ket{i}$ respresented by $n,J,F,m_F$
and $n',J',F,m_{F}'$ for the final states $\ket{f}$ and upon the laser
polarization ($\epsilon=0,\pm 1$ for $\pi$ and $\sigma^\pm$ transitions
respectively). Using the Wigner-Eckart theorem \cite{Sobelman}, the matrix
elements $\bra{i}e{\bf r}\ket{f}$ can be expressed as a product of a real {\it
Clebsch-Gordan coefficient} $\bra{F,m_F}F',1,m_F',\epsilon\rangle$ and a
reduced matrix element $\bra{F}|e{\bf r}|\ket{F'}$:
\begin{equation} \label{equ:WignerEckart}
\bra{i}e{\bf r}\ket{f}=\bra{F}|e{\bf
  r}|\ket{F'}\bra{F,m_F}F',1,m_F',\epsilon \rangle.
\end{equation}
This reduced matrix element can be further simplified by factoring out the $F$
and $F'$ dependence into a Wigner $6j$ symbol, leaving a {\it fully reduced
matrix element} $\bra{J}|e{\bf r}|\ket{J'}$ that depends only on the
electronic orbital wavefunctions:
\begin{equation}
\bra{F}|e{\bf r}|\ket{F'}=\bra{J}|e{\bf 
  r}|\ket{J'}(-1)^{F'+J+I+1}\sqrt{(2F'+1)(2J+1)} 
  \left( \begin{array}{ccc} J & J' & 1 \\ F' & F & I \end{array} \right)_{6j} .
\end{equation}
The remaining fully reduced matrix element $\bra{J}|e{\bf r}|\ket{J'}$
can be calculated from the lifetime of the transition $J \rightarrow
J'$ via the expression \cite{Steck87}
\begin{equation} \label{equ:lifetime}
\frac{1}{\tau_{JJ'}}=\frac{\omega_{if}^3}{3 \pi \epsilon_0 \hbar c^3}
               \frac{2J+1}{2J'+1} |\bra{J}|e{\bf r}|\ket{J'}|^2.
\end{equation}
On the basis of Eq. \ref{equ:lightshiftgeneral}, the {\it Wigner-Eckart}
theorem and Eq. \ref{equ:lifetime} one can derive a general expression for the
light-shift of any atomic hyperfine state $\ket{J,F,m_F}$ coupling to the
manifold of final states $\{\ket{J',F',m_F'}\}_f$, which is given by

\begin{equation} \label{equ:lightshiftgeneral2}
\Delta E(J,F,m_F) = \frac{3 \pi c^2 I}{2}
\sum_{J',F',m_F'} \frac{(2J'+1)(2F'+1)}{\tau_{JJ'} \Delta_{FF'}'
\omega_{FF'}^3}  \left(
\begin{array}{ccc} J & J' & 1 \\ F' & F & I \end{array} \right)_{6j}^2
|\bra{F,m_F}F',1,m_F',\epsilon \rangle|^2.
\end{equation}
For simplicity we have introduced the effective detuning
$\Delta_{FF'}'$:
\begin{equation} \label{equ:effectivedetuning}
\frac{1}{\Delta_{FF'}'}=\frac{1}{\omega_{FF'}-\omega}+\frac{1}{\omega_{FF'}+
      \omega}, 
\end{equation}
where $\omega_{FF'}$ denotes the atomic transition frequency and
$\omega$ is the frequency of the classical light field.

\subsubsection{Light-shift of the 5$^2S_{1/2}$ and 5$^2P_{3/2}$ state}

\begin{figure}[t]
\centerline{\scalebox{1}{\includegraphics[width=7cm]{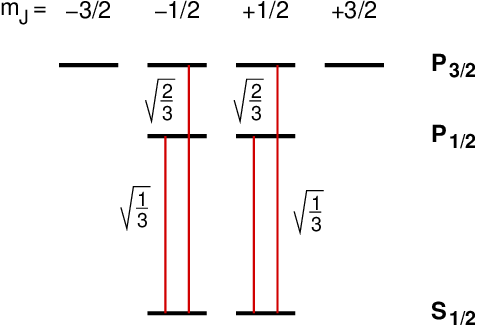}}}
\caption{Reduced level scheme in $^{87}$Rb. Dipole matrix elements for
$\pi$-polarized light are shown as multiples of $\bra{J}|e{\bf r}|\ket{J'}$.
}\label{pict:ClebschGordan}
\end{figure}
In the case of a resolved fine-structure, but unresolved hyperfine structure,
one can consider the atom in spin-orbit coupling, neglecting the coupling of
the nuclear spin. The interaction with the laser field can thus be considered
in the electronic angular momentum configuration of the D lines, $J= 1/2
\rightarrow J'= 1/2, 3/2$. 

In this situation, illustrated in Fig. \ref{pict:ClebschGordan}, one can first
calculate the light shifts of the two electronic ground states $m_J=\pm 1/2$
with the simplified relation
\begin{equation} \label{equ:lightshiftground1}
\Delta E(J,m_j,\omega) = \frac{-|\hat{E}|^2}{4 \hbar} \sum_{J'}
  \frac{|\bra{J}|e {\bf r}|\ket{J'}|^2}{\Delta_{JJ'}} \sum_{m_J'}
  |\bra{J,m_J}J',1,m_J' ,\epsilon\rangle|^2.
\end{equation}
\begin{figure}[t]
\centerline{\scalebox{1}{\includegraphics[width=12cm]{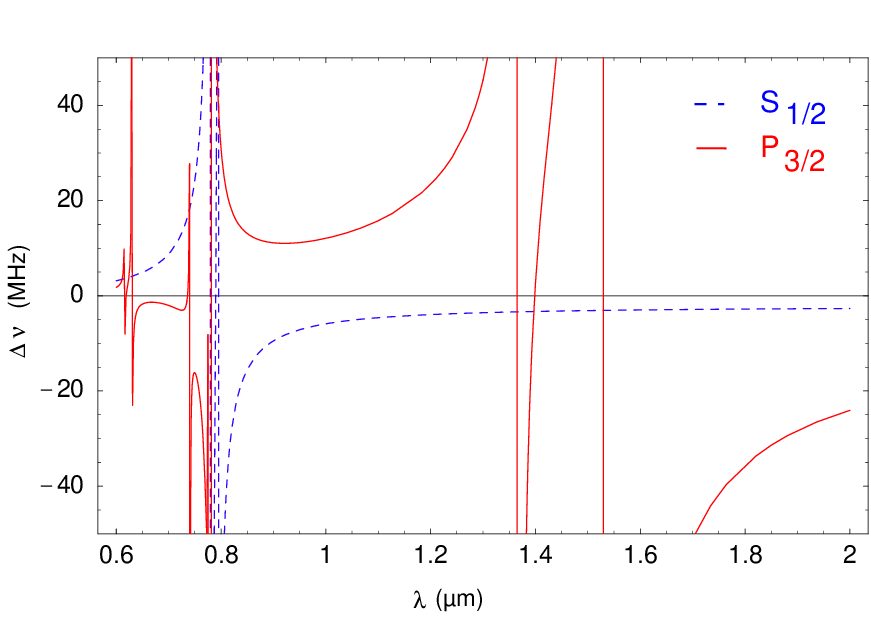}}}
\caption{5$^2S_{1/2}$ ground- (dashed blue line) and 5$^2P_{3/2}$
excited-state (solid red line) light-shift $\Delta \nu$ as a function of the
dipole laser wavelength $\lambda$. At 1.4 $\mu$m both states have the same
light shift. In contrast to the ground state, the excited state Zeeman sub
levels $m_F'$ experience different light shifts. Here, we assumed equal
occupation of the excited state Zeeman sub-levels and calculated the mean
light shift.}\label{pict:magicpoint}
\end{figure}
On the basis of this equation, one can derive a general result for the
potential of a ground state with total angular momentum $J$ and magnetic
quantum number $m_J$, which is valid for both linear and circular polarization
as long as all optical detunings stay large compared to the ground- and
excited-state hyperfine structure splitting: 
\begin{equation} \label{equ:lightshiftRb}
U_{dip}({\bf r}) = \frac{3\pi c^2}{2} \left( \frac{\Gamma_{1/2} (1-\epsilon
g_J m_J)}{3 \omega_{1/2}^3 \Delta_{1/2} } + \frac{\Gamma_{3/2} (2+\epsilon g_J
m_J)}{3 \omega_{3/2}^3 \Delta_{3/2}} \right) I({\bf r}).
\end{equation}
Here $g_J=2$ is the well known Lande factor for alkali atoms and $\epsilon$
characterizes the laser polarization ($\epsilon=0,\pm 1$ for linearly and
$\sigma^{\pm}$ polarized light). The detunings $\Delta_{1/2}, \Delta_{3/2}$ of
the laser frequency are referred to the atomic transitions $S_{1/2},J
\rightarrow P_{1/2}$ and $S_{1/2},J \rightarrow P_{3/2}$ (the $D_1$ and $D_2$
line).

For {\it linear polarization}, both electronic ground states $m_J = \pm 1/2$
are shifted by the same amount because of simple symmetry reasons. After
coupling to the nuclear spin, the resulting $F, m_F$ states have to remain
degenerate like the two original $m_J$ states. Consequently, all magnetic
sublevels show the same light shifts.

For {\it circular polarization}, the far-off-resonance light field lifts the
degeneracy of the two magnetic sublevels of the electronic $^2S_{1/2}$ ground
state. In this sense, the circularly polarized light acts like a {\it
fictitious magnetic field}.

With the help of Eq. \ref{equ:lightshiftgeneral2} and the knowledge of the
lifetimes of all coupling energy levels, it is possible to calculate the light
shift of any Zeeman sub-level in the hyperfine structure of the states
$5^2S_{1/2}$, $5^2P_{1/2}$ and $5^2P_{3/2}$. Due to the coupling of the
excited states $5^2P_{1/2}$ and $5^2P_{3/2}$ to even higher states (see
Fig. \ref{pict:Rb87termschema}), their energy shifts may have the same sign as
the ground state energy shift. Thus, the dipole force can be attractive even
for an atom in an excited state (see Fig. \ref{pict:magicpoint}), in contrast
to the simplified two-level model of Fig. \ref{pict:twolevellightshift}. There
exists a {\it magic} wavelength for a linear polarized dipole laser field, for
which the excited state $5^2P_{3/2}$ and the ground state are equally shifted
(see Fig. \ref{pict:magicpoint}). This wavelength amounts to approximately
$\lambda=1.4$ $\mu$m, enabling state-insensitive optical cooling in the dipole
potential.

For the photon scattering rate $\Gamma_{Sc}$ of Rubidium, the same line
strength factors are relevant as for the ground state dipole potential, since
absorption and light shifts are determined by the same transition matrix
elements. For linear polarization one explicitely obtains
\begin{equation} \label{equ:scatterrategeneral}
\Gamma_{Sc}({\bf r}) = \frac{\pi c^2}{2 \hbar} \left( \frac{\Gamma_{1/2}^2}{
  \omega_{1/2}^3 \Delta_{1/2}^2 } + \frac{2 \Gamma_{3/2}^2}{\omega_{3/2}^3
  \Delta_{3/2}^2} \right) I({\bf r}). 
\end{equation}

\subsubsection{Red-detuned focused-beam trap}

\begin{figure}[t]
\centerline{\scalebox{1}{\includegraphics[width=10cm]{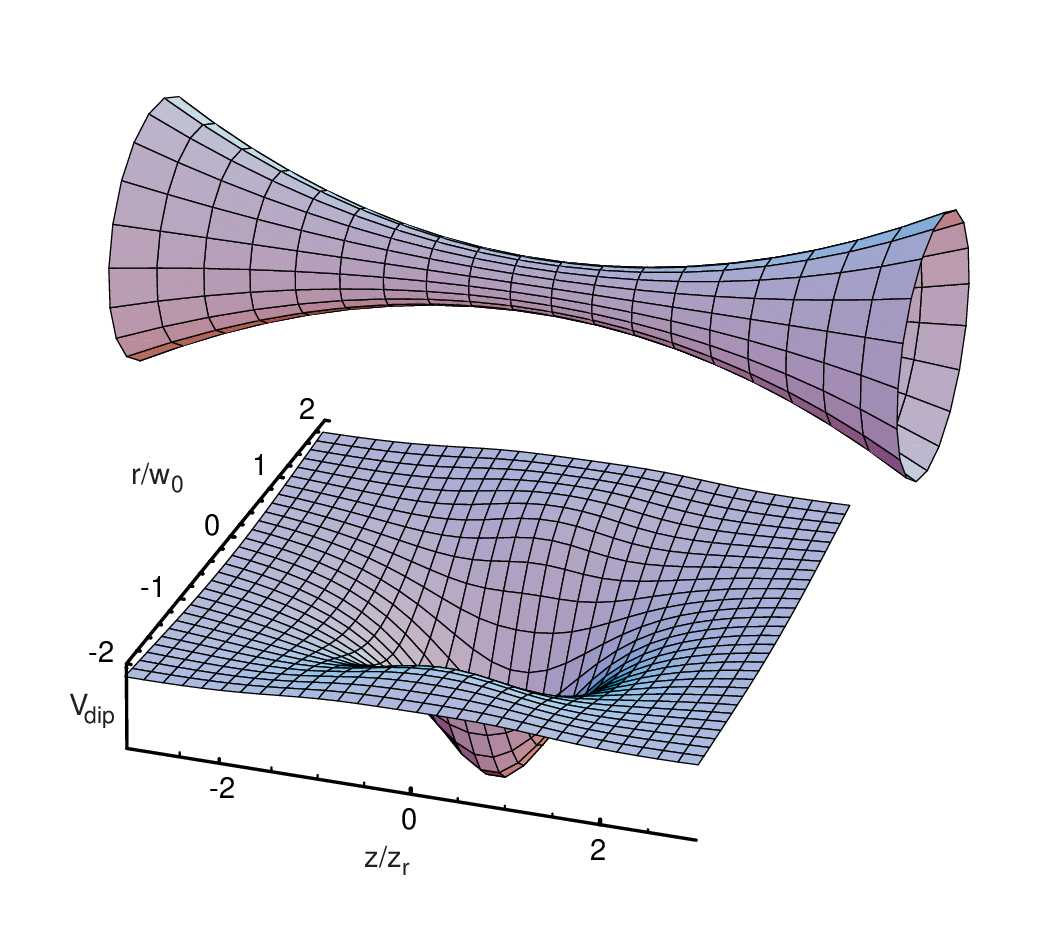}}}
\caption{Schematical drawing of a focused Gaussian laser beam and cross
  section of the corresponding trapping profile for a red detuned dipole laser.
}\label{pict:dipoletrap}
\end{figure}
A focused Gaussian laser beam tuned far below the atomic resonance frequency
represents the simplest way to create a dipole trap. The spatial intensity
distribution of a focused Gaussian beam with power $P$ propagating along the
z-axis is described by
\begin{equation} \label{equ:Gaussianbeam}
I(r,z)= \frac{2 P}{\pi w^2(z)} exp\left(\frac{-2 r^2}{w^2(z)}\right)
\end{equation}
where $r$ denotes the radial coordinate. The $1/e^2$ radius $w(z)$ depends on
the axial coordinate $z$ via
\begin{equation} \label{equ:waist}
w(z)=w_0 \sqrt{1+(\frac{z}{z_R})^2}
\end{equation}
where the focal radius $w_0$ is called beam waist and $z_R=\pi
w_0^2/\lambda$ denotes the Rayleigh length. From the intensity distribution
one can derive the optical potential using Eq. \ref{equ:lightshiftgeneral},
\ref{equ:lightshiftgeneral2}, or \ref{equ:lightshiftRb}. The trap depth
$\hat{U}$ is given by $\hat{U}=|U(r=0,z=0)|$. The Rayleigh length $z_R$ is
larger than the beam waist by a factor of $\pi w_0/\lambda$. Therefore the
potential in the radial direction is much steeper than in the axial direction.

If the mean kinetic energy $k_B T$ of an atomic ensemble is much smaller than
the potential depth $\hat{U}$, the extension of the atomic sample is radially
small compared to the beam waist and axially small compared to the Rayleigh
range. In this case, the optical potential can be well approximated by a
simple cylindrically symmetric harmonic oscillator
\begin{equation} 
U(r,z)=-\hat{U}\left[ 1 - 2\left(\frac{r}{w_0}\right)^2 -
  \left(\frac{z}{z_R}\right)^2 \right].
\end{equation}
The oscillation frequencies of a trapped atom are given by $\omega_R= (4
\hat{U} / m w_0^2)^{1/2}$ in the radial direction, and $\omega_z= (2 \hat{U}/
m z_R^2)^{1/2}$ in the axial direction. For a trap depth of $\hat{U} = 1$ mK,
a beam waist of $w_0=3.5 \mu$m, the oscillation frequencies are
$\omega_R/2\pi=26.2$ kHz and $\omega_z/2\pi=1.3$ kHz.

\subsection{Trap loading}

The standard way to load atoms into a dipole trap is to start from a
magneto-optical trap (MOT) \cite{Monroe90} or an optical molasses, by simply
overlapping the dipole trap with the atomic cloud in the MOT, before the
latter is turned off. Therefore, efficient cooling techniques are an essential
requirement since the attainable depths of optical dipole traps are generally
below 1 mK. Once atoms are captured, further cooling can be applied to
compensate possible heating mechanisms which would otherwise lead to a loss of
atoms. In this section, I briefly discuss methods which are of relevance for
trapping a single atom in a microscopic optical dipole trap.

\subsubsection{Doppler cooling}

Doppler cooling proposed by H\"ansch \& Schawlow in 1975 \cite{Haensch75} is
based on cycles of near-resonant absorption of a photon and subsequent
spontaneous emission resulting in a net atomic momentum change per cycle of
one photon momentum $\hbar k$ with $k= 2\pi/\lambda$ denoting the wavenumber
of the absorbed photon. Cooling is counteracted by heating due to the momentum
fluctuations by the recoil of spontaneously emitted photons
\cite{Minogin87}. Equilibrium between cooling and heating determines the
lowest achievable temperature. For Doppler cooling of two-level atoms in
standing waves (``optical molasses''), the minimum temperature $T_D$ is given
by
\begin{equation}
T_D = \frac{\hbar \Gamma}{2 k_B},
\end{equation}
where $\Gamma$ is the spontaneous decay rate of the cooling transition. For
Rubidium 87 the Doppler temperature is 146 $\mu$K, which is sufficiently low
to load atoms into a dipole trap. The first demonstration of dipole trapping
\cite{Chu86} used Doppler cooling for loading the trap and keeping the atoms
from beeing heated out.

\subsubsection{Polarization gradient cooling}

The Doppler temperature is a somewhat artificial limit since it is based on
the simplifying assumption of a two-level atom. Atoms with a more complex
level structure like Rb can be cooled far below $T_D$ in standing waves with
spatially varying polarizations \cite{Lett88}, whereby the cooling mechanisms
are based on optical pumping between ground state Zeeman sublevels
\cite{Dalibard89}.

With polarization-gradient molasses one can prepare atomic samples at
temperatures on the order of $\sim 10 T_{rec}$ with the recoil temperature
\begin{equation} \label{equ:recoil}
T_{rec} = \frac{\hbar^2 k^2}{m}
\end{equation} 
being defined as the temperature associated with the kinetic energy gain by
emission of one photon. Polarization-gradient cooling allows efficient loading
of a dipole trap, either by cooling inside a magneto-optical trap (MOT)
\cite{Monroe90,Steane91} or by cooling in a short molasses phase before
transfer into the dipole trap.

Besides enhancing the loading efficiency, polarization-gradient cooling can be
applied directly to atoms trapped in a dipole potential by subjecting them to
near-resonant polarizations gradients \cite{Boiron98}. Since the
cooling mechanism relies on the modification of the ground-state Zeeman
sublevels by the cooling light, a necessary condition for efficient cooling is
the independency of the trapping potential from the Zeeman substate
\cite{Garraway00}, which can be easily fulfilled in dipole traps.

\subsection{Heating and losses}

It is a well-known experimental fact in the field of laser cooling and
trapping that collisional processes between atoms can lead to substantial trap
loss. Here I will discuss the most important features of heating and cold
collisions.

\subsubsection{Heating rate}

A fundamental source of heating is the spontaneous scattering of trap photons,
which due to its random nature causes fluctuations of the radiation force. In
a far-detuned optical dipole trap, the scattering is completly elastic (or
quasi-elastic if a Raman process changes the atomic ground state).

In general both absorption and spontaneous re-emission processes show
fluctuations and thus both contribute to heating \cite{Minogin87}. At large
detunings, where scattering processes follow Poisson statistics, the heating
due to fluctuations in absorption corresponds to an increase of the thermal
energy by exactly the recoil energy $E_{rec}=\frac{k_B T_{rec}}{2}$ per
scattering event. This first contribution occurs in the propagation direction
of the light field and is thus anisotropic. The second contribution is due to
the random direction of the photon recoil in spontaneous emission. This
heating also increases the thermal energy by one recoil energy $E_{rec}$ per
scattering event, but distributed over all three dimensions. In most cases of
interest the trap mixes the motional degrees on a time scale faster than or
comparable to the heating. The overall heating thus corresponds to an increase
of the total thermal energy by $2 E_{rec}$ in a time
$\bar{\Gamma}_{sc}^{-1}$. Therefore the heating power is given by
\begin{equation} \label{equ:heatingpower}
P_{heat}=2 E_{rec} \bar{\Gamma}_{sc} = k_B T_{rec}  \bar{\Gamma}_{sc}. 
\end{equation}

In thermal equilibrium, the mean kinetic energy per atom in a
three-dimensional trap is $\bar{E}_{kin}=3 k_B T/2$. Introducing the parameter
$\kappa \equiv \bar{E}_{pot}/\bar{E}_{kin}$ as the ratio of potential and
kinetic energy, one can express the mean total energy $\bar{E}$ as
\begin{equation} \label{equ:totalenergy}
\bar{E}= \frac{3}{2}(1+\kappa)k_B T.
\end{equation}

The relation between mean energy and temperature (Eq. \ref{equ:heatingpower})
allows to reexpress the heating power resulting from photon scattering
(Eq. \ref{equ:totalenergy}) as a heating rate 
\begin{equation} \label{equ:heatingrate}
\dot{T}=\frac{2/3}{1+\kappa} T_{rec} \bar{\Gamma}_{sc},
\end{equation}
describing the corresponding increase of temperature with time. 

\subsubsection{Collisions}

\begin{figure}[t]
\centerline{\scalebox{1}{\includegraphics[width=10cm]{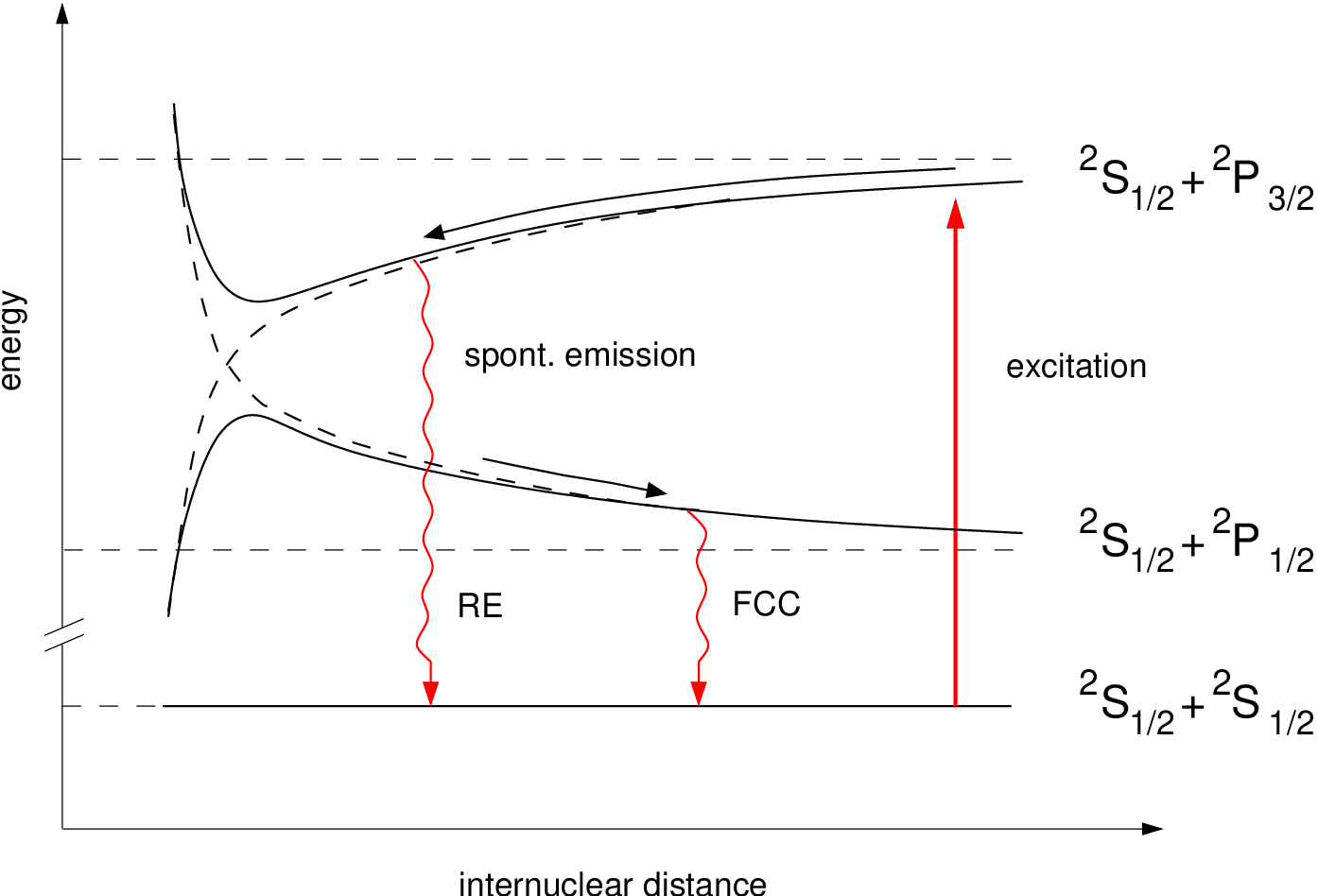}}}
\caption{Schematic diagram of processes leading to trap loss in an optical
  dipole trap during the loading stage in the presence of near resonant
  cooling light. } \label{pict:collision}
\end{figure}

In general, trap loss becomes important if the colliding atoms are not in
their absolute ground state. In an inelastic process the internal energy can
be released into the atomic motion, causing escape from the trap. Due to the
shallowness of optical dipole potentials, even the collisional release of the
relatively small amount of energy from transitions in the hyperfine structure
of the atomic ground state $^2S_{1/2}$ will always lead to trap loss.

Trap loss can also occur as a result of light-assisted binary collisions
involving atoms in the excited state (see Fig. \ref{pict:collision}). In {\it
the radiative escape process} (RE) the atoms gain kinetic energy from their
mutual attraction and when a spontaneous photon is emitted that has less
energy than the one that was initially absorbed. The energy difference appears
as kinetic energy and can be enough to eject the atoms out of the trap. In a
{\it fine-structure changing collision} (FCC) the atoms are excited to the
$^2S_{1/2} +$$^2P_{3/2}$ molecular potential. When they reach the
short-distance region they may change the molecular state during the collision
leading to the repulsive molecular potential $^2S_{1/2} +$$^2P_{1/2}$. The
atoms gain kinetic energy from the transition, which is sufficient for
ejection of one or both atoms out of the trap.

\subsubsection{Collisional blockade in microscopic optical dipole traps}
\label{subsection:CollisionalBlockade}

\begin{figure}[t]
\centerline{\scalebox{1}{\includegraphics[width=15.5cm]{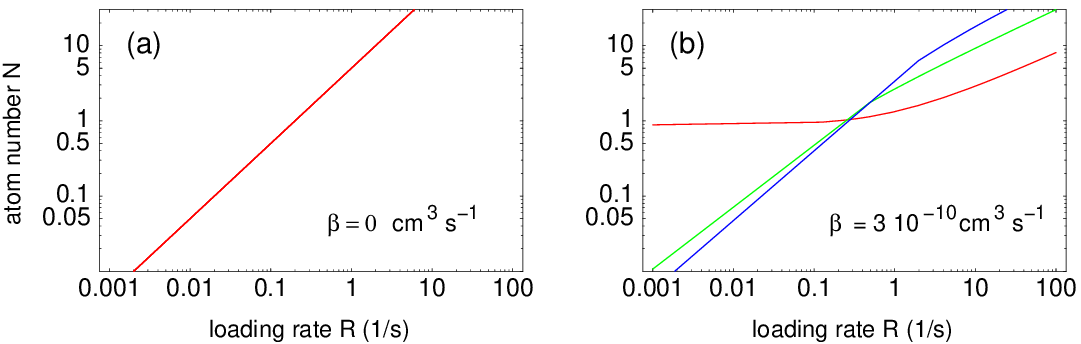}}}
\caption{Average number $N$ of trapped atoms as a function of the loading rate
$R$ for different values of the waist $w_0$ (red line: $w_0 = 4 \mu$m; green
line: $w_0 = 8 \mu$m; blue line: $w_0 = 12 \mu$m), {\bf (a)} without and {\bf
(b)} with two-body collisions. In presence of light-induced two-body
collisions a blockade regime is seen for a beam waist of $w_0 \leq$ 4 $\mu$m.}
\label{pict:collblockade} 
\end{figure}

During the loading stage of an optical dipole trap, light-assisted binary
collisions due to the presence of MOT light \cite{Kuppens00} dominate the trap
loss. When the dipole trap volume is small enough, this binary collisions
prevent loading of more than one atom \cite{Schlosser01,Schlosser02}.

The number $N$ of trapped atoms can be described by the equation
\begin{equation} \label{equ:traploading}
\frac{dN}{dt} = R - \gamma N - \beta' N (N-1),
\end{equation}
where $R$ is the loading rate, $\gamma$ the {\it single-particle loss}
coefficient taking into account collisions with the background gas in the
vacuum apparatus, and $\beta$ the {\it two-body loss} coefficient which
describes cold binary collisions.  

Since this equation is written for the atom number and not for the atomic
density, the value of $\beta'$ is inversely proportional to the volume of the
trap $\beta'=\beta/V$. The volume is found by approximating the trapped sample
of atoms as a cylinder with radius and length determined by the size of the
trapping beam waist and the temperature of the atoms. The volume is then given
by
\begin{equation} \label{equ:trapvolume}
V= \pi w_{0}^2 z_R \ln\left(\frac{1}{1-\eta}\right) \sqrt{\frac{\eta}{1-\eta}},
\end{equation}
where $\eta = k_B T / |\hat{U}|$. For $^{87}Rb$ the total loss rate $\beta$
due to RE and FCC was measured in a far-off-resonance dipole trap experiment
\cite{Kuppens00} to be $3 \times 10^{-10} ... 10^{-9} $cm$^3$ s$^{-1}$.

In order to get quantitative predictions for the mean atom number during the
loading stage of a dipole trap, one can solve the differential equation
\ref{equ:traploading}. For the single particle loss parameter $\gamma$ I will
use $0.2$ s$^{-1}$, which is consistent with our observations (see
sec. \ref{sec:atomstatistics}). Figure \ref{pict:collblockade} shows in a
log-log scale the mean number of trapped atoms as a function of the loading
rate. For a trapping-beam waist $w_0 \le4 \mu$m and in presence of cooling
light the mean number of trapped atoms is locked to one. This blockade effect
was observed the first time by Schlosser et al. \cite{Schlosser02} in a
microscopic optical dipole trap.

\subsubsection{Atom number distribution}

\begin{table}[h]{ \bf \caption{\label{tab:markov} Physical processes and
      elementary probabilities which lead to a 
      change in the atom number of +1, -1 and -2.}}
    \begin{center}
    \begin{tabular*}{120mm}[c]{c@{\extracolsep{\fill}}cc}
        \hline\hline
        physical process & probability & change in N \\ \hline
	loading & $R dt$ & +1 \\
	background collisions & $\gamma N dt$ & -1 \\
	two-body collisions & $\beta' N (N-1)/2$ & -2 \\[0.5ex]
        \hline\hline  
    \end{tabular*}
    \end{center}
\end{table}

The solution of the differential equation \ref{equ:traploading} which
describes the loading behaviour allows to calculate only the mean number of
trapped atoms. But it does not give insight into the atom number
statistics. This information can be gathered by modelling the loading process
by a {\it Markov process}.

In this modell the overall probability to have $N$ single atoms at a certain
time in the dipole trap is represented by a state vector $\vec{r}$. The change
of this state vector per unit time is given by a transfer matrix $M$ which
describes the relevant physical processes which govern the loading and loss
behaviour (see Tab.\ref{tab:markov}). In general this matrix is of dimension
$(N+1)\times(N+1)$. For simplicity I will restrict myself to the case of
maximal 5 atoms in the trap and therefore the probability to have 0,1,2,3,4 or
5 atoms in the trap is given by the state vector
$\vec{r}=(p_0,p_1,p_2,p_3,p_4,p_5)$ and the transfer matrix by
\begin{equation}
M = \left(\begin{array}{cccccc}
  A_1   & \gamma & \beta' & 0 & 0 & 0 \\
  R  & A_2  & 2 \gamma & 3 \beta' & 0 & 0\\
  0 & R  & A_3  & 3\gamma  & 6 \beta' & 0 \\
  0 & 0 & R & A_4 & 4 \gamma & 10 \beta'  \\
  0 & 0 & 0 & R  & A_5  & 5 \gamma \\
  0 & 0 & 0 & 0 & R & A_6
\end{array}\right),
\end{equation}
where $A_N = 1-(R+N\gamma+N(N-1)\beta'/2)$ for $N=0,1,..,5$. 

The stationary atom number distribution $\vec{r}_{stat}$ is then given 
by the solution of the system of linear algebraic equations which can be
written as
\begin{equation}
(M-E)\vec{r}_{stat}=\vec{0},
\end{equation}
where $E$ is the unity matrix. 

\begin{figure}[h]
\centerline{\scalebox{1}{\includegraphics[width=12cm]{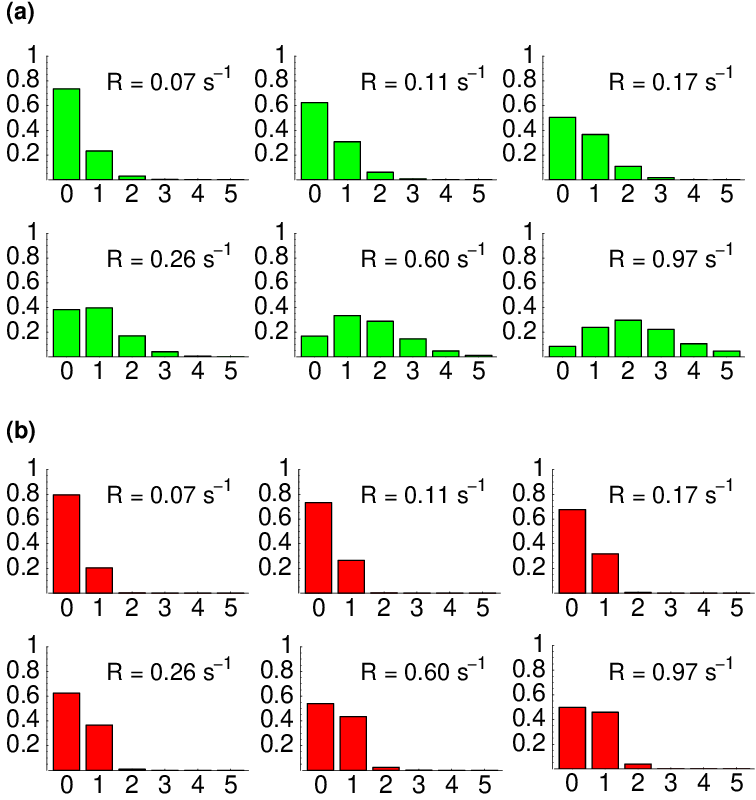}}}
\caption{Stationary atom number distributions for different loading rates $R$
  relevant for the present experiment. Each histogram shows the probability to
  have 0..5 atoms in the dipole trap. {\bf (a)} for a waist of $10 \mu$m we
  expect poissonian occupation statistics. {\bf (b)}, for a beam waist of $3.5
  \mu$m the number of trapped atoms is locked to one. The measured histograms
  in Fig. \ref{pict:scanMOT_histo} clearly prove this ``blockade effect''.}
\label{pict:collblockadeHISTO}
\end{figure}

Now one can calculate the stationary atom number distributions for different
loading rates $R$ relevant for the present experiment (see
Fig. \ref{pict:collblockadeHISTO}). Here we assume the situation when
near-resonant colling light is present during the loading stage of the dipole
trap. For a dipole laser beam waist of 10 $\mu$m we expect poissonian
occupation statistics. In this regime there is a high probability to trap more
than one atom at a time. When the dipole laser beam waist is $3.5 \mu$m the
probability to load more than one atom is remarkably suppressed leading to
subpoissonian occupation statistics. This blockade regime appears up to a
maximum loading rate of 1 atom/s.
\clearpage

%------------------------------------------------------------------------------

\section{Experimental setup}
 
In order to load only single atoms into a dipole trap the waist of the
focussed Gaussian dipole laser beam should be smaller than $4 \mu$m. Moreover,
flexibility in optical access and no optical components inside the vaccum
chamber have been the main design criteria of our experimental setup. The use
of a glass cell provides the best optical access for laser beams and imaging
optics. The fluorescence light scattered by a single atom in the focal spot of
the dipole laser beam is collected with a confocal microscope. This simple
setup \cite{Saucke02} allows to completely suppress stray light from the
dipole laser and reflexions of the cooling beams into the detection optics.

\subsection{Vacuum system}

To guarantee long storage times of atoms in a dipole trap our experiment must
be performed in an ultra high vacuum (UHV) environment. Therefore we set up a
compact UHV steel chamber pumped only by a ion pump (Varian, StarCell, 24
l/s). To have free access to the experimental region for laser beams and
detection optics we connected a commercial spectroscopy glass cell made by
Hellma with outer dimensions of $25\times25\times70$ mm$^3$ without
antireflexion coatings directly to a milled hole in a UHV steel flange and
sealed it with Indium wire. This simple and inexpensive setup yields a
background gas pressure of better than $3\times10^{-11}$ mbar. A Rubidium
dispenser inside the UHV chamber operated at a current of 2.5 A serves as the
source of atoms. Under this operating conditions the residual Rb gas pressure
is below $10^{-10}$ mbar and allows to store atoms in the dipole trap up to
several seconds. Details concerning the vacuum setup can be found in
\cite{Saucke02}.
\begin{figure}[h]
\centerline{\scalebox{1}{\includegraphics[width=8cm]{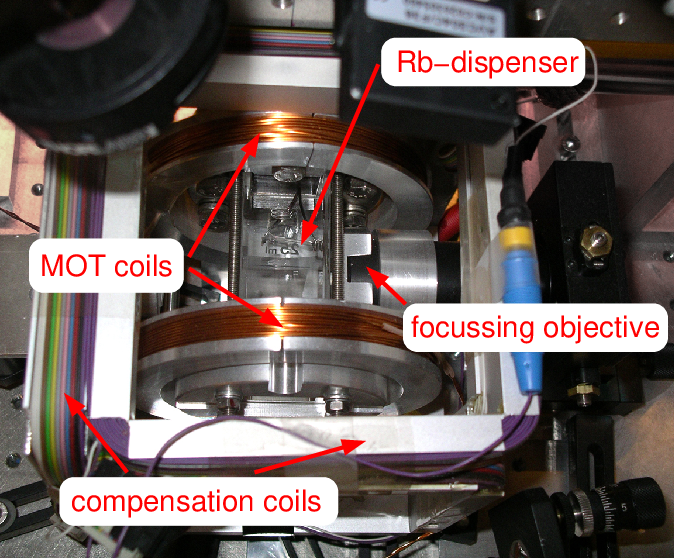}}}
\caption{Detailed picture of the experimental setup.}
\label{pict:photodetail}
\end{figure}

\subsection{Laser system}

The application of laser cooling and the coherent manipulation of a single
atom require the lasers to be stabilized onto or near atomic hyperfine
transitions of the $D2$ line with a wavelength of 780 nm. For this purpose we
use diode lasers locked to the Rubidium spectrum by using Doppler free
saturation spectroscopy. The lock signal is generated by a heterodyne lock-in
technique in the radio frequency domain \cite{Haensch71,Camparo85}. This
technique guarantees a longterm frequency stability of less than 2 MHz. The
bandwidth of the lasers is reduced to 0.6 MHz by operating them as grating
stabilized external cavity lasers \cite{Ricci95}. The mode structure of all
lasers is monitored online with a confocal scanning Fabry-Perot interferometer
(FPI) with a free spectral range of 375 MHz. This allows to identify mode
jumps of the diode lasers and reject inappropriate experimental data.

\begin{itemize}

\item The {\bf cooling laser} has to be red-detuned from the cycling
transition 5$^2S_{1/2},F=2 \rightarrow$ 5$^2P_{3/2},F'=3$ up to 5 natural
linewidths ($\Gamma=6$ MHz) and will be used for the coherent manipulation
(see chapter \ref{chapter:AtomicStateDetection}) of a single atom. For obvious
reasons this applications require switching properties down to 10 ns which are
realized in this experiment with acousto-optic modulators (AOM) in double-pass
configuration \cite{Saucke02,Vrana04} driven by a tuneable RF signal. A laser
beam passing the AOM twice shifts the laser frequency by $2\times 150 ... 250$
MHz. Therefore the cooling laser is locked to the crossover transition
5$^2S_{1/2}, F=2 \rightarrow$ 5$^2P_{3/2}, F'=1,2$ which is red-detuned by
345.5 MHz from the cycling transition.

\item {\bf Repump laser:} Although the 5$^2S_{1/2}, F=2 \rightarrow$
  5$^2P_{3/2}, F'=2$ transition is detuned from the cooling transition by 45
  natural linewidths, there is a finite probability of excitation to the
  hyperfine level 5$^2P_{3/2}, F'=2$, from where a spontaneous decay to the
  5$^2S_{1/2}, F=1$ ground state can occur. To ensure efficient laser cooling
  a seperate repump laser transfers the atomic population back into the
  5$^2S_{1/2}, F=2$ level. The repump laser is frequency stabilized via a beat
  signal to a master laser \cite{Saucke02,Greiner00,Schuenemann99} which is
  locked to the atomic transition 5$^2S_{1/2}, F=1 \rightarrow$ 5$^2P_{3/2},
  F'=1$. The repump laser frequency can be varied on one hand with the tunable
  beat signal on the other hand with the driving frequency of an AOM in
  double-pass configuration. This setup offers a great flexibility to tailor
  switchable laser pulses at different frequencies.

\item {\bf Dipole trap laser:} The far-off-resonace dipole trap is generated
  by a focused Gaussian laser beam of a single mode laser diode (SDL) at 856
  nm with a maximum output power of 200 mW. Because the dipole laser operates
  at a detuning of 61 nm from atomic resonance it is sufficient to stabilize
  the laser wavelength on the order of a few 0.1 nm. This is achieved only by
  stabilizing the temperature of the laser diode within 0.05 K.

\end{itemize}

\begin{figure}[h]
\centerline{\scalebox{1}{\includegraphics[]{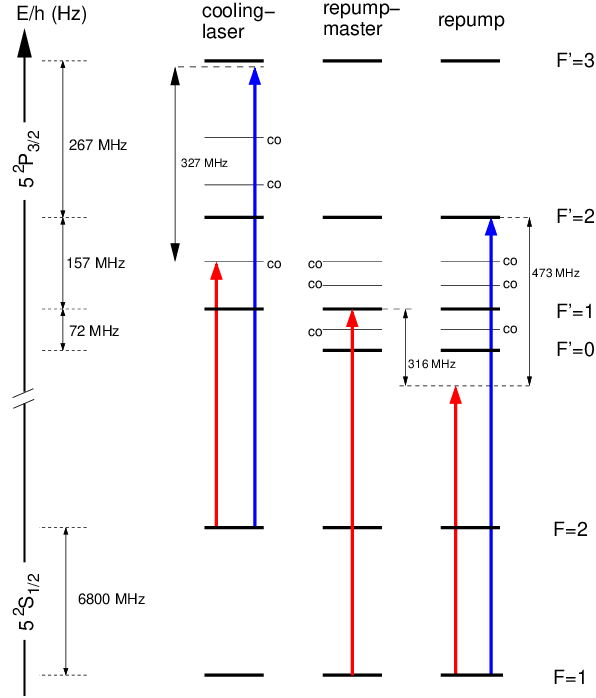}}}
\caption{Laser frequencies used in the present experiment for laser cooling
  and trapping of a single $^{87}$Rb atom. Red (blue) arrows: laser
  frequencies before (after) AOM.}
\label{pict:laserwellen}
\end{figure}

\subsection{Magneto optical trap}
 
\begin{figure}[t]
\centerline{\scalebox{1}{\includegraphics[width=12cm]{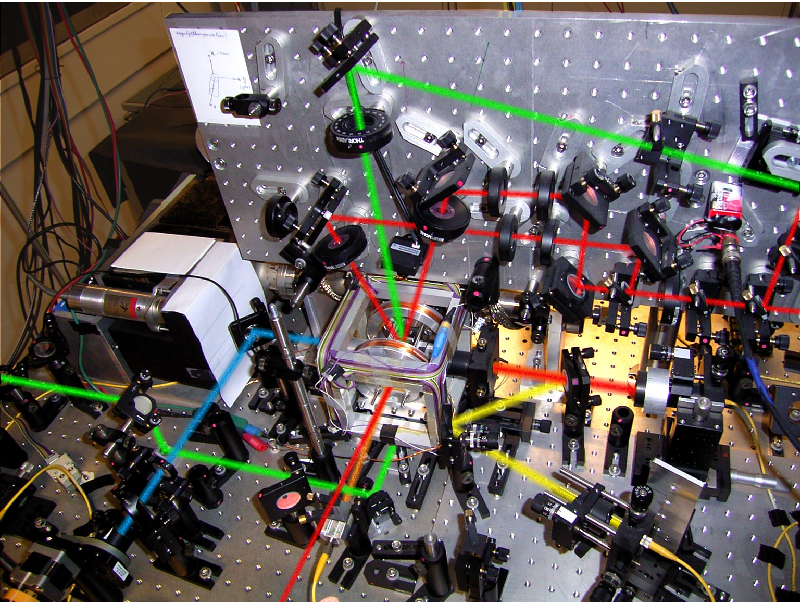}}}
\caption{{\bf Photograph of the experimental apparatus}. On the vertical base
plate above the vacuum chamber optics for the MOT and the cycling laser can be
seen. The optics on the right hand side of the MOT chamber is for focussing
the dipole laser beam collecting atomic fluorescence light.}
\label{pict:photosetup}
\end{figure}

Because of the small potential depth of our optical dipole trap - typically
0.5 ... 1 mK - atoms have to be precooled in a magneto optical trap (MOT) to
the micro-Kelvin regime \cite{Raab87,Metcalf99} before they can be transferred
into the dipole trap. The basic idea of a MOT is to use dissipative light
forces which introduce an effective friction force to slow down and cool atoms
from a thermal atomic gas. At the same time an inhomogeneous magnetic field is
applied which introduces a spatial dependence of the light force leading to a
confinement of the atom cloud.
\begin{figure}[h]
\centerline{\scalebox{1}{\includegraphics[width=5cm]{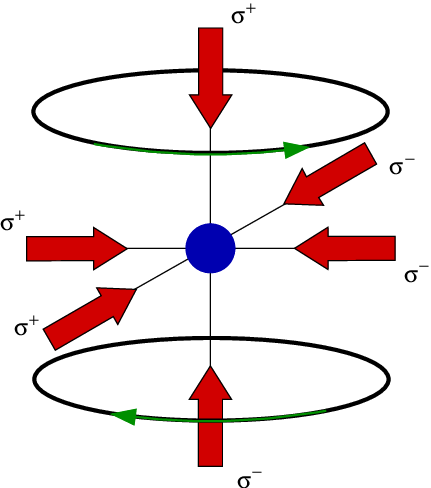}}}
\caption{Geometry of a magneto optical trap (MOT). The magnetic quadrupole
  field for the MOT is generated by an anti-Helmholtz coil pair. Six
  circularly polarized laser beams are overlapped at the trap center.}
\label{pict:MOT}
\end{figure}
The schematic setup for a MOT is shown in Fig. \ref{pict:MOT}. Six
red-detuned, retroreflected, and circularly polarized laser beams with a
diameter of 2 mm are directed onto the center of the magnetic quadrupole field
which is produced by two magnetic coils in anti-Helmholtz configuration. The
coils can generate magnetic field gradients of up to $\partial B / \partial z
= 11$ G/cm at a maximum current of 2 A. In contrast to a standard MOT the two
beams within the plane of the coils make an angle of 34$^\circ$, instead of
the usual 90$^\circ$ (see Fig. \ref{pict:photosetup}). Sufficiently slow atoms
are captured in the MOT from the thermal $^{87}$Rb background gas, whereby the
pressure can be adjusted by varying the current through the Rubidium
dispenser. Typically the MOT confines $3\times10^{4}$ atoms in a 1-mm-diameter
cloud.

\subsection{Dipole trap and detection optics}

The detection of extremely low levels of fluorescence light scattered by a
single atom requires detection optics which allow to collect atomic
fluorescence from a large solid angle. On the other hand we want to suppress
$\ket{\pi}$-polarized single photons from the transition
5$^2P_{3/2},F'=0,m_{F'}=0 \rightarrow$ 5$^2S_{1/2}, F=1, m_F=0$ reducing the
fidelity of the entangled atom-photon state we are interested in to
generate. The smaller the solid angle of the collected atomic fluorescence
light the higher the entanglement fidelity. Furthermore, we want to make use of
the collisional blockade effect which can be achieved with a trapping beam
waist smaller than 4 $\mu$m.

\begin{figure}[t]
\centerline{\scalebox{1}{\includegraphics[width=13cm]{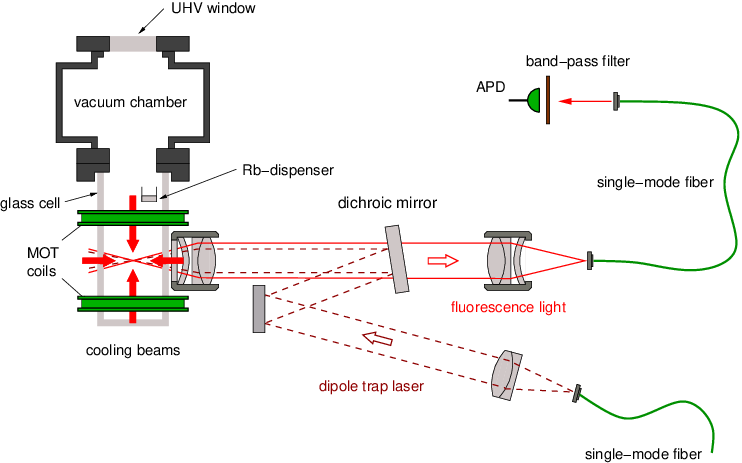}}}
\caption{{\bf Experimental setup of the dipole trap and fluorescence
    detection.} The dipole laser is focussed at the intersection of three
    counterpropagating laser beams for optical cooling. Fluorescence light is
    collected with a confocal microscope into a single-mode optical fiber and
    detected with a silicon APD.}
\label{pict:setup}
\end{figure} 

\subsubsection{Dipole trap}

To fulfill these requirements, we use a commercial achromatic laser objective
(LINOS), mounted outside the vacuum chamber with a working distance of 30 mm
and a numerical aperture of $NA=0.38$, for focussing the dipole laser beam and
imaging atomic fluorescence light from the dipole trap region at the same
time. Before the dipole laser beam is focussed down, it is spatially filtered
by a single mode optical fiber, improving the beam quality of the diode laser
beam significantly. An additional plane glass plate in between the glass cell
of the UHV chamber and the focussing objective is inserted to improve the
imaging properties of the objective \cite{Saucke02}. This addition especially
reduces the aberration of the light beam in the focal point, enabling
diffraction limited imaging. For the trapping beam we achieve a minimum waist
of $w_0=3.5 \pm 0.2$ $\mu$m with a quality factor of $M^2=1.1$. Here I
emphasize, that this waist is used without exception in all experiments of
this thesis. To adjust and to stabilize the depth of the dipole potential the
power of the dipole laser beam can be controlled with an acousto-optic
modulator (AOM) in single-pass configuration. For a laser power of 44 mW, a
wavelength of 856 nm and a beam waist of the trapping beam we calculate a
depth of the dipole potential of 1 mK and photon scattering rate of 24
s$^{-1}$.

\subsubsection{Detection optics} 

Fluorescence light from the dipole trap region is collected with the same
focusing objective in a confocal arrangement and separated from the trapping
beam with a dichroic mirror. The waist of the detection beam is
$w_0=2.2\pm0.2$ $\mu$m and is overlapped tranversally and longitudinally with
the waist of the trapping beam. This simple setup allows to collect
fluorescence light from a single atom with an effective numerical aperture of
$NA=0.29$ and guarantees a maximal entanglement fidelity F=0.99 because
$\pi$-polarized photons are hardly collected \cite{Volz}. To suppress stray
light from the cooling beams the fluorescence light is coupled into a single
mode optical fiber for spatial filtering. Finally, it is detected with single
photon sensitivity with a silicon avalanche photodiode (APD) operated in the
Geiger mode \cite{Weber00} with a typical dark count rate of 250 counts/s. If
one assumes isotropic emission of unpolarized atomic fluorescence light
(corresponding to the situation of a single atom in the light-field
configuration of the MOT) we calculate an overall detection efficiency for
single photons of $2\times 10^{-3}$ including transmission losses of optical
components and the quantum efficiency $\eta_{q}=0.5$ of our APDs
\cite{Saucke02}.

The APD signal resulting from the detection of a photon is converted into a
TTL pulse and sent to a timer card which integrates the number of detected
photons in time intervals down to 10 ms. This signal serves as a real time
monitor of the dipole trap and its dynamics (see Fig. \ref{pict:trace}) and is
continously displayed on a computer screen and recorded into a data file.
\clearpage

\section{Observation of single atoms in a dipole trap}
\label{section:singleatoms}

\begin{figure}[h]
\centerline{\scalebox{1}{\includegraphics[width=10cm]{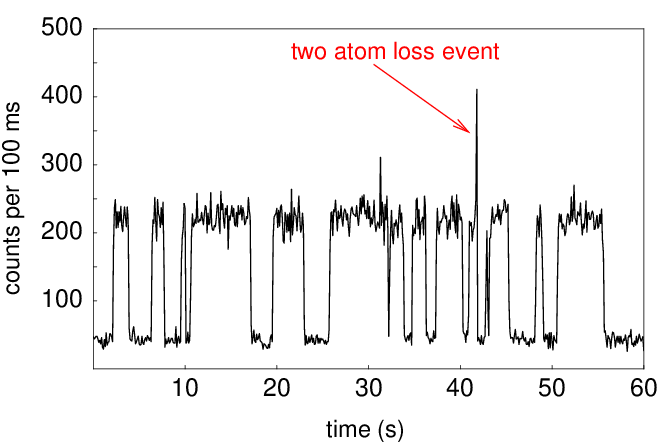}}}
\caption{{\bf Single atom detection}. Number of photons counted by
an avalanche photodiode per 100 ms. Due to the small trap volume and light
induced two-body collisions we observe either one or no atom at a time.}
\label{pict:trace}
\end{figure}

The dipole trap is loaded by simply overlapping it with the atomic cloud in
the MOT, where the atoms are precooled into the $\mu$K regime. Typically the
cooling laser is red detuned to the hyperfine transition 5$^2S_{1/2},F=2
\rightarrow$ 5$^2P_{3/2},F=3$ by $3..5 \Gamma$ and the repump laser on
resonance with the hyperfine transition 5$^2S_{1/2},F=1 \rightarrow$
5$^2P_{3/2},F=2$ of the D2 line. When both the dipole trap and the pre-cooling
MOT are turned on together, we observe characteristic steps in the detected
fluorescence signal (Fig. \ref{pict:trace}), corresponding to individual atoms
entering and leaving the dipole trap. By changing the magnetic field gradient
of the MOT we can adjust the density of the atomic cloud and therefore the
loading rate of atoms into the dipole trap from $R=1$ atom/15 s to $R = 1$
atom/s (see Fig. \ref{pict:scan_MOT_result}, inset (a)). If a second atom
enters the trap both are immediately lost. This effect that atoms leave the
trap in pairs is a result of light-induced binary collisions and will be
investigated in detail in \ref{sec:atomstatistics}.

\begin{figure}[t]
\centerline{\scalebox{1}{\includegraphics[width=10cm]{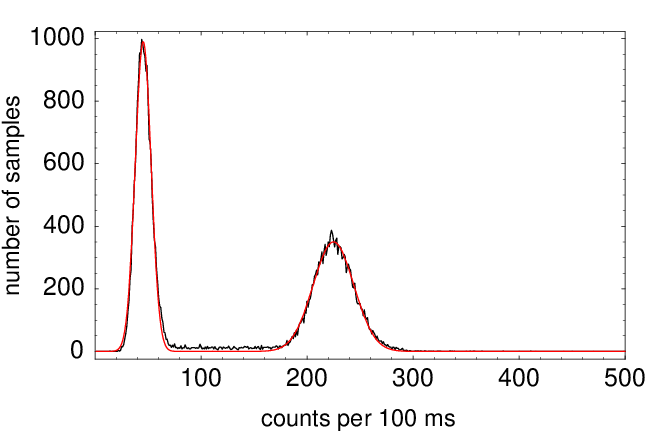}}}
\caption{Single atom detection. Histogram of photon counts per 100 ms of
Fig. \ref{pict:trace}.}
\label{pict:traceHISTO}
\end{figure}

The fluctuations of the photon count rate originate in the poissonian counting
statistics. A typical histogram of photon counts is shown in
Fig. \ref{pict:traceHISTO}. The dark count rate of the APD and background gas
fluorescence cause a first peak in the histogram at $N_0=45$ counts/100 ms and
a rms width of $\sigma_0=7$ counts/100 ms, whereby the dark count rate of the
APD (400 counts/s) dominates the background. Atomic fluorescence light
radiated by cold atoms in the surrounding MOT determine the residual
contribution. Cooling laser light reflected by optical components is
completely suppressed in the present setup. The second peak in
Fig. \ref{pict:traceHISTO} at $N_1=225$ counts/100 ms and a rms width of
$\sigma_1=19$ counts/100 ms is caused by resonance fluorescence of a single
atom in the dipole trap. Typical photon counting rates are 500 - 1800 s$^{-1}$
per atom depending on the detuning and intensity of the cooling laser. For a
maximum counting rate of 1800 s$^{-1}$ per atom we determine a minimum time
interval of 10 ms to be able to distinguish 1 atom from the background
counts. As a criterion we require that the difference of count rates of ``1''
atom and ``0'' atom, $\Delta N=N_1 - N_0$, is 4 times larger than the
statistical fluctuations of $N_1$. Provided the fluctuations of the photon
count rate are caused only by the poissonian counting statistics the expected
error of this method is 4 percent.

\subsection{Trap lifetime} \label{subsect:traplifetime}

The lifetime of single atoms in the dipole trap depends on one hand on
collisions with atoms from the background gas, whereby the collision rate is
determined by the residual background gas pressure in the vacuum chamber. For
a gas pressure below $10^{-10}$ mbar we expect a lifetime up to several
seconds. On the other hand, this value can drop by many orders of magnitude
during the loading stage because light-induced binary collisions dominate the
trap loss.
\begin{figure}[t]
\centerline{\scalebox{1}{\includegraphics[width=10cm]{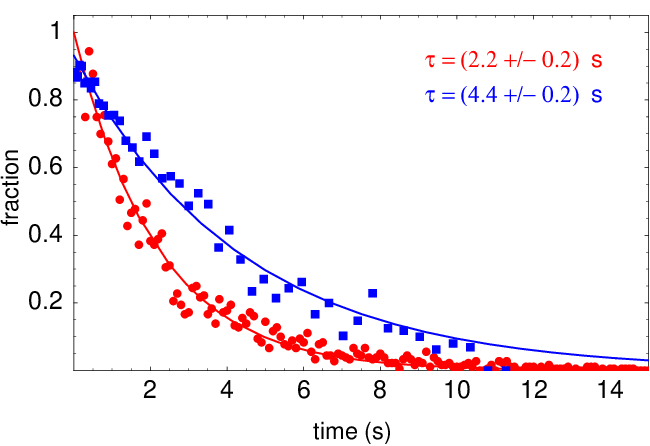}}}
\caption{Fraction of single atoms in the dipole trap as a function of time
with (red data points) and without (blue data points) cooling light. A fit of
the experimental data with a simple exponential decay yields a 1/e lifetime of
$2.2 \pm 0.2$ s and $4.4 \pm 0.2$ s, respectively.}
\label{pict:tracelifetime}
\end{figure}

\begin{figure}[h]
\centerline{\scalebox{1}{\includegraphics[width=11.5cm]{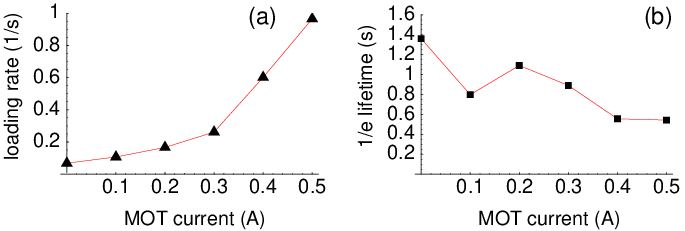}}}
\caption{{\bf (a)} Loading rate $R$ and {\bf (b)} $1/e$ lifetime in presence of
cooling light as a function of the MOT current. Increasing the MOT current
reduces the trap lifetime due to an increase of the loading rate.}
\label{pict:scan_MOT_result}
\end{figure}

Depending on the trap depth and the loading rate we observe a mean $1/e$
lifetime between 0.5 and 2.2 s (see Fig. \ref{pict:scan_MOT_result}, inset
(b)). Here the lifetime is extracted directly by histogramming the length of
fluorescence steps. Because the cooling and the repump laser of the MOT are
present during the loading stage of the dipole trap, light-induced two-body
collisions dominate the trap loss and therefore the measured lifetime.

In a different experiment the characteristic lifetime of single atoms limited
only by background gas collisions is measured. Therefore, we switch on the
cooling and repump laser of our MOT and wait for a single atom. If the
observed fluorescence exceeds a threshold of 1200 counts/s the MOT lasers are
switched off. After a variable delay time $t$, the MOT lasers are switched on
again and atomic fluorescence light is collected for 50 ms. If the measured
fluorescence still exceeds the threshold, the atom has stayed in the trap. If
we do not observe atomic fluorescence light, the atom is lost due to
collisions with the background gas. This experimental sequence is repeated for
a range of delay times $t$ and for each delay value many times. Finally, we
derive a histogram, where the probability to redetect a single atom in the
dipole trap is given as a function of the delay time and fit this data set
with an exponential decay yielding a $1/e$ lifetime of $4.4\pm0.2$ s.

\clearpage

\subsection{Atom number statistics} \label{sec:atomstatistics}

\begin{figure}[h]
\centerline{\scalebox{1}{\includegraphics[width=10.5cm]{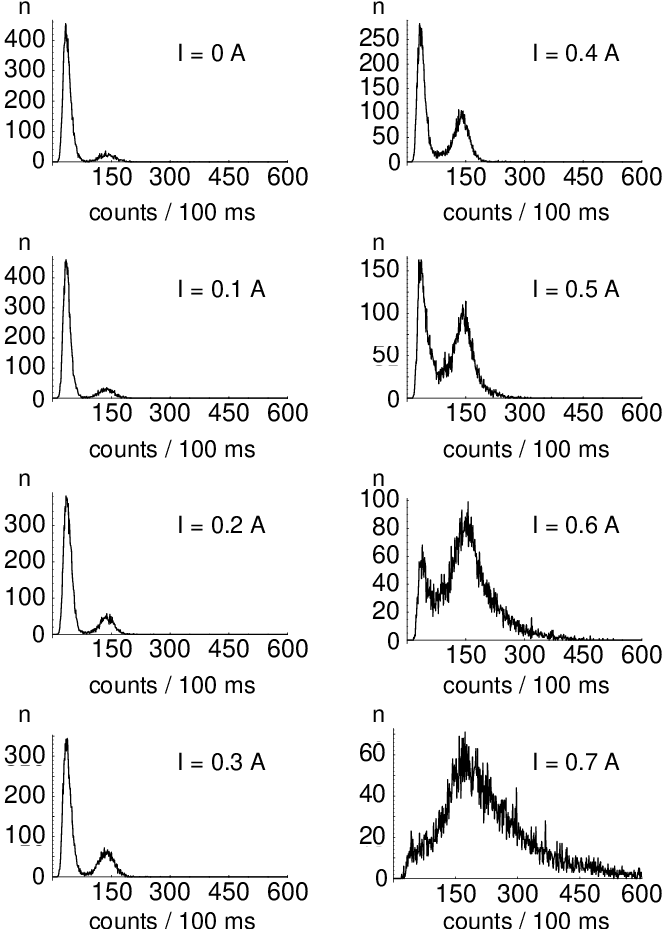}}}
\caption{Histograms of detected photon counts in 100 ms for different MOT
  currents (I). Up to $0.5$ A only single atoms are observed in the dipole
  trap.}
\label{pict:scanMOT_histo}
\end{figure}

One important fact concerning the fluorescence detection in our experiment is,
that photon count rates corresponding to more than one atom never occur up to
a maximum MOT current of 0.5 A corresponding to a maximum loading rate of
$R=1$ s$^{-1}$. This can be seen directly from the histogramms of photon
counts in Fig. \ref{pict:scanMOT_histo}.

\clearpage

\begin{figure}[h]
\centerline{\scalebox{1}{\includegraphics[]{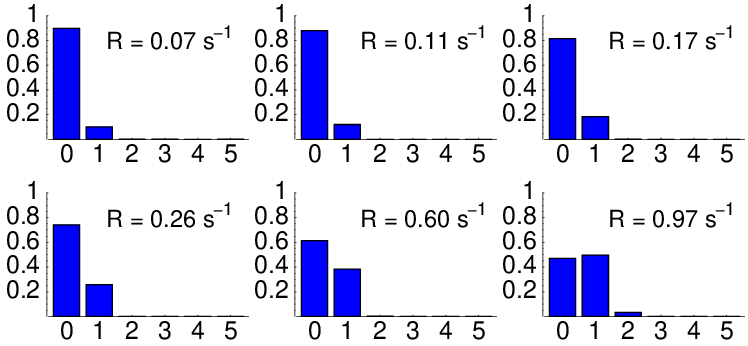}}}
\caption{Measured atom number distributions for different loading rates. Due
  to the small dipole trap volume and the presence of light-induced two-body
  collisions during the loading stage the number of trapped atoms is locked to
  one up to a loading rate of 1 atom/s.}
\label{pict:measuredHISTO}
\end{figure}

To compare the measured atom number distributions directly with the calculated
distributions for a trapping beam waist of $3.5 \mu$m (see
Fig. \ref{pict:collblockadeHISTO}), the continuous photon counting histograms
are converted by numerical integration into bar-histograms where the
probability to load atoms into the dipole trap is plotted as a function of the
atom number N (see Fig. \ref{pict:measuredHISTO}). As expected, our
measurement results confirm the ``blockade effect'' \cite{Schlosser02},
locking the maximum number of trapped atoms to one.

\subsection{Temperature measurement of a single atom}

In the present experiment a single optically trapped atom is cooled during
fluorescence detection by three-dimensional polarization gradients in an
optical molasses. This leads to a final kinetic energy on the order of 100
$\mu$K \cite{Garraway00}. Information about the kinetic energy can be gathered
by the spectral analysis of the emitted resonance fluorescence, because
Doppler effects due to the motion in the confining potential lead to a line
broadening in the emitted fluorescence spectrum.

For low excitation intensities the fluorescence spectrum of a two-level atom
exhibits an elastic peak centered at the incident laser frequency
$\omega_{L}$, while for higher intensities an inelastic component becomes
dominant, with contributions at the frequencies $\omega_{L}$ and $\omega_{L}
\pm \Omega_{0}$ \cite{Mollow69}, where $\Omega_{0}$ denotes the Rabi
frequency. This so so-called ``Mollow triplet'' arises from the dynamical
Stark splitting of the two-level transition and has been observed in a number
of experiments \cite{Schuda74,Wu75,Hartig76}, using low-density atomic beams
or a single trapped and laser-cooled Ba$^+$ ion
\cite{Stalgies96}. Surprisingly, there are only a few experimental
investigations of the {\it coherent} scattering process, with a frequency
distribution determined by the exciting laser. Subnatural linewidths were
demonstrated with atomic beam experiments \cite{Hartig76,Gibbs76}, atomic
clouds in optical molasses \cite{Westbrook90,Jessen92} and a single trapped
and laser-cooled Mg$^+$ ion \cite{Hoeffges97a,Hoeffges97b}.

For our laser cooling parameters the fluorescence spectrum is dominated by
elastic Rayleigh scattering \cite{Tannoudji98}. Hence, the emitted fluorescence
light exhibits the frequency distribution of the exciting laser (0.6 MHz FWHM)
field broadened by the Dopper effect. Position-dependent atomic transition
frequencies in the dipole trap due to the inhomogeneous AC-Stark shift give no
additional broadening because the spectrum of the elastically scattered
fluorescence light is determined only by the frequency distribution of the
exciting light field and not by the atomic transition frequencies.

The resolution achieved with narrow-band Fabry-Perot filter resonators for
spectral analysis is limited to approximately 1 MHz and allows to measure an
effective line broadening on the order of several 10 kHz. This resolution is
sufficient to determine the kinetic energy with sufficient accuracy.
 
\subsubsection{Experimental setup}

\begin{figure}[t]
\centerline{\scalebox{1}{\includegraphics[width=12cm]{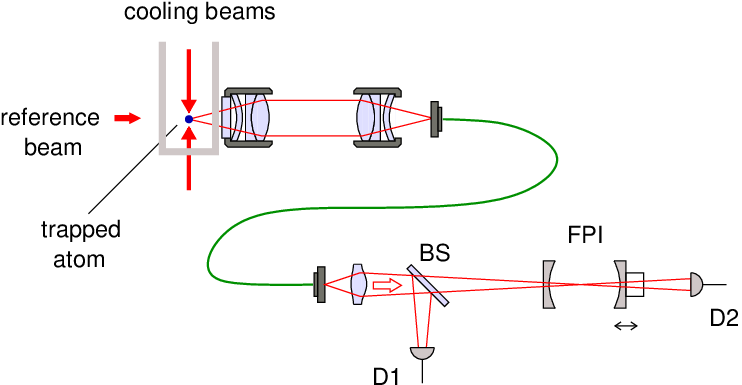}}}
\caption{Setup for the measurement of the resonance fluorescence spectrum of
  light scattered by a single atom in an optical dipole trap. The fluorescence
  and excitation laser light is collected confocally to the trapping beam (see
  Fig. \ref{pict:setup}) and analyzed with a scanning Fabry-Perot
  Interferometer (FPI). The reference laser is used to monitor length drifts
  of the resonator and to determine the instrumental function of the FPI.}
  \label{pict:temperatureSETUP} 
\end{figure}

The scattered fluorescence spectrum is analyzed via a scanning Fabry-Perot
interferometer (FPI) with a frequency resolution of $0.45$ MHz (full width
half maximum), a transmission of $40\%$ and a finesse of 370. To measure the
spectrum only at times we trap single atoms, a part of the fluorescence light
is monitored separately with a reference APD (D1) (see
Fig. \ref{pict:temperatureSETUP}). Since the broadening of the atomic emission
spectrum due to the Doppler effect is expected to be a small effect, the
instrumental function of the spectrometer and the exciting laser line width
have to be known accurately. In order to achieve this, we shine a fraction of
the exciting light (reference beam) into the collection optics (see
Fig. \ref{pict:temperatureSETUP}). This way, both reference and scattered
light are subject to the identical spectrometer instrumental function whereby
the reference laser spectrum is also used to monitor length drifts of the
filter cavity. In the experiment the spectrum of the reference beam and the
fluorescence light scattered by a single atom in the dipole trap were recorded
alternately. After each measurement a compensation of the cavity drift was
performed by referencing the cavity frequency to the maximum transmission of
the reference laser.

\subsubsection{Experimental results}

\begin{figure}[t]
\centerline{\scalebox{1}{\includegraphics[width=10cm]{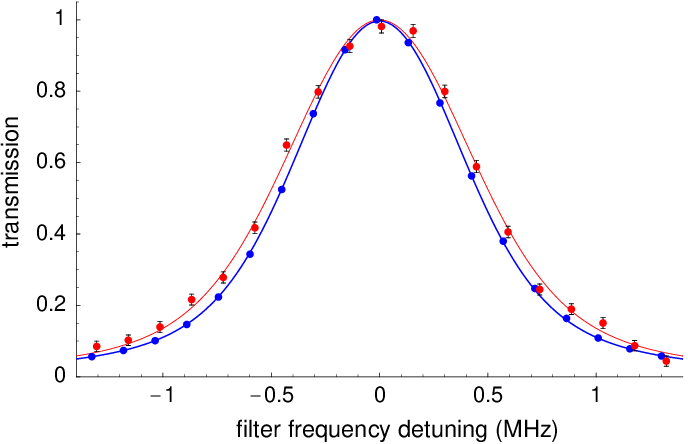}}}
\caption{Measured spectra of the single atom resonance fluorescence (red data
  points) and the excitation light (blue data points). The spectra exhibit a
  width of $0.94\pm0.01$ MHz and $1.04\pm0.1$ MHz (FWHM) for the excitation
  and the fluorescence light, respectively. Experimental parameters: $I_{CL}=
  80$ mW/cm$^2$, $I_{RL}=12$ mW/cm$^2$, $\Delta_{CL} = - 2 \pi\times19$ MHz,
  $\hat{U}=(0.62\pm0.06)$ mK.}
\label{pict:temperatureRESULT}
\end{figure}

With this procedure we obtained the two (normalized) data sets in
Fig. \ref{pict:temperatureRESULT}. As expected, the fluorescence spectrum
scattered by a single atom exhibits a ``subnatural'' linewidth of $1.04 \pm
0.01$ MHz (FWHM) because the elastic Rayleigh contribution dominates the
scattering process. The exciting laser light field exhibits a linewidth of
$0.94\pm 0.01$ MHz (FWHM) which is the convolution of the transmission
function of the Fabry-Perot filter with the spectral width of the excitation
laser. The depicted error bars reflect the statistical error from the
individual count rates of each data point. For the reference laser this error
is too small to be visible in this graph.

For an atom at rest the resonance fluorescence spectrum should show the same
linewidth as the exciting light field. Any finite kinetic energy of the atom
will lead to a broadening of the atomic emission spectrum and therefore can be
used for a determination of the ``temperature''.  To extract a mean kinetic
energy from the measured spectra in Fig. \ref{pict:temperatureRESULT}, we
assume the same stationary Gaussian velocity distribution in all directions.
According to this assumption we convolve a Gaussian distribution with the
measured reference laser line profile. The resulting function is fitted to the
data points of the fluorescence spectrum with the variance of the Gaussian
profile being the only free fit parameter \cite{footnote1}. From this fitted
variance we directly obtain the mean kinetic energy $E_{kin}$ of a single atom
in the dipole trap of
\begin{equation} \label{equ:temperatureRESULT}
E_{kin}=\frac{1}{2}m\langle \Delta v^2\rangle=(110 \pm
15)^{+14}_{-25}\mbox{ }\mu \mbox{K}\cdot k_B,
\end{equation}
with a statistical error of $\pm 15 \mu$K. $k_B$ denotes the Boltzmann
constant, $m$ the atomic mass and $\langle \Delta v^2 \rangle$ the
mean quadratic velocity.

The calculation of the mean kinetic energy contains a systematic error because
the cooling beams have different angles relative to the axes defined by the
dipole trap and the detection optics. The overall Doppler broadening of the
elastic scattered fluorescence light depends on these angles. Therefore, an
upper bound for this error is estimated by assuming that the atoms scatter
light only from the beams which would give the highest or lowest velocities,
respectively. Within the experimental errors, the measured temperature is
equal to or smaller than the Doppler temperature of $^{87}$Rb (146 $\mu$K).

\section{Conclusion and discussion}

A single $^{87}$Rb atom is loaded from a magneto-optical-trap (MOT) into an
optical dipole trap that operates at a detuning of 61 nm from atomic
resonance. Atoms stored in this far-off-resonance optical dipole trap have a
very small scattering rate and therefore negligible photon recoil
heating. Confinement times up to 5 s are achieved with no additional
cooling. For loading the dipole trap, the MOT cooling lasers are switched on,
and atomic fluorescence light is collected with a microscope objective and
detected with a single photon avalanche diode (APD). Well seperated
equidistant steps in the detected fluorescence signal allow to monitor the
number of trapped atoms in a ``noninvasive'' way and in real time. Because of
the small dipole trap volume (beam waist $w_0 = 3.5 \mu$m), cold binary
collisions assisted by cooling light lock the maximum number of trapped atoms
to one. Measurements for different loading rates clearly prove this ``blockade
effect''. The overall detection efficiency of single photons is 0.2 percent
including transmission losses and the quantum efficiency of the APD.

Using a scanning filter-cavity we determined the spectrum of the emitted
single atom resonance fluorescence. Due to Rayleigh-scattering the measured
atomic fluorescence spectrum is dominated mainly by the spectral profile of
the exciting light field. In addition we observe a Doppler broadening of the
scattered atomic fluorescence spectrum, which allowed us to determine an upper
bound of the mean kinetic energy of the trapped atom corresponding a
temperature of 110 $\mu$K.

%------------------------------------------------------------------------------
%------------------------------------------------------------------------------

\chapter{Single photons from single atoms}

\section{Introduction}

The experimental results from the previous chapter (see Fig. \ref{pict:trace})
suggest that we observe only single atoms in our dipole trap. This assumption
can be verified by a detailed statistical analysis of the measured stream of
photon counts. In particular, nonclassical features of the resonance
fluorescence are observed in the second-order correlation function
$g^{(2)}(\tau)$, whereby, most prominently, the so-called {\it antibunching}
behaviour indicates the single particle character of the radiating source.

To prove that only a single atom is stored in our trap we have set up a
Hanbury-Brown-Twiss (HBT) experiment \cite{Hanbury56a,Hanbury56b} and studied
the statistical properties of the detected fluorescence light. The measured
second-order correlation function exhibits strong photon antibunching
verifying the presence of a single atom. In addition the two-photon
correlations show the internal quantum dynamics of the population occupation
of the atomic hyperfine levels involved in the excitation process, whereby the
observed oscillations can be explained with a four-level model. To compare the
theoretical model with the experimental data we numerically solve optical
Bloch equations and calculate the second order correlation function of the
emitted fluorescence light. We find good agreement with the measured data.

Due to the complex structure of photon-pair correlations which depend on the
internal and external dynamics of the atom-light interaction, I will first
introduce theoretical aspects relevant for the understanding of the present
experiment before I will discuss measured experimental data.

\section{Theoretical framework}
\subsection{Second-order correlation function}

For the scattered light field of atomic resonance fluorescence described by
the electric field operators $\mathbf{E^{+}}$ and $\mathbf{E^{-}}$, the
two-photon correlation function $g^{2}(\tau)$ is given, according to Glauber
\cite{Glauber63} (see also \cite{Loudon}), by
\begin{equation} \label{equ:g2}
g^{(2)}(\tau) = \frac{\langle\mathbf{E^{-}}(t)\mathbf{E^{-}}(t+\tau)
\mathbf{E^{+}}(t+\tau)\mathbf{E^{+}}(t)\rangle}{{\langle \mathbf{E^{-}}(t)
\mathbf{E^{+}}(t) \rangle}^2}, 
\end{equation}
where $\tau \ge 0$. For almost monochromatic light fields and a small
detection probability, this function is the conditional probability of
detecting a photon at time $t+\tau$, given the previous detection of another
photon at time $t$, normalized by the factorized value for statistically
independent photons. Whereas classical fields show correlation functions
$g^{(2)}(0) \ge 1$ and $g^{(2)}(\tau) \le g^{(2)}(0)$, the nonclassical
resonance fluorescence of a single atom exhibits the so-called {\it
antibunching} behaviour, i.e., $g^{(2)}(0) =0$ and $g^{(2)}(\tau) >
g^{(2)}(0)$. This condition signals sub-Poissonian emission probability.

\subsection{Two-level atom}

The resonance scattering from an atom can be described by the electromagnetic
field operators in the Heisenberg picture. For a two-level atom at position
$\mathbf{R}$ with a dipole transition, the electric field operator
$\mathbf{E^+}(\mathbf{r},t)$ at a point $\mathbf{r}$ is given in the far-field
region by
\begin{equation} \label{equ:E}
\mathbf{E}^+(\mathbf{r},t)=\mathbf{E}^{+}_{free}(\mathbf{r},t)
+\mathbf{E}^{+}_{S}(\mathbf{r},t),
\end{equation}
where $\mathbf{E}^{+}_{free}(\mathbf{r},t)$ is the field unperturbed by the
atom. This part contains the vacuum fluctuations and the externally applied
(laser) light. The point of detection $\mathbf{r}$ is chosen such that the
excitation light field is negligible and therefore the freely-propagating part
can be dropped. The second contribution in Eq. (\ref{equ:E}) is the source
field part $\mathbf{E}^{+}_{S}(\mathbf{r},t)$ which represents the field
scattered by the atom. It is given by \cite{Loudon}
\begin{equation} \label{equ:sf}
\mathbf{E}^{+}_{S}(\mathbf{r},t)=-\frac{\omega_0^2 |\bra{g}\mathbf{d}\ket{e}|
  \sin\theta}
{4\pi\epsilon_0c^2|\mathbf{r-R}|}  \vec{\epsilon}_S
\hat{\pi}(t-\frac{|\mathbf{r-R}|}{c}).   
\end{equation}
Here, $\omega_0$ is the angular frequency of the atomic transition $\ket{g}
\rightarrow \ket{e}$, $\bra{g}\mathbf{d}\ket{e}$ is the corresponding dipole
matrix element, and $\mathbf{R}$ is the center-of-mass coordinate of the
atom. The operator $\hat{\pi}(\mathbf{r},t)$ contains the spatio-temporal
dependence of the light field and is defined by $\hat{\pi} =
\ket{g}\bra{e}$. For the negative part of the electric field the corresponding
operator is defined by $\hat{\pi}^{\dagger} = \ket{e}\bra{g}$
respectively. The angle $\theta$ is subtended between the dipole matrix
element and the direction of observation, $\mathbf{r-R}$. The linear
polarization of the emitted field, $\vec{\epsilon}_S$, is in the plane
perpendicular to the direction of observation, and parallel to the projection
of the dipole matrix element into this plane.

The coherence properties of the light emitted by a single two-level atom in
free space can now be calculated with the use of the source-field expression
(\ref{equ:sf}), and we get for the second-order correlation function of the
emitted fluorescence light
\begin{equation} \label{equ:g2_sf}
g^{(2)}(\tau)=\frac{\langle \hat{\pi}^{\dagger}(t)
  \hat{\pi}^{\dagger}(t+\tau) \hat{\pi}(t+\tau)\hat{\pi}(t)
  \rangle}{\langle\hat{\pi}^{\dagger}(t) \hat{\pi}(t) \rangle^2}, 
\end{equation}
where the time delay $\tau$ is always positive. This relation contains the
expectation value of two observables at different times. With the help of the
{\it quantum regression theorem} \cite{Lax66} the two-time expectation values
can be reduced to particular one-time expectation values. Performing some
calculations \cite{Scully}, the second-order correlation function is found to
be:
\begin{equation} \label{equ:g2_2level}
g^{(2)}(\tau)=\frac{\rho_{ee}(\tau)}{\rho_{ee}(\infty)}.
\end{equation}
Here $\rho_{ee}(\tau)$ is the element of the atom's density matrix that
represents the population of the excited atomic state at the time $\tau$. This
element can be calculated from optical Bloch equations (OBE) \cite{Metcalf99}
with the initial conditions $\rho_{ee}(0)=\rho_{eg}(0)=\rho_{ge}(0)=0$, and
$\rho_{gg}(0)=1$, which describe the state of the atom immediately after the
emission of a photon. The theoretically predicted correlation function is
given by \cite{Carmichael76}
\begin{equation} \label{g2_2level_A}
g^{(2)}(\tau)=1-e^{-3\Gamma\tau/4}[\cos(\Omega_R\tau)+
              \frac{3\Gamma}{4\Omega_R} \sin(\Omega_R\tau)],
\end{equation}
where $\Omega^2_R=\Omega_{0}^{2}+\Delta^2 -(\Gamma/4)^2$, with the Rabi
frequency $\Omega_0$ at resonance, the natural linewidth $\Gamma$, and the
detuning $\Delta$.

\begin{figure}[t]
\centerline{\scalebox{1}{\includegraphics[]{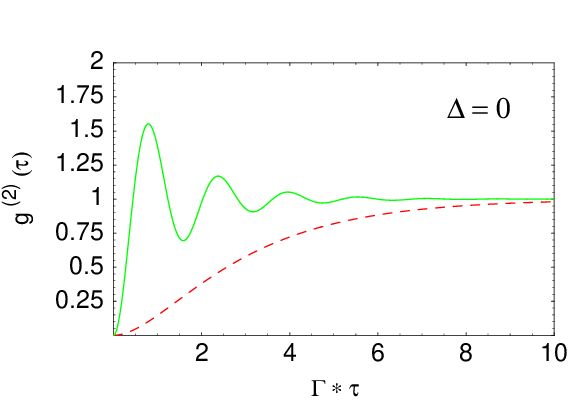}}}
\caption{Second-order correlation function $g^{(2)}(\tau)$ of a two-level atom
  versus the dimensionless delay time $\Gamma \tau$ for $\Omega_R/\Gamma=4$
  (green solid line) and $\Omega_R/\Gamma=0.4$ (red dashed line,
  respectively).}
\label{pict:g2_2level}
\end{figure}

In Fig. \ref{pict:g2_2level}, the second-order correlation function is plotted
as a function of the time delay $\tau$ for different values of the generalized
Rabi frequency $\Omega_R$ of the driving field. For $\tau=0$ the two-photon
correlation function $g^{(2)}(\tau)=0$, which corresponds to the phenomenon of
{\it photon antibunching}.  Once a photon is emitted, the atom is found in the
ground state and it takes the driving field some time to reexcite the atom to
the upper level, from which the next photon can be emitted. On average, this
delay is of the order of the Rabi period $\Omega_{R}^{-1}$. For a weak driving
field, $\Omega_R/\Gamma<1$, $g^{(2)}(\tau)$ increases monotonically from 0 to
1 as $\tau$ is increased. For a strong driving field, $\Omega_R/\Gamma>1$,
i.e. the generalized Rabi frequency $\Omega_R$ and thus the frequency of
populating and depopulating the excited state is greater than the decay rate
$\Gamma$. Therefore $g^{(2)}(\tau)$ shows an oscillatory dependence on
$\tau$. The magnitude of these oscillations decreases as $\tau$ is increased
and $g^{(2)}(\tau)$ approaches unity as $\tau \rightarrow \infty$. The upper
limit of $g^{(2)}(\tau)$ for the resonance fluorescence of a two-level atom is
2.

Both antibunching and sub-Poissonian statistics in resonance fluorescence have
been observed in quantum optical experiments with atoms
\cite{Kimble77,Short83}; however corrections for the fluctuating atom numbers
in the atomic beam had to be taken into account. With a single trapped Mg$^+$
ion Diedrich {\it et al.} \cite{Diedrich87} observed antibunching of an
individual two-level atom. In that experiment antibunching in the resonance
fluorescence of one, two, and three trapped ions was observed and the
antibunching property decreased as predicted for increasing ion numbers, since
for an increasing number of independent atoms the photon counts become more
and more uncorrelated.

Aside from antibunching and sub-Poissonian statistics, Schubert {\it et al.}
\cite{Schubert92} observed two-photon correlations in the resonance
fluorescence of a single Ba$^+$ ion with a maximum much larger than what is
possible with two-level atoms, as well as photon antibunching with much larger
time constants of the initial photon anticorrelation. A detailed study of the
internal eight-level dynamics on the basis of optical Bloch equations allowed
a quantitative description of the observed $g^{(2)}(\tau)$ functions
\cite{Schubert95}.

\subsection{Four-level model}

\begin{figure}[h]
\centerline{\scalebox{1}{\includegraphics[]{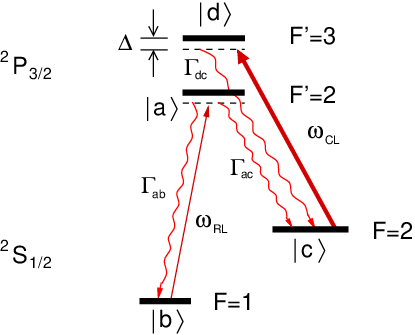}}}
\caption{Effective level structure in $^{87}$Rb used for fluorescence
  detection. The cooling laser field (CL) couples $F=2 \rightarrow F'=3$ and
  $F=2 \rightarrow F'=2$, whereas the repump laser field (RL) couples only $F=1
  \rightarrow F'=2$.}\label{pict:4level}
\end{figure}

For the fluorescence detection of a single atom in the dipole trap we use the
MOT cooling laser (CL) red detuned to the unperturbed hyperfine transition
5$^2S_{1/2}, F=2 \rightarrow$ 5$^2P_{3/2}, F'=3$ (Fig. \ref{pict:4level}) by
$\Delta= 4..5 \Gamma$. To avoid optical pumping to the hyperfine ground level
$F=1$ during the loading process we additionally shine in a repump laser (RL)
on resonance with the hyperfine transition 5$^2S_{1/2}, F=1 \rightarrow$
5$^2P_{3/2}, F'=2$. Because the atom is stored in a dipole trap, the AC
Stark-effect additionally shifts the cooling and repump laser to the red of
resonance and leads to significant optical pumping to $F=1$. Hence, we expect
a more detailed multi-level model - which includes also this pumping effect -
to correctly describe the internal dynamics of the atom-light interaction.

In general it is quite complicated to describe the situation of a laser-cooled
single atom in a dipole trap, because the six counter-propagating circularly
polarized cooling laser beams of the MOT form an intensity lattice in space
and a discrete set of points with zero intensity. The form of this
interference pattern is specific to the set of phases chosen. Due to the
finite kinetic energy corresponding to a temperature of approximately 110
$\mu$K (see preceding chapter) the atom oscillates in a classical picture with
an amplitude corresponding to several optical wavelengths. During this
oscillatory movement the atom experiences both a changing intensity and
polarization. This situation suggests to simplify the internal atomic dynamics
neglecting the Zeeman substructure of the involved hyperfine levels and to
treat the exciting cooling and repump light as two unpolarized laser fields
with an average intensity of six times the single beam intensity.

\subsubsection{Optical Bloch equations}

In the following I will focus on an atomic four-level system as indicated in
Fig. \ref{pict:4level} and the interaction with two classical light fields.
The Hamiltonian $H$ of the system reads
\begin{equation}
H=H_0+H_{int},
\end{equation}
where $H_0$ denotes the Hamiltonian of the free atom and $H_{int}$ the
interaction with the repump laser field $\mathbf{E}_{RL}$ of angular frequency
$\omega_{RL}$ and the cooling laser field $\mathbf{E}_{CL}$ of angular
frequency $\omega_{CL}$ respectively.

The equation of motion for the atomic density matrix $\rho$ of the system is
given by
\begin{equation} \label{equ:liouville}
\dot{\rho} = \frac{-i}{\hbar}[H,\rho] + R, 
\end{equation}
where the brackets $[\hbox{ },\hbox{ }]$ are defined by $[H,\rho]=H \rho-\rho
H$ and the relaxation term $R$ represents the spontaneous decay process. In
the rotating-wave approximation (RWA) we obtain for the Hamiltonian $H$ the
matrix representation in the basis of the bare atomic states $|a\rangle$,
$|b\rangle$, $|c\rangle$ and $|d\rangle$ corresponding to the hyperfine states
$\ket{F'=2}$, $\ket{F=1}$, $\ket{F=2}$ and $\ket{F'=3}$:
\begin{equation}
H = \hbar \left(
\begin{array}{cccc}
         \omega_a & -\frac{\Omega_{1}}{2} e^{-i \omega_{RL} t} &
         -\frac{\Omega_{2}}{2} e^{-i \omega_{CL} t} & 0 \\ 
         -\frac{\Omega_{1}}{2} e^{i \omega_{RL} t} & \omega_b & 0 & 0 \\ 
         -\frac{\Omega_{2}}{2} e^{i \omega_{CL} t} & 0 & \omega_c & 
	 -\frac{\Omega_{3}}{2} e^{i \omega_{CL} t} \\ 
	 0 & 0 & -\frac{\Omega_{3}}{2} e^{-i \omega_{CL} t} & \omega_d
\end{array} \right).
\end{equation}
We assume the repump laser to excite the $F'=2$ level, whereas the cooling
laser excites both hyperfine levels $F'=2$ and $F'=3$. The strength of these
couplings are described by the on-resonance Rabi frequencies $\Omega_1$,
$\Omega_2$ and $\Omega_3$ respectively. They are defined by
\begin{equation} \label{equ:Rabifrequencies}
\Omega_1=\Gamma \sqrt{\frac{I_{RL}}{2 I_{s}^{12}}}, \quad \Omega_2=\Gamma
\sqrt{\frac{I_{CL}}{2 I_{s}^{22}}} \quad \mbox{and} \quad \Omega_3=\Gamma
\sqrt{\frac{I_{CL}}{2 I_{s}^{23}}},
\end{equation}
where $\Gamma = 2\pi \times 6 \times 10^6$ Hz is the decay rate of the
transition $5^{2}P_{3/2} \rightarrow 5^{2}S_{1/2}$. For an {\it isotropic}
excitation light field (i.e. a pumping field with equal components in all {\it
three} possible polarizations) the saturation intensities of the hyperfine
transitions $F=1 \rightarrow F'=2$, $F=2 \rightarrow F'=2$ and $F=2
\rightarrow F'=3$ can be calculated \cite{Steck87} to $I_{s}^{12}=6.01$
mW/cm$^2$, $I_{s}^{22}=10.01$ mW/cm$^2$ and $I_{s}^{23}=3.58$
mW/cm$^2$. $I_{CL}$ and $I_{RL}$ denote the intensities of the cooling and
repump laser, respectively.

The relaxation term $R$ in the master equation represents spontaneous decay
\cite{Wangsness53,Shore} from the excited hyperfine levels $a$ and $d$. In
matrix representation we obtain
\begin{equation}
R = \left(
\begin{array}{cccc}
  -(\Gamma_{ab}+\Gamma_{ac})\rho_{aa} & -\gamma_{ab}\rho_{ab} &
  -\gamma_{ac}\rho_{ac} & -\gamma_{ad}\rho_{ad} \\ -\gamma_{ab}\rho_{ba} &
  \Gamma_{ab}\rho_{aa} & 0 & -\gamma_{bd}\rho_{bd} \\ 
 -\gamma_{ac}\rho_{ca} & 0 &
  \Gamma_{ac}\rho_{aa}+\Gamma_{dc}\rho_{dd} & -\gamma_{dc}\rho_{cd}
   \\ -\gamma_{ad}\rho_{da} & -\gamma_{bd}\rho_{db} & -\gamma_{dc}\rho_{dc} & -\Gamma_{dc}\rho_{dd}
\end{array} \right),
\end{equation}
where $\Gamma_{ab}$, $\Gamma_{ac}$, and $\Gamma_{dc}$ are the energy and
$\gamma_{ab}$, $\gamma_{ac}$, $\gamma_{ad}$, $\gamma_{bd}$, and $\gamma_{dc}$
are the respective phase relaxation rates. Energy relaxation from $c
\rightarrow b$ is neglected.

The energy relaxation rates appear in the diagonal elements of $R$ and
describe population loss of all involved levels due to spontaneous
emission. In detail, $\Gamma_{dc}$ is given by the spontaneous emission rate
$\Gamma$ of the $^2P_{3/2}$ excited state in $^{87}$Rb, whereas $\Gamma_{ac}$
and $\Gamma_{ab}$ refer to spontaneous decay from $a \rightarrow c$ and $a
\rightarrow b$. The ratio of $\Gamma_{ab}/\Gamma_{ac}$ is given by the
branching ratio of the respective hyperfine transitions and obeys the relation
\begin{equation}
\Gamma=\Gamma_{ac}+\Gamma_{ab}.
\end{equation}
In the case of an isotropically polarized excitation light field the
respective branching ratio is 1/2. Population loss acts also on the nondiagonal
matrix elements of $R$. In general, the phase relaxation rate of a transition
$n \rightarrow m$ is given by \cite{Shore}
\begin{equation}
\gamma_{nm}=\frac{1}{2}(\Gamma_{n}+\Gamma_{m}),
\end{equation}
where $\Gamma_{n}$ denotes the sum of any population loss of level $n$, and
$\Gamma_{m}$ of $m$ respectively. In our case the phase relaxation rates are
given by
\begin{eqnarray}
\gamma_{ab}=\frac{1}{2}(\Gamma_{ab}+\Gamma_{ac}), & \quad 
\gamma_{ac}=\frac{1}{2}(\Gamma_{ab}+\Gamma_{ac}), & \quad 
\gamma_{ad}=\frac{1}{2}(\Gamma_{ab}+\Gamma_{ac}+\Gamma_{dc}) \\
\gamma_{bd}=\frac{1}{2}\Gamma_{dc}, & \quad
\gamma_{dc}=\frac{1}{2}\Gamma_{dc}.
\end{eqnarray}
From Eq. (\ref{equ:liouville}) we derive optical Bloch equations (OBE) for
the four-level system as shown in Fig. \ref{pict:4level}. As this set of
16 coupled ordinary differential equations entails no new physics, they are not
shown explicitly. 

To solve the OBE by numerical integration \cite{NumericalRecipes} it is
convenient to suppress any explicit time dependence in the coefficients of the
OBE. Hence we introduce new variables $\hat{\rho}_{ik}$ ($i,k = a,b,c,d$)
which are given in matrix representation by
\begin{equation}
\hat{\rho} = \left(
\begin{array}{cccc}
  \rho_{aa} & \rho_{ab} e^{i \omega_{RL} t} & \rho_{ac} e^{i \omega_{CL} t} &
  \rho_{ad} \\ \rho_{ba} e^{-i \omega_{RL} t} & \rho_{bb} & \rho_{bc} e^{-i
  (\omega_{RL}-\omega_{CL}) t} & \rho_{bd} e^{-i \omega_{RL} t}\\ \rho_{ca}
  e^{-i \omega_{CL} t} & \rho_{cb} e^{i (\omega_{RP}-\omega_{CL}) t} &
  \rho_{cc} & \rho_{cd} e^{-i \omega_{CL} t} \\ \rho_{da} & \rho_{db} e^{i
  \omega_{RL} t} & \rho_{dc} e^{i \omega_{CL} t} & \rho_{dd} \\
\end{array} \right).
\end{equation}

\subsubsection{Photon correlation function}

The photon correlation function as defined in Eq. (\ref{equ:g2}) is a two-time
expectation value of the electric field operators ${\bf E^-}$ and ${\bf E^+}$
of the scattered light field. Hence, features of the photon correlation
function originate from the structure of $\mathbf{E}$. For the present case
of a four-level atom Eq. (\ref{equ:E}) can be generalized by assuming that
each dipole transition of the atom contributes to the field a term like
$\mathbf{E}^+_{S}$ in Eq. (\ref{equ:sf}). With the help of the quantum
regression theorem \cite{Lax66} the resulting correlation function is given by
\begin{equation}  \label{correlation_rho}
g^{(2)}(\tau) = \frac{\rho_{aa}(\tau) +
\rho_{dd}(\tau)}{ \rho_{aa}(\infty) +
\rho_{dd}(\infty)}, 
\end{equation} 
the ratio of the excited state populations at time $\tau$ and in the steady
state ($\tau=\infty$). To evaluate this expression, the optical Bloch
equations have to be solved for both steady-state and time-dependent cases.

The time-dependent solution depends on the initial conditions for the density
matrix, and it is this point where different correlation functions can be
distinguished \cite{Schubert95}. Physically, the initial conditions describe
the state of the atom after the emission of the first photon and consequently
depend on the properties of that photon. Thus the correlation function depends
on the wavelength of the first photon, but does not depend on the properties
of the second photon.

In the HBT experiment performed (see next section) we do not distinguish from
which transition the first photon came from. In this case the initial
conditions for the numerical calculation of the density matrix elements
$\rho_{aa}(\tau)$ and $\rho_{dd}(\tau)$ can be determined from the
steady-state solution. All matrix elements vanish, $\rho_{ij}(0)=0$, except
\begin{eqnarray}
\rho_{bb}(0) & = &\frac{\Gamma_{ab}\rho_{aa}(\infty)}{(\Gamma_{ab}+\Gamma_{ac})
    \rho_{aa}(\infty) 
    + \Gamma_{dc} \rho_{dd}(\infty)} \quad \hbox{and} \\
\rho_{cc}(0) & = &
    \frac{\Gamma_{ac}\rho_{aa}(\infty)+\Gamma_{dc}\rho_{dd}(\infty)} 
    {(\Gamma_{ab}+\Gamma_{ac})
    \rho_{aa}(\infty) 
    + \Gamma_{dc} \rho_{dd}(\infty)}.
\end{eqnarray}
These initial conditions characterize the state of the atom immediately after
the emission of any photon: the atom is in the $^2S_{1/2}$ levels $\ket{b}$
and $\ket{c}$, whose relative population is given by the probabilities for
spontaneous emission on the transitions $\ket{a} \rightarrow \ket{b}$,
$\ket{a} \rightarrow \ket{c}$ and $\ket{d} \rightarrow \ket{c}$.
 
\begin{figure}[h]
\centerline{\scalebox{1}{\includegraphics[width=12cm]{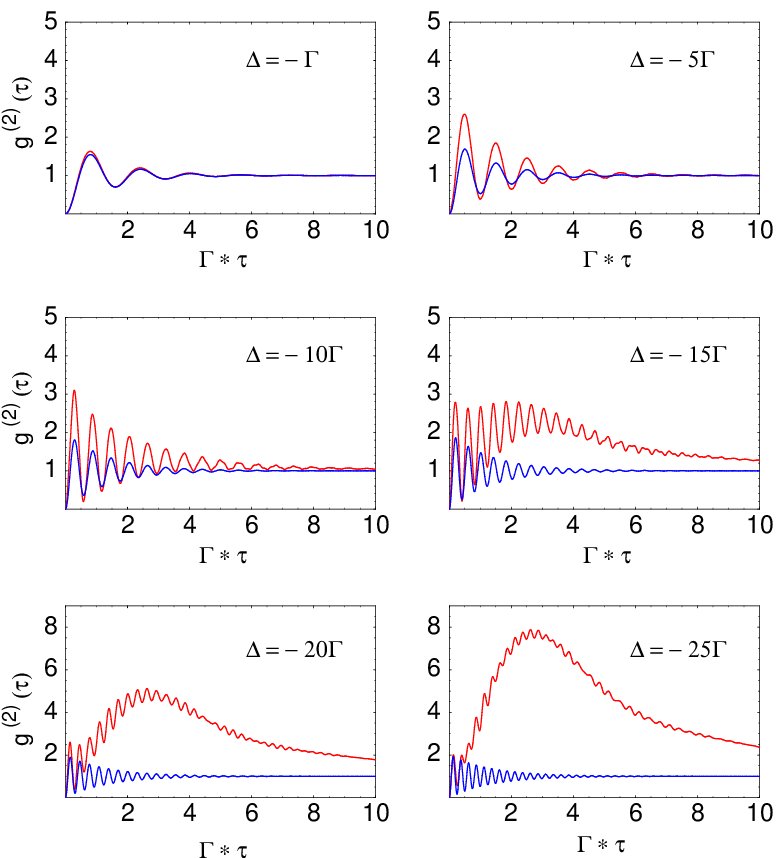}}}
\caption{Calculated intensity correlation function $g^{(2)}(\tau)$ for
  different detunings $\Delta$ of the cooling laser. Blue line: two-level
  model; red line: four-level model. Experimental parameters: $I_{CL}=100$
  mW/cm$^2$, $I_{RL}=12$ mW/cm$^2$} \label{pict:g2_24level}
\end{figure}

Following the described procedure we calculated the second-order correlation
function $g^{(2)}(\tau)$ for different detunings $\Delta$ of the cooling laser
(see Fig. \ref{pict:g2_24level}). The repump laser is supposed to be on
resonance. Increasing $\Delta$ leads to significant optical pumping to the
$F=1$ hyperfine ground level and to a breakdown of the two-level model. To
demonstrate this effect, we calculated the photon correlation function for a
two-level and four-level model for equal experimental parameters. For small
detunings of the cooling laser, the second-order correlation functions based on
a two-level and four-level model are indistinguishable. 
\clearpage

For larger detunings, corresponding to the situation of a single $^{87}$Rb
atom in a MOT or an optical dipole trap ($\Delta=-5..-10 \Gamma$), the
photon-pair correlation function of the four-level model shows deviations from
the predictions of a simple two-level model. Most prominently the amplitudes
of the Rabi oscillations exceed the upper limit of 2, which is predicted by a
two-level model. Increasing the detuning of CL further increases the coupling
to the level $\ket{a}$. Hence, the multi-level structure becomes more and more
dominant. On the one hand this gives rise to several generalized Rabi
frequencies, on the other hand to significant optical pumping to $F=1$. For
large detunings of the cooling laser ($\Delta>-20\Gamma$) the envelope of the
correlation function shows the predicted feature of a three-level
$\Lambda$-type atom \cite{Pegg86}. The calculated correlation functions in
Fig. \ref{pict:g2_24level} demonstrate this behavior very well.

\subsection{Motional effects} \label{subsect:motion}

Since the correlation function of the emitted light depends on the incident
field, the motion of the atom will modify the photon correlations. Moving
through the dipole trap the atom crosses various spots of a light interference
pattern with different intensity and polarization. In our experiment this
interference pattern is formed by the intersection of three pairs of
counterpropagating $\sigma^+$-$\sigma^-$ polarized laser beams used for laser
cooling and fluorescence detection. The polarization of resonance fluorescence
is determined by the magnetic orientation of the atom, which in turn depends
on the local light field and changes on the time scale of atomic motion over
an optical wavelength $\lambda$. Thus, in addition to correlations of the
total intensity (\ref{equ:g2}) one expects polarization effects to be visible
in two-photon correlations \cite{Gomer98}.

Our measurements of the photon statistics are only sensitive to total
intensity correlations caused by a standing light field pattern. Hence, I will
restrict the following theoretical considerations only to total intensity
correlations. Because of the entanglement of internal and external degrees of
freedom, a proper description of the atomic dynamics is a non-trivial problem
\cite{Hodapp95,Marksteiner96}. I do not intend here to present a sophisticated
theory but rather to describe the atomic motion in a simple phenomenological
way.

\subsubsection{Fokker-Planck equation}

The trapping force in a dipole trap with a superimposed 1D
$\sigma^+$-$\sigma^-$ polarization gradient cooling includes two partial
forces, i.e., the gradient force of the dipole potential (see Eq.
\ref{equ:dipforce}) and the radiation pressure force \cite{Garraway00} and can
be expressed as a damped harmonic oscillation with spring constant $\kappa$
and friction parameter $\alpha$. In a simple model random fluctuations of the
friction force may be characterized by a diffusion constant $D=k_B T/\alpha$,
where $T$ is the temperature (which has the meaning of an average kinetic
energy of the trapped atom) and $k_B$ denotes the Boltzmann constant. Treating
the dipole potential as an effective harmonic potential, the theory of Brownian
motion \cite{Metcalf99} can be used to derive a Fokker-Planck equation for the
atomic motion:
\begin{equation} \label{equ:Fokker}
\frac{\partial}{\partial t}f=\frac{\kappa}{\alpha}\frac{\partial}{\partial z}
(z f) + \frac{kT}{\alpha} \frac{\partial^2}{\partial z^2} f.
\end{equation}
Here the function $f=f(z,z_0,t)$ describes the probability density for the
atom to be at time $t$ at position $z$ for the initial atomic position
$z(t=0)=z_0$.

For atomic motion in potential-free space (situation in an optical molasses),
the Fokker-Planck equation reduces to a diffusion equation
\begin{equation} \label{equ:Diffusion}
\frac{\partial}{\partial t}f=D\frac{\partial^2}{\partial z^2} f.
\end{equation}
The probability function $f$ which solves this differential equation has the
form
\begin{equation}
f(z,z_0,t)=\frac{1}{\sqrt{2\pi} \sigma(t)} e^{-\frac{(z-z_0)^2}{2\sigma^2(t)}},
\end{equation}
where $\sigma(t)=\sqrt{2Dt}$ describes the temporal spread of the atomic
positional probability, which is an expanding Gauss if there is no potential.

\subsubsection{Total intensity correlation function}

In order to gain physical insight how the diffusive motion of a single atom in
an intensity modulated standing light field will influence the total intensity
correlation function, I will start with an atom modeled by a classical emitter
with an induced dipole moment proportional to the local light field. Then the
corresponding total correlation function is given by
\begin{equation}
g_{tot}^{(2)}(\tau)=\frac{\int\int_{-\infty}^{\infty} I(z_0) I(z) 
                    f(z_0,z_0,\infty) f(z,z_0,\tau) dz dz_0}{\langle
                    I(t)\rangle^2},
\end{equation}
where the position dependent fluorescence intensity is given by $I(z)$ and the
probability density for the atom to be at time $t$ at $z$, if its initial
position was $z(t=0)=z_0$, is $f(z,z_0,t)$. 

Let's assume that the atom moves ``diffusively'' in potential-free space in a
one-dimensional light field configuration, produced by two plane waves
counterpropagating along $z$ with the same frequency, equal amplitudes and
equal linear polarizations ($I(z)\propto \cos^2(kz)$). Then the total
second-order correlation function $g^{(2)}_{tot}(\tau)$ of the emitted
fluorescence light is given by \cite{Gomer98}
\begin{equation} \label{equ:g2_diffusion}
g^{(2)}_{tot}(\tau)=1+\frac{1}{2} e^{-2 k^2 \sigma^2(\tau)}.
\end{equation}
This correlation function contains information on the character of the atomic
motion. For instance, diffusion ($\sigma \propto \sqrt{t}$) is indicated by an
exponential decay whereas ballistic motion ($\sigma \propto t$) yields a
Gaussian decay. All correlations vanish after the atom has moved more than a
distance of $\lambda/2$. The corresponding decay time constant $\tau_0$ is
then given by
\begin{equation}
\tau_0=\frac{1}{4 k^2 D},
\end{equation}
where $k=2\pi/\lambda$. As the diffusion constant $D$ depends on the atomic
temperature $T$ and the friction parameter $\alpha$, the preceding relation
allows to determine the atomic temperature provided the friction parameter was
calculated \cite{Garraway00} or measured.

\section{Experimental determination of the photon statistics}

\subsection{Setup}

\begin{figure}[b]
\centerline{\scalebox{1}{\includegraphics[width=9cm]{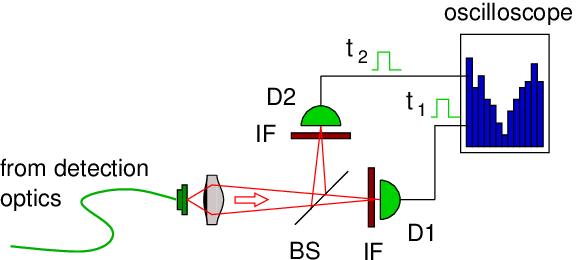}}}
\caption{Hanbury-Brown-Twiss setup to measure the photon pair correlation
function $g^{(2)}(\tau)$. The fluorescence light is sent through a beam
splitter BS onto two single photon detectors D1, D2 to record detection time
differences $\tau = t_1 - t_2$.}
\label{Bild:hbtsetup}
\end{figure}

The correlation function $g^{(2)}(\tau)$ of the collected fluorescence light
is measured in a standard Hanbury-Brown-Twiss configuration with two single
photon detectors ($D1$ and $D2$) behind a beam splitter (BS,
Fig. \ref{Bild:hbtsetup}). To suppress cross talk in the optical setup,
infront of each detector we insert a bandpass filter (IF) blocking unwanted
fluorescence light caused by the breackdown flash of the other silicon
avalanche photo diode which follows the detection of a photon
\cite{Kurtsiefer01}. The differences of detection times $\tau = t_1 - t_2$ of
photon pair events are then recorded in a storage oscilloscope with a time
resolution of 100 ps. However, the minimum bandwidth of the measurements is
limited by the 1.4 ns timing jitter of the detectors.

The normalized distribution of time differences $\tau$ is equivalent to the
second order correlation function $g^{(2)}(\tau)$ as long as $\tau$ is much
smaller than the mean time difference between two detection events
\cite{Reynaud83}. For correct normalization of the measured $g^{(2)}(\tau)$ we
divide the coincidences in each time bin $\Delta\tau$ by $r_1 \times r_2
\times \Delta\tau \times T_{int}$, where $r_1$ and $r_2$ are the mean count
rates of the two detectors, and $T_{int}$ is the total integration time with
an atom in the trap.

\begin{figure}[h]
\begin{center}
\includegraphics[]{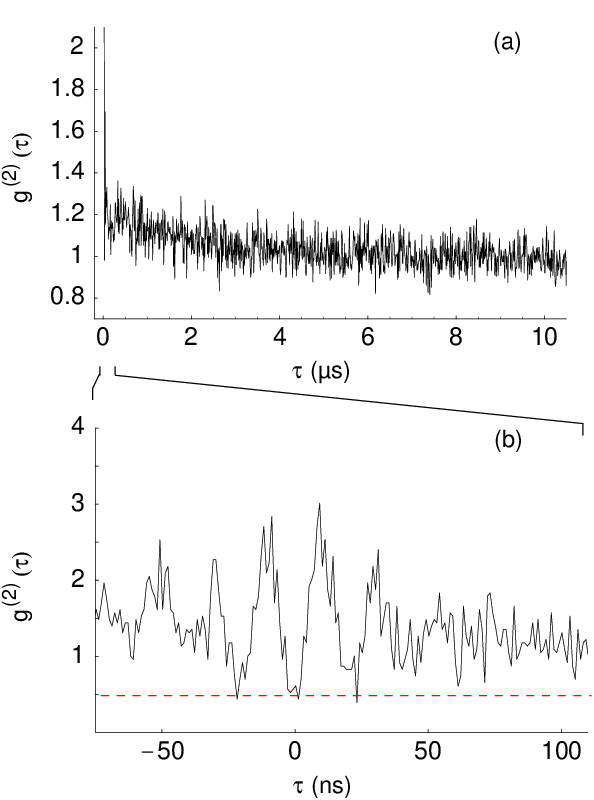}
\caption{Second-order correlation function $g^{(2)}(\tau)$ of the resonance
  fluorescence of a single $^{87}$Rb atom. {\bf (a)} On long
  timescales, $g^{(2)}(\tau)$ shows a monotonous decay for $\mid\tau\mid
  \le 2..3 \mu$s. {\bf (b)} On short timescales, clear photon anti-bunching at
  $\tau=0$ and oscillations due to Rabi flopping are observed. The red dashed
  line indicates accidental coincidences due to the dark count rate of the
  detectors. Experimental parameters: $I_{CL}=103$ mW/cm$^2$, $I_{RL}=12$
  mW/cm$^2$, $\Delta/2\pi=-31$ MHz, $\hat{U}=0.38$ mK.} \label{Bild:hbtmessung}
\end{center}
\end{figure}
\clearpage

\subsection{Experimental results}

In a first experiment we measured the second-order correlation function of the
light collected from the center of our dipole trap potential in the continous
loading mode of the trap (cooling and repump lasers are present during the
whole measurement time). In this operating regime we observe characteristic
steps in the detected fluorescence signal (see Fig. \ref{pict:trace}). To
minimize background contributions, photon correlations are acquired only at
times we observe fluorescence exceeding a threshold of 1200 counts per second
(cps), i.e. when an atom was inside the trap.

Fig. \ref{Bild:hbtmessung} shows the resulting correlation function
$g^{(2)}(\tau)$ for a dipole potential of $\hat{U}=0.38 \pm 0.04$ mK, a total
cooling laser intensity $I_{CL}\approx 103 $mW/cm$^2$ and detuning
$\Delta/2\pi$ of -31 MHz respectively. On a $\mu$s timescale the correlation
function shows an exponential decay from the asymptotic value 1.24 at $\tau=0$
to 1.0 for large $\tau$ with a time constant of 1.8 $\mu$s. This decay is
caused by the fluorescence of an atom undergoing diffusive motion in the
intensity-modulated light field of the cooling beam configuration (see
\ref{subsect:motion}).

\begin{figure}[h]
\begin{center}
\includegraphics[width=10cm]{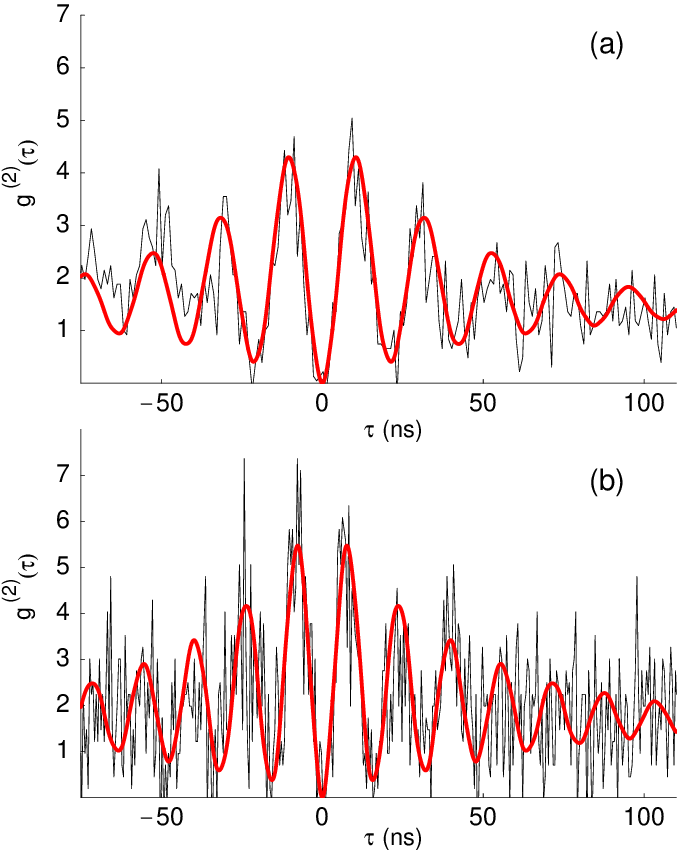}
\caption{Intensity correlation function $g^{(2)}(\tau)$ (background corrected)
of the resonance fluorescence from a single $^{87}$Rb atom in the dipole trap
for two different trap depths. Red solid line: calculation. Experimental
parameters: $I_{CL}=103$ mW/cm$^2$, $I_{RL}=12$ mW/cm$^2$, $\Delta/2\pi=-31$
MHz, {\bf (a)} $\hat{U}=0.38$ mK, {\bf (b)} $\hat{U}=0.81$ mK.}
\label{Bild:hbttheory}
\end{center}
\end{figure}

On short timescales, most prominently the uncorrected minimum value
$g^{(2)}(0)=0.52 \pm 0.14$ at zero delay $\tau=0$ clearly proves photon
anti-bunching of the emitted fluorescence. Including accidental coincidences
due to a dark count rate of 300 $s^{-1}$ of each detector (dashed red line in
Fig. \ref{Bild:hbtmessung}), we derive a corrected minimum value
$g^{(2)}_{corr}(\tau=0)=0.02 \pm 0.14$. Within our experimental errors this is
compatible with perfect photon anti-bunching verifying the presence of only
one single atom in the dipole trap. Furthermore, the signature of
Rabi-oscillations is observed due to the coherent interaction of the cooling
and repump laser fields with atomic hyperfine levels involved in the
excitation process. The oscillation frequency is in good agreement with the
simple two-level model \cite{Carmichael76} and the oscillation amplitude is
damped out on the expected timescale of the $5 ^{2}P_{3/2}$ excited state
lifetime. Increasing the dipole laser power from 16.7 mW to 35.5 mW without
changing the laser cooling parameters increases the detuning of the cooling
laser to the hyperfine transition $5 ^{2}S_{1/2}(F=2) \rightarrow 5
^{2}P_{3/2}(F'=3)$ due to an increase of the AC Stark-shift of the atomic
levels in the dipole trap laser field. This effect was observed as the
expected increase of the oscillation frequency from 47.5 MHz to 62.5 MHz (see
Fig. \ref{Bild:hbttheory}).

\begin{figure}[h]
\centerline{\scalebox{1}{\includegraphics[width=10cm]{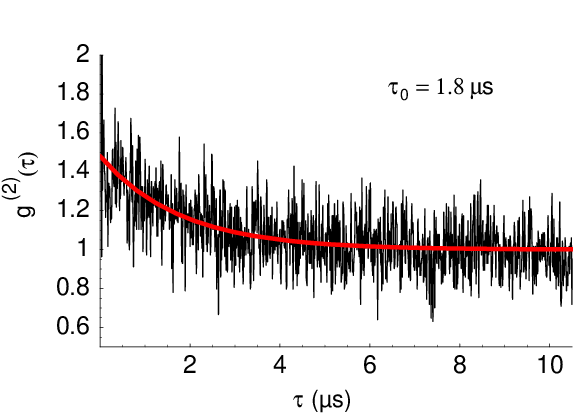}}}
\caption{Background corrected second-order correlation function
  $g^{(2)}(\tau)$ in the resonance fluorescence of a single atom in the dipole
  trap and fit with model-function (\ref{equ:g2_diffusion}). Experimental
  parameters: $I_{CL}=103$ mW/cm$^2$, $I_{RL}=12$ mW/cm$^2$,
  $\Delta/2\pi=-31$ MHz, $\hat{U}=0.38$ mK.}
\label{pict:hbtTheorie2}
\end{figure}

In contrast to a two-level atom, the measured and background corrected
correlations in Fig. \ref{Bild:hbttheory}(b) show a maximum value of 6 close
to $\tau=0$. This increase of the oscillation amplitude is a consequence of
the atomic multi-level structure and can be explained very well by the
four-level model presented in the theory section of this chapter.

For a detailed interpretation of the measured $g^{(2)}(\tau)$ functions on a
short timescale we calculated $g^{(2)}(\tau)$ on the base of the four-level
model for given experimental parameters and multiplied it with the function
$1+Ae^{-\tau/\tau_0}$, describing the additional decay contribution due to
motion in the intensity modulated light field of the cooling beam
configuration. The parameters $A$ and $\tau_0$ we determined from an extra fit
to the background corrected correlation function on the $\mu$s time scale (see
Fig. \ref{pict:hbtTheorie2}). The light-shift of the atomic hyperfine levels
is calculated on the basis of Eq. \ref{equ:lightshiftgeneral2} for both the
ground levels $F=1$ and $F=2$ as well as the relevant hyperfine levels $F'=2$
and $F'=3$ of the $5^2P_{3/2}$ state and included into the four-level
model. Furthermore a reduction of the light-shift potential is incorporated in
the calculations due to a finite kinetic energy of the atom of approximately
100 $\mu$K. For two different depths of the optical trapping potential we find
good agreement of the measured second order correlation function
$g^{(2)}_{\tau}$ with the calculated one (see Fig. \ref{pict:hbtTheorie2}).

\section{Conclusion and discussion}

To prove that only single atoms are stored in our dipole trap, the statistical
properties of the detected fluorescence light were studied with a
Hanbury-Brown-Twiss setup. The measured second-order correlation function
$g^{(2)}(\tau)$ of the detected fluorescence light exhibits strong photon
antibunching verifying the presence of a single atom. In addition the measured
two-photon correlations show the internal and external dynamics of the atomic
hyperfine levels involved in the excitation process. Due to the AC Stark-shift
of the atomic levels in the dipole potential and the resulting increase of the
detuning of the cooling light, significant population is continuously
transferred to the second hyperfine ground state 5$^2S_{1/2}, F=1$. Thus the
atom-light interaction can not be modeled with a simple two-level model. An
atomic four-level model is developed and its predictions are compared with
measured experimental data. Within our experimental errors we find good
agreement of the calculated second order correlation function with the
measured correlation function.

%-----------------------------------------------------------------------------
%-----------------------------------------------------------------------------

\chapter{Detection of atomic superposition states}
\label{chapter:AtomicStateDetection}

\section{Introduction}

To verify entanglement between the spin state of a single atom and the
polarization of a spontaneously emitted photon, one has to perform correlated
local measurements of the internal quantum states of the atom and the
photon. The polarization state of a single photon can be measured relatively
simply by a combination of a polarization filter and a single photon
detector. However, the measurement of the spin state of a single atom is not
trivial and therefore the primary experimental challenge of the present work.

In experiments with trapped ions, the state of an atomic qubit encoded in the
internal atomic level structure can be measured with almost perfect detection
efficiency on the basis of the ``electron shelving'' method
\cite{Dehmelt75}. This technique has the relevant property that depending on
the internal quantum state, the atom scatters fluorescence light from an
incident laser light field. Hence, the presence or absence of atomic
fluorescence light indicates directly the quantum state of the atom.

In this chapter I will show how the experimental techniques of coherent
population trapping (CPT) and stimulated Raman adiabatic passage (STIRAP) can
be used for phase-sensitive probing of coherent superposition states of a
single atom. Therefore I will first give a short theoretical review on
coherent population trapping and adiabatic population transfer in three- and
four-level systems, before I present experiments that allow to detect a single
Rubidium 87 atom in a coherent superposition
$\ket{m_F=-1}+e^{i\phi}\ket{m_F=+1}$ of the Zeeman sublevels $m_F=\pm1$ of the
hyperfine ground state $^2$S$_{1/2}$, $F=1$.

\section{Theoretical framework}

Atomic states that do not couple to an incident light field are called dark
states and are used in the present work in various ways to prepare and analyze
atomic states. In general one can distinguish three kinds of dark states. (1)
The atom is in a state that can couple to a light field due to selection rules
of electromagnetic dipole transitions. But the incident light field is detuned
far-off-resonance and therefore couples negligibly to the incident light. (2)
The atom is in a state that can not couple to a light field due to atomic
selection rules. (3) The atom is in a coherent superposition of two ground
states and exposed to respective resonant light fields. But due to destructive
interference of the excitation amplitudes the absorption is canceled and the
atom remains dark.

Dark states of the first two kinds can be populated relatively simply by
optical pumping and are insensitive to experimental parameters, provided the
atomic states do not mix due to a strong perturbation effect, e.g. the
interaction with a strong electric or magnetic field. Coherent dark states of
the third case are much more sensitive to experimental parameters because they
depend on the phase coherence of the incident light fields and on the temporal
evolution of the respective atomic energy levels. The understanding for the
preparation and the analysis of coherent dark states is essential for the
state selective detection of the atomic qubit. Hence, I will give a short
review on coherent population trapping and its applications.

\subsection{Coherent population trapping - dark states}

\begin{figure}[h]
\centerline{\scalebox{1}{\includegraphics[]{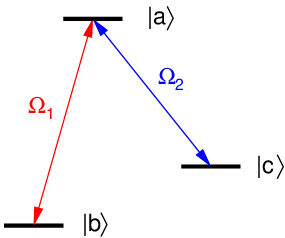}}}
\caption{Three-level system coupled by two lasers of angular frequencies
  $\omega_1$ and $\omega_2$, respectively. Due to destructive interference of
  the respective transition amplitudes the linear superposition
  $\ket{b}-e^{i\phi}\ket{c}$ of the ground states does not couple to the
  resonant light fields (see Eq. \ref{equ:darkstategeneral}).}
\label{pict:STIRAP_lambda}
\end{figure}

We consider now coherent population trapping in a three-level atom interacting
with two classical light fields of angular frequencies $\omega_1$ and
$\omega_2$ as shown in Fig. \ref{pict:STIRAP_lambda}. We assume the atom has
only three energy levels in a so-called $\Lambda$ configuration in which two
lower levels $\ket{b}$ and $\ket{c}$ are coupled to a single upper level
$\ket{a}$.

The Hamiltonian $H$ for the system, in the rotating-wave approximation, is
given by
\begin{equation}
H = H_0+H_1,
\end{equation}
where
\begin{eqnarray}
H_0 = \hbar (\omega_a \ket{a}\bra{a} + \omega_b \ket{b}\bra{b} +
      \omega_c \ket{c}\bra{c}), \\ 
H_1 = -\frac{\hbar}{2}(\Omega_1 e^{-i (\omega_{1}+\phi_1) t} \ket{a}\bra{b} +
                       \Omega_2 e^{-i (\omega_{2}+\phi_2) t} \ket{a}\bra{c})
		       + H.c.
\end{eqnarray}
Here $\Omega_1 \exp{(-i \phi_1)}$ and $\Omega_2 \exp{(-i \phi_2)}$ are the
Rabi frequencies associated with the coupling of the field modes to the atomic
transitions $\ket{a}\rightarrow\ket{b}$ and $\ket{a}\rightarrow\ket{c}$,
respectively. The atomic wave function of the system can be written in the
form
\begin{equation}
\ket{\psi(t)}= c_a(t) e^{-i \omega_a t}\ket{a} + 
               c_b(t) e^{-i \omega_b t}\ket{b} +
               c_c(t) e^{-i \omega_c t}\ket{c},
\end{equation}
where the $c_i(t)$ are the complex amplitudes of the respective atomic basis
states $\ket{i}$. Its dynamics is governed by the time-dependent
{Schr\"odinger} equation $i \hbar\ket{\dot{\psi}(t)}=H \ket{\psi(t)}$.

We now assume the initial atomic state to be a superposition of the two lower
levels $\ket{b}$ and $\ket{c}$
\begin{equation}
\ket{\psi(0)}=\cos{\theta}\ket{b}+\sin{\theta} e^{-i \psi}\ket{c}.
\end{equation}
For $\Omega_1=\Omega_2$, $\theta=\pi/4$, and $\phi_1-\phi_2-\psi=\pm\pi$ it
can be verified \cite{Scully} that
\begin{equation} \label{equ:darkstategeneral}
\ket{\psi_d(0)}=\frac{1}{\sqrt{2}}(\ket{b}-e^{i\phi} \ket{c})
\end{equation}
is one of three eigenstates of the Hamiltonian $H$, whereby the phase
$\phi=\phi_1-\phi_2$ is given by the phase difference between $\Omega_1$ and
$\Omega_2$. The mentioned eigenstate has a zero eigenvalue and it is the only
eigenstate which does not include a contribution of the excited level
$\ket{a}$. Due to destructive interference of the transition amplitudes the
atomic population is {\it trapped} in a coherent superposition of the lower
states and there is no absorption or scattering even in the presence of
resonant light fields. So this state is called ``dark''. However, the
orthogonal state $\ket{b}+e^{i\phi} \ket{c}$ couples to the incident light
fields $\Omega_1$ and $\Omega_2$ because of constructive interference of
$\Omega_1$ and $\Omega_2$. Hence, this state is called ``bright''.

\subsection{Stimulated Raman adiabatic passage}

\begin{figure}[t]
\centerline{\scalebox{1}{\includegraphics[]{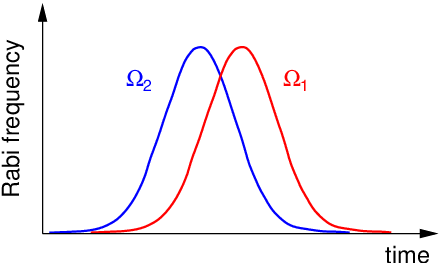}}}
\caption{Time dependence of the Rabi frequencies required for efficient
  population transfer from the initial state $\ket{b}$ to the final state
  $\ket{c}$ under adiabatic following conditions \cite{Kuklinski89}.}
\label{pict:STIRAP_pulse}
\end{figure}

An interesting and powerful application of coherent population trapping in
three-state atoms is the transfer of population with a counter-intuitive
sequence of pulses \cite{Oreg84}. This technique called stimulated Raman
adiabatic passage (STIRAP) allows, in principle, a complete coherent
population transfer from a single initial to a single final quantum state
\cite{Oreg84,Kuklinski89,Coulston91,Laine96,Vitanov97a,Vitanov97b}. The
underlying physical mechanism is the existence of an adiabatically decoupled
(or dark) state, which at early and late times coincides with the initial and
target quantum state respectively. This process is robust against moderate
variations of the pulse form, area and delay between the pulses and therefore
is well suited for experimental applications.

Consider again the three-level system in
Fig. \ref{pict:STIRAP_lambda}. Provided we start at $t=0$ with the atom in
state $\ket{b}$ and $\Omega_1=0$ with $\Omega_2$ finite and then proceed to
turn $\Omega_2$ off while adiabatically turning $\Omega_1$ on (see
Fig.\ref{pict:STIRAP_pulse}), we will end up with the atom in the state
$\ket{c}$. This is made clear by realizing that the atom is in the
time-dependent dark state \cite{Kuklinski89}
\begin{equation} \label{equ:darkstateSTIRAP}
\ket{\psi_d(t)}= \cos{\theta(t)} \ket{b} - \sin{\theta(t)} e^{i\phi}
\ket{c},
\end{equation}
where
\begin{equation}
\tan{\theta(t)}=\frac{\Omega_1(t)}{\Omega_2(t)},
\end{equation}
and $\phi$ is the relative phase between $\Omega_1$ and $\Omega_2$. When the
pulse $\Omega_2$ precedes the pulse $\Omega_1$, the mixing angle $\theta$ is
initially zero. Therefore the trapped state $\ket{\psi_d(t)}$ coincides
initially with state $\ket{b}$ while $\bra{c}\psi_d(t=0)\rangle=0$. When the
pulse areas are sufficiently large, $\Omega_{eff}T\gg 1$, where
$\Omega_{eff}=\sqrt{\Omega_1^2+\Omega_2^2}$ is the effective two-photon Rabi
frequency and $T$ is the interaction time, non-adiabatic coupling to
intermediate state $\ket{a}$ is small. According to
(\ref{equ:darkstateSTIRAP}) the population then remains in the dark state
$\ket{\psi_d(t)}$ and evolves into $\ket{c}$, depending on the evolution of
the mixing angle $\theta$.

\subsection{Tripod STIRAP}

An interesting and powerful extension of STIRAP is tripod-STIRAP which allows
to create or probe, in a robust way, a superposition of atomic states
\cite{Unanyan98,Theuer99,Unanyan99,Vewinger03}.

\begin{figure}[t]
\centerline{\scalebox{1}{\includegraphics[width=5.5cm]{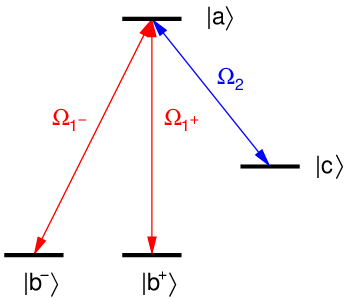}}}
\caption{Four-level system of tripod-STIRAP coupled by three lasers. The Rabi
  frequencies of the pump and Stokes laser are $\Omega_1$ and $\Omega_2$,
  respectively.}
\label{pict:TripodSTIRAP}
\end{figure}

In this case the coupling scheme consists of four atomic levels coupled by
three lasers (see Fig. \ref{pict:TripodSTIRAP}). The Hamiltonian of a resonant
tripod system has four adiabatic states \cite{Unanyan98} that are parametrized
by two mixing angles $\theta(t)$ and $\Phi(t)$, defined by
\begin{equation}
\tan{\theta(t)}=\frac{\Omega_1(t)}{\Omega_2(t)}, \quad \hbox{and} \quad
\tan{\Phi(t)}=\frac{\Omega_{1^-}(t)}{\Omega_{1^+}(t)},
\end{equation}
where $\Omega_{1^-}$, $\Omega_{1^+}$ and $\Omega_{2}$ are the Rabi frequencies
of the coupling lasers and
$\Omega_1(t)=\sqrt{\Omega_{1^+}^2(t)+\Omega_{1^-}^2(t)}$. Two of the adiabatic
states are orthogonal degenerate dark states, i.e., states without components
of the intermediate state $\ket{a}$,
\begin{eqnarray}
\ket{\psi_d^1(t)}&=&\cos{\theta(t)}\left(\sin{\Phi(t)}\ket{b^-} +
                                \cos{\Phi(t)}e^{i\phi_1}
				\ket{b^+}\right)
				- \sin{\theta(t)}e^{i\phi_2}\ket{c}, \\
\ket{\psi_d^2(t)}&=&\cos{\Phi(t)}\ket{b^-} -
                    \sin{\Phi(t)}e^{i\phi_1}\ket{b^+},
\end{eqnarray}
where $\phi_1$ and $\phi_2$ are the relative phase between $\Omega_{1^-}$ and
$\Omega_{1^+}$, and $\Omega_{1^-}$ and $\Omega_{2}$ respectively.

In the present experiment, there is only one linearly polarized pump laser,
which produces two coincident and copropagating $\sigma^+$ and $\sigma^-$
polarized fields with the same intensity; hence $\Omega_{1^-}=\Omega_{1^+}$
and $\Phi=\pi/4$. Moreover, because the pump field $\Omega_1$ is delayed with
respect to the Stokes field $\Omega_2$, we have $\theta(0)=0$ and
$\theta(+\infty)=\pi/2$. Hence, in the adiabatic limit, the atom evolves along
the adiabatic path $\ket{\psi_d^1(t)}$ \cite{Unanyan98} from the initial state
\begin{equation}
\ket{\psi_i}=\frac{1}{\sqrt{2}}\left(\ket{b^-}+e^{i\phi_1}\ket{b^+}\right)
\end{equation}
to the final state $\ket{c}$. If the atom was initially prepared in the
orthogonal state $1/\sqrt{2} (\ket{b^-}-e^{i\phi_1}\ket{b^+})$, the
population will remain in this dark state during the STIRAP pulse sequence and
will not be transferred to $\ket{c}$.

\subsubsection{Choice of the atomic measurement basis}

For $\phi_1=0$, the associated bright and dark states 
\begin{equation}
\frac{1}{\sqrt{2}}(\ket{b^-}+\ket{b^+}) \quad \hbox{and} \quad
\frac{1}{\sqrt{2}}(\ket{b^-}-\ket{b^+}),
\end{equation} 
form an orthonormal basis for the two ground states $\ket{b^-}$ and
$\ket{b^+}$. This basis is denoted in the following chapter the $\sigma_x$
basis, provided that the ground states $\ket{b^-}$ and $\ket{b^+}$ are
identified with the respective eigenstates $\ket{\su}$ and $\ket{\sd}$ of
$\sigma_z$. To measure the atomic qubit in the complementary $\sigma_y$ basis
one has to set the relative phase $\phi_1$ of $\Omega_{1^-}$ and
$\Omega_{1^+}$ to $\pi/2$. We find, that the corresponding set of orthogonal
bright and dark states
\begin{equation}
\frac{1}{\sqrt{2}}(\ket{b^-}+i\ket{b^+}) \quad \hbox{and} \quad
\frac{1}{\sqrt{2}}(\ket{b^-}-i\ket{b^+}).
\end{equation}
are the eigenstates of $\sigma_y$.

\section{Phase-sensitive probing of Zeeman superposition states}

\begin{figure}[h]
\centerline{\scalebox{1}{\includegraphics[width=9.5cm]{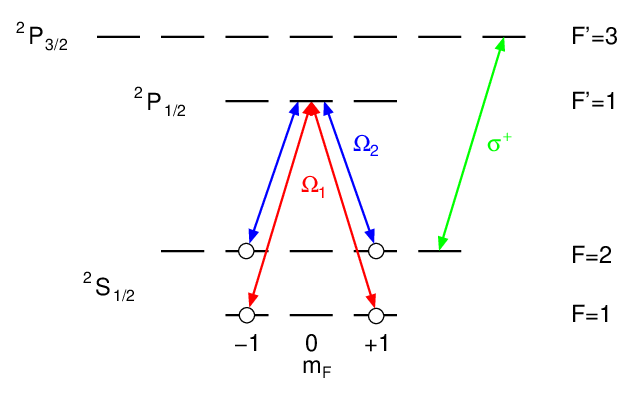}}}
\caption{Coupling scheme for the analysis of coherent superposition
  states. The STIRAP process transfers the superposition
  $\frac{1}{\sqrt{2}}(\ket{1,-1}+e^{i\phi}\ket{1,+1})$ to the $F=2$ ground
  state whereas the orthogonal state
  $\frac{1}{\sqrt{2}}(\ket{1,-1}-e^{i\phi}\ket{1,+1})$ remains in $F=1$. To
  distinguish the hyperfine ground states, a $\sigma^+$ polarized detection
  laser pulse resonant to the cycling transition $^2S_{1/2},F=2 \rightarrow$
  $^2P_{3/2},F'=3$ is used.}
\label{pict:STIRAP_termschema}
\end{figure}

\subsection{Introduction}

For the detection of atomic superposition states we set up a STIRAP-laser
system which couples a superposition of the Zeeman sub-levels $m_F=-1$ and
$m_F=+1$ of the hyperfine ground level $F=1$ to a superposition of Zeeman
sublevels of $F=2$. Because the adiabatic population transfer has to evolve
via a dark state that contains only the initial levels (here $\ket{1,-1}$ and
$\ket{1,+1}$) and the target state $F=2$ (a certain superposition of Zeeman
sublevels), a proper readout of the atomic phase information can only be
realized if the intermediate level of the STIRAP process has the same or less
Zeeman manifold as the hyperfine ground level $F=1$.

In the present experiment the atomic state detection is performed in two
steps.
\begin{itemize}
\item A linearly polarized STIRAP-laser pulse, propagating along the
  quantization axis $z$ (parallel to the observation direction of atomic
  fluorescence light), couples to a superposition
  $\frac{1}{\sqrt{2}}(\ket{F=1,m_F=-1}+e^{i\phi}\ket{F=1,m_F=+1})$ and
  transfers the atomic population adiabatically via the intermediate level
  $^2P_{1/2}, F=1, m_F=0$ to the final level $F=2$ while the orthogonal state
  $\frac{1}{\sqrt{2}}(\ket{F=1,m_F=-1}-e^{i\phi}\ket{F=1,m_F=+1})$ remains in
  $F=1$. The relative phase $\phi=2\alpha$ between $\ket{F=1,m_F=-1}$ and
  $\ket{F=1,m_F=+1}$ is determined by the linear polarization angle $\alpha$
  of the pump field $\Omega_1$ with respect to the $x$ axis.
\item After the state-selective population transfer the atom is in a
superposition of the two hyperfine ground states $F=1$ and $F=2$. To
discriminate these states we apply a detection laser pulse resonant to the
cycling transition $^2S_{1/2}, F=2 \rightarrow$ $^2P_{3/2},F'=3$. Provided
the atom is in $F=2$ it scatters photons from this laser beam, and with each
scattering event the atom acquires an additional photon momentum $\hbar\vec
k$. After approximately 40 to 50 scattering events the atom is pushed out of
the trap. Finally, to read out the atomic state the cooling and repump lasers
of the MOT are switched on and the atomic fluorescence light is integrated for
30..60 ms to decide if the atom is still in the trap or not.
\end{itemize}

\begin{figure}[h]
\centerline{\scalebox{1}{\includegraphics[width=7cm]{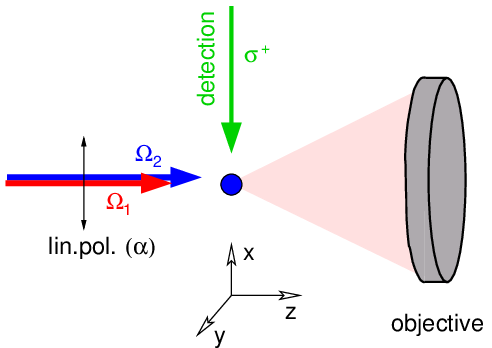}}}
\caption{Geometry of the experiment. The two circularly polarized pump laser
  components (here denoted by $\Omega_1$) are generated by a linearly
  polarized laser propagating in the $z$ direction, whose polarization forms
  an angle $\alpha$ with the $x$ axis. The propagation of the circularly
  polarized detection pulse is parallel to the $x$ axis.}
\label{pict:STIRAP_setup1}
\end{figure}

The experimental confirmation of this sequence is accomplished in two
steps. In a first simple experiment we verify that we can distinguish atomic
populations in the hyperfine ground levels $F=1$ and $F=2$. Then we prepare
atomic Zeeman superposition states via CPT. Finally, to read out the phase
$\phi$ of such superpositions we apply a STIRAP technique.

\subsection{Hyperfine state preparation and detection}

\begin{figure}[h]
\centerline{\scalebox{1}{\includegraphics[width=12cm]{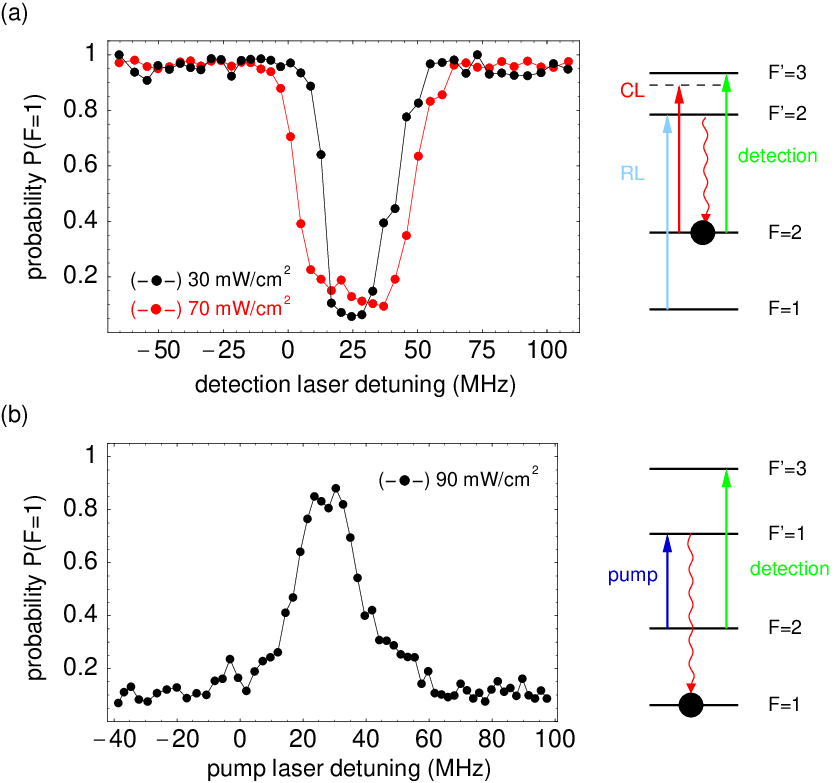}}}
\caption{Probability to detect the atom in the hyperfine ground state $F=1$
  after application of a hyperfine state selective detection laser pulse as a
  function of (a) the detection, (b) the pump laser detuning provided the atom
  is initially prepared in (a) $F=2$, (b) $F=1$ respectively. The relative
  detunings refer to transitions undisturbed by the AC-Stark effect of the
  dipole laser beam.}
\label{pict:PushPump_result}
\end{figure} 

The analysis of coherent superposition states requires the ability to
discriminate in a second step atomic population in $F=1$ from $F=2$ with high
efficiency. This task can be accomplished by scattering atomic fluorescence
light from the closed transition $^2S_{1/2}, F=2 \rightarrow$ $^2P_{3/2},
F'=3$ by applying resonant circularly polarized laser light. Provided the atom
is prepared in $F=1$, then the atomic population is shelved in this ``dark''
state because the incident laser field does not couple to this level and the
atom scatters no light. In contrast, if the atom is prepared in any Zeeman
sub-level of $F=2$ it will be pumped within a few scattering cycles to the
outer Zeeman sub-level $m_F=\pm2$ and will scatter many photons from the ideal
two-level transition $^2S_{1/2}, F=2, m_F=\pm2 \rightarrow$ $^2P_{3/2}, F'=3,
m_F'=\pm3$ whereby off-resonant excitation to $^2P_{3/2}, F'=2$ is suppressed
by atomic selection rules. The presence or absence of atomic fluorescence
light indicates in which hyperfine ground state the atom is.

In contrast to electromagnetically trapped ions, shelving can not be applied
in the usual way to optically trapped atoms. The main reason for this is, that
atoms are removed from a dipole trap by the transfer of few photon
recoil-momentum kicks $\hbar \vec{k}$. Therefore, it is not possible to
scatter atomic fluorescence light for at least 10 ms, which is the minimum
integration time to discriminate ``one'' atom in the trap from the case where
``no'' atom is present whereby the detected fluorescence rate is given by the
dark count of the single photon detector (see chapter
\ref{chapter:diptrap}). But combining this push-out effect with a redetection
sequence of the atom realizes a destructive state detection
\cite{Vrana04}. Provided that the atom is detected after application of the
detection pulse we know that the population was in $F=1$. However, if the atom
is not detected we know that the atom was in $F=2$.

To demonstrate this property, the atom is prepared in the hyperfine ground
state $F=2$ by optical pumping. For this purpose we switch on the cooling and
repump laser of our MOT and wait for a single atom. If the atomic fluorescence
exceeds a certain threshold value, the cooling laser is switched off 4 ms
before the repump laser. As the repump laser serves to depopulate the $F=1$
level the whole atomic population is pumped to the ``dark'' state $F=2$. Then
we apply a 10 $\mu$s long circularly polarized detection laser pulse with a
variable detuning and finally we switch on again the cooling and repump-laser
and collect fluorescence light from the dipole trap region for 55 ms, in order
to check whether the atom is still in the trap or not.

\begin{figure}[t]
\centerline{\scalebox{1}{\includegraphics[]{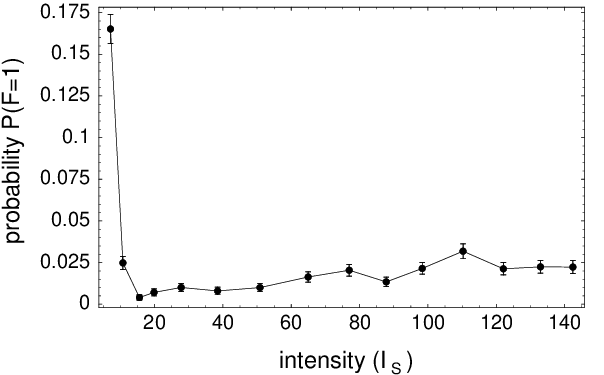}}}
\caption{Probability to detect a single atom in the hyperfine ground state
  $F=1$ as a function of the detection laser intensity in units of the
  saturation intensity $I_S$ after the atom was pumped to the hyperfine ground
  state $F=2$. Optimal hyperfine state detection is achieved for 20..40
  saturation intensities $I_S$ and a pulse duration of 10 $\mu$s.}
\label{pict:cycling_powerscan_2}
\end{figure}

In Fig. \ref{pict:PushPump_result} (a) the probability of redetection is
plotted as a function of the detection laser detuning for different laser
intensities. If the detection pulse is resonant to the light-shifted
transition $^2S_{1/2}, F=2 \rightarrow$ $^2P_{3/2}, F'=3$ the atom is removed
from the dipole trap. If the detection pulse is too intense (red data points)
the maximum detection efficiency on resonance will drop due to an increase of
nonresonant excitation to $^2P_{3/2}, F'=2$ and subsequent decay to $F=1$. This
effect is directly observed as power broadening of the line shape. In a second
experiment (see Fig. \ref{pict:PushPump_result} (b)) the atom is prepared in
the hyperfine ground state $F=1$ by an additional pump laser pulse before
applying the push-out and redetection sequence. When the pump laser is on
resonance with the light-shifted hyperfine transition $^2S_{1/2}, F=2
\rightarrow$ $^2P_{3/2}, F'=1$ the redetection probability reaches its maximum
because the atom is more efficiently pumped into the dark state $F=1$ . Out of
resonance the excitation probability to $^2P_{3/2}, F'=1$ decreases, such that
the atomic population basically remains in $F=2$ and the atom is kicked out of
the trap.

\begin{figure}[h]
\centerline{\scalebox{1}{\includegraphics[width=12cm]{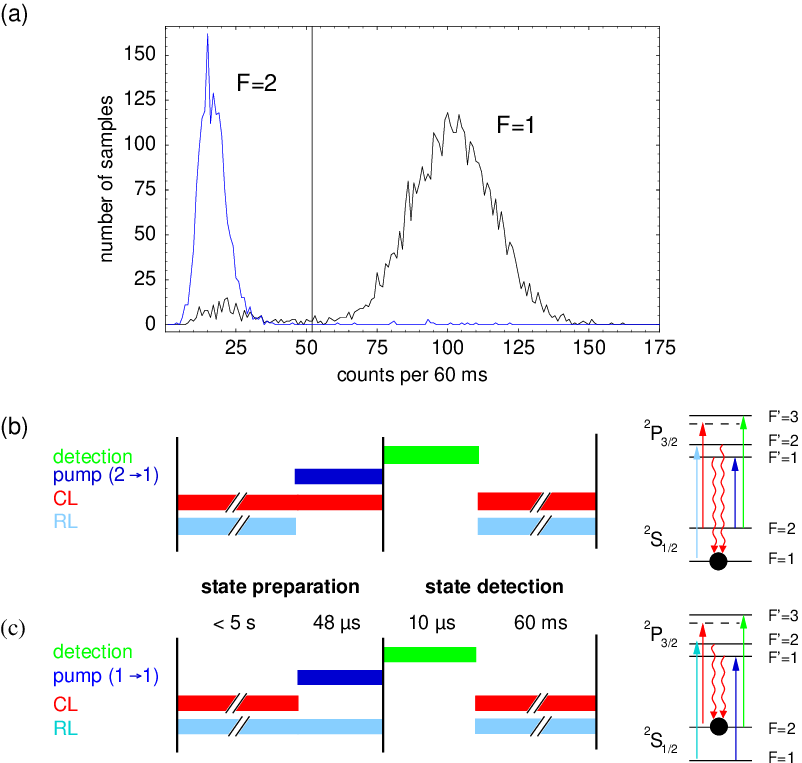}}}
\caption{(a) Histogramm of photon counts per 60 ms during redetection after
  application of a hyperfine state selective detection laser pulse. Atoms
  prepared in $F=1$ (black line) survive the detection pulse, whereas atoms in
  $F=2$ (blue line) are removed from the dipole trap. (b), (c) timing sequence
  and partial level scheme of $^{87}$Rb for state detection in $F=1$ and
  $F=2$, respectively. (Experimental parameters: detection laser intensity = 40
  saturation intensities $I_S$).}
\label{pict:cycling_messung}
\end{figure}
\clearpage

In both measurements the duration and intensity of the pump and detection laser
pulses were not optimized to achieve efficient state detection within the
shortest period of time. In a sequence of adjustment measurements we scanned
systematically the power (see fig. \ref{pict:cycling_powerscan_2}) and
duration of the push-out pulse and we find that single atoms prepared in $F=2$
are removed from the dipole trap with a probability of $0.99\pm0.01$ (see
fig. \ref{pict:cycling_messung} (a)), whereas atoms in $F=1$ survive this
pulse and are redetected in the dipole trap with a probability of
$0.95\pm0.01$. These numbers are not corrected by accidental loss of the atom
during the redetection sequence due to collisions with ``hot'' atoms from the
background gas (see \ref{subsect:traplifetime}) or with cold atoms from the
surrounding optical molasses and they also include incomplete preparation of
the atom due to inefficient optical pumping.

\subsubsection{Conclusion}

The application of a detection laser pulse resonant to the light-shifted
cycling transition $^2S_{1/2}, F=2, m_F=2 \rightarrow$ $^2P_{3/2}, F'=3,
m_F'=3$ and subsequent redetection of the atom in the dipole trap allows to
discriminate the hyperfine ground states $F=1$ and $F=2$ of Rubidium 87 with a
minimum efficiency of 0.95. This value gives the upper bound of the maximum
visibility achievable in the two-particle correlation measurements used for
the verification of atom-photon entanglement.

\subsection{Preparation of Zeeman superposition states}

The key element for the detection of the atomic qubit in complementary
measurement bases is the state-selective adiabatic population transfer from a
coherent superposition of Zeeman sublevels $\ket{m_F=-1}$ and $\ket{m_F=+1}$
of the hyperfine ground state $F=1$ to the hyperfine ground state $F=2$. To
verify this important property of our STIRAP-scheme it is necessary to prepare
in a first step a well defined superposition
$\frac{1}{\sqrt{2}}(\ket{m_F=-1}+e^{i\phi}\ket{m_F=+1})$ of the Zeeman
sublevels $m_F=-1$ and $m_F=+1$. This task can be realized by two different
techniques. One the one hand one could use directly the projective
polarization measurement of a spontaneously emitted photon, whereby the
polarization state of the photon is initially entangled with the magnetic
quantum number $m_F=\pm1$. Because this task requires a rather complicated
control of many experimental parameters including the well defined preparation
of the entangled state it seems much more simple to populate directly a
coherent superposition
$\frac{1}{\sqrt{2}}(\ket{m_F=-1}+e^{i\phi}\ket{m_F=+1})$ by optical pumping
into a coherent dark state. However, also this technique has its difficulties,
because coherent population trapping in the Zeeman substructure of a given
atomic state is extremely sensitive to residual magnetic fields. The reason
for this sensitivity is the time dependence of Zeeman dark-states due to
Larmor precession. In order to populate a stable Zeeman dark state by optical
pumping it is therefore necessary to reduce the magnetic field below a
threshold value.

Because the atom is stored in an optical dipole trap which is located inside a
UHV vacuum chamber, it is not possible to measure the magnetic field exactly
at the position of the atom with usual Hall- or magnetic flux probes. A way to
avoid this difficulties is to use the atom itself as a probe. Here I will
show, how the Zeeman-splitting of the hyperfine ground state $^2S_{1/2}, F=1$
allows to minimize the magnetic field and therefore enables the population of
coherent Zeeman-superposition states by optical pumping.

\subsubsection{Time-dependence of Zeeman superposition states}

\begin{figure}[t]
\centerline{\scalebox{1}{\includegraphics[width=10cm]{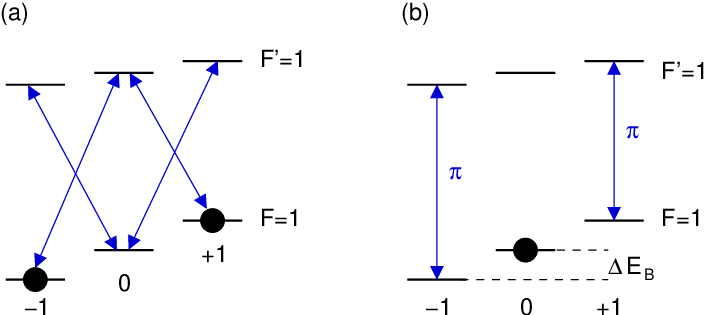}}}
\caption{Zeeman-splitting of a hyperfine transition $F=1 \rightarrow F'=1$. If
  the linear pump laser polarization is (a) perpendicular to the magnetic
  field, the dark state
  $\frac{1}{\sqrt{2}}(\ket{m_F=-1}-e^{i\phi}\ket{m_F=+1})$ is not stable due
  to Larmor-precession, (b) parallel to the magnetic field, the dark state
  $\ket{m_F=0}$ is an energy eigenstate of the system and stable.}
\label{pict:Hanle_term1}
\end{figure}

Consider the simplified case of an atomic transition consisting of a single
hyperfine ground-state $F=1$ and a single excited hyperfine state $F'=1$. For
laser light whose linear polarization direction is orthogonal to the
quantization axis $z$ and parallel to $y$ ($\phi=0$), the state
\begin{equation}
\ket{S_0}=\frac{1}{\sqrt{2}}(\ket{m_F=-1}-\ket{m_F=+1})
\end{equation}
is dark.

Provided, the level-shift of the bare atomic states $\ket{F=1,m_F=-1}$ and
$\ket{F=1,m_F=+1}$ due to interaction with a magnetic field is small compared
to the hyperfine-splitting, the Zeeman-splitting $\Delta E_B$ is given in
first order perturbation theory by
\begin{equation}
\hbar\omega_L = \Delta E_B = \mu_B g_F m_F |B_z|,
\end{equation}
where $\omega_L$ is the Larmor frequency, $\mu_B$ denotes Bohrs magneton,
$g_F$ the hyperfine-Lande-factor, $m_F$ the magnetic moment and $|B_z|$ the
absolute value of the magnetic field in z-direction. The time-evolution of the
state $\ket{S_0}$ is then given by
\begin{equation}
\ket{S(t)}=\frac{1}{\sqrt{2}}(\ket{m_F=-1}-e^{+2i\omega_L
t}\ket{m_F=+1}).
\end{equation}

After a characteristic time $T=\pi/2\omega_L$, the atom precessed into the
orthogonal quantum state, which is a bright state for the incident linearly
polarized laser field and the atom scatters photons until it is pumped again
into the dark state $\ket{S_0}$. The magnitude of the magnetic field
determines the time-evolution of the dark state and therefore the rate of
scattered photons.

Similar considerations are valid for a linearly polarized laser whose
polarization direction is parallel to the x-axis. In the case of
$\pi$-polarized light - the linear polarization direction now is parallel to
the magnetic field - the atom is pumped in the energy eigenstate
$\ket{F=1,m_F=0}$. However, this state is stable in time, independent of the
magnitude of the magnetic field pointing along z. Provided the linear
polarization direction of a pump laser field - resonant to the hyperfine
transition $F=1 \rightarrow F'=1$ - is perpendicular to the magnetic field,
the rate of scattered photons increases, as the magnitude of the magnetic
field increases. For a ``zero'' field the rate is minimal. This effect,
similar to the Hanle-effect, is used in the present experiment to minimize the
magnetic field for efficient preparation of atomic superposition states.

\subsubsection{Magnetic field compensation}

\begin{figure}[h]
\centerline{\scalebox{1}{\includegraphics[width=9cm]{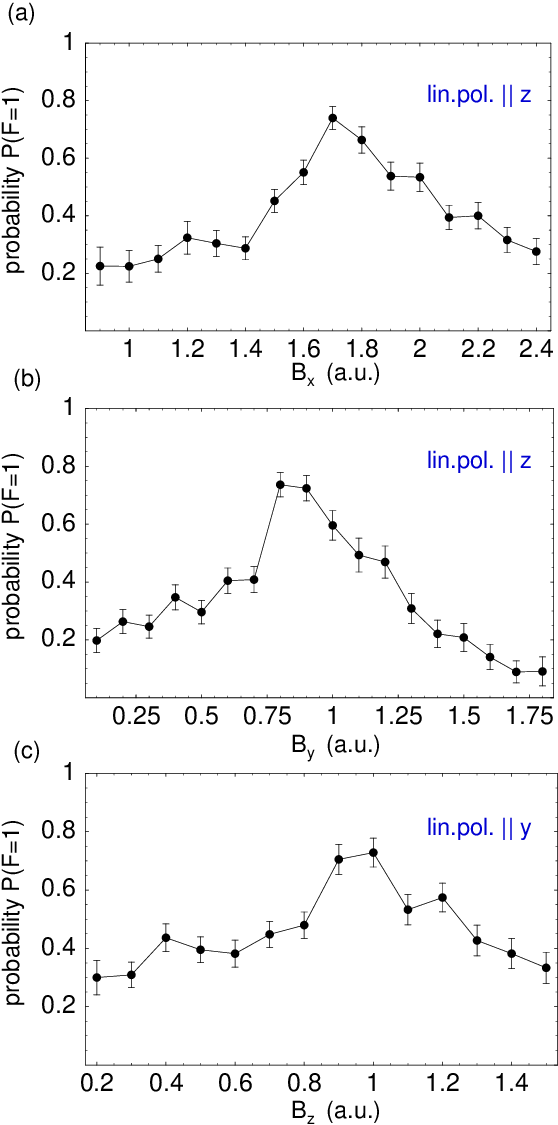}}}
\caption{Probability to detect the atom in the dipole trap after application
  of a 1.8 ms pump pulse as a function of the magnetic compensation field in
  x-,y- and z-direction. If the respective magnetic field component is
  minimal, the atom is effectivly pumped into the stable Zeeman dark-state
  $F=1, m_F=0$, and scatters no light. Therefore, the atom has a high survival
  probability.}
\label{pict:Hanle_result}
\end{figure}

To minimize the magnetic field we shine on the optically trapped $^{87}$Rb
atom a 1.8 ms linearly polarized dichromatic laser pulse resonant to the
hyperfine transitions $^2S_{1/2}, F=1$ $\rightarrow$ $^2P_{3/2}, F'=1$ and
$^2S_{1/2}, F=2$ $\rightarrow$ $^2P_{3/2}, F'=1$ and vary the magnetic field
orthogonal to the given linear polarization direction of the pump beams by
adjusting the current in the compensation coils. Then we redetect the atom in
the dipole trap. If the magnetic field is minimal the atom scatters only few
photons during the 1.8 ms and therefore has the highest probability to
survive. This procedure is performed for all three componenents $B_x$, $B_y$
and $B_z$ of the magnetic field vector by rotating the linear polarization of
the pump beam and scanning the respective orthogonal magnetic field
component. From the measured experimental data in Fig. \ref{pict:Hanle_result}
and with the knowledge of the geometry of the compensation coils we can
determine an upper bound of the residual magnetic field of 300 mGauss. This
value is confirmed by the observation of a Larmor-precession frequency of 370
kHz \cite{Volz} corresponding to a magnetic field of 132 mGauss.
\clearpage

\begin{figure}[h]
\centerline{\scalebox{1}{\includegraphics[width=8cm]{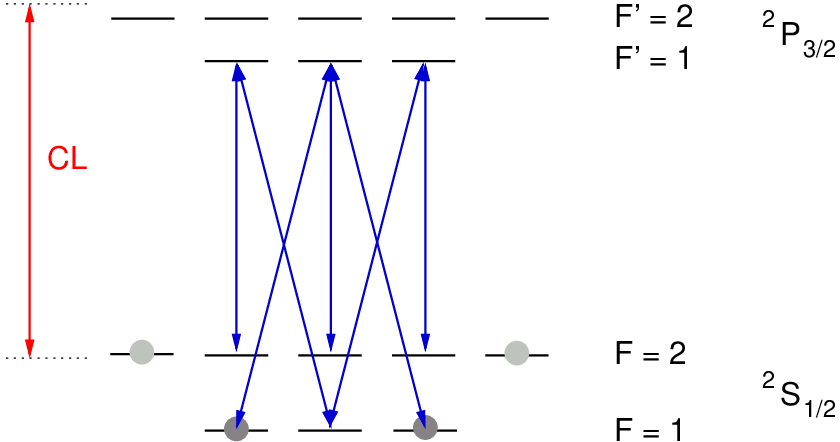}}}
 \caption{Preparation of a Zeeman-superposition state
 $\frac{1}{\sqrt{2}}(\ket{1,-1}-\ket{1,+1})$ via coherent population trapping
 (CPT). To avoid a dark state in $F=2$ we additionally apply the unpolarized
 cooling laser (CL) of our MOT.} 
 \label{pict:exp_zeeman_pumpen}
\end{figure}

\subsubsection{State preparation}

To prepare an atomic superposition state
\begin{equation}
\ket{S_{prep}}=\frac{1}{\sqrt{2}}(\ket{1,-1}-\ket{1,+1})
\end{equation}
via coherent population trapping (CPT) we apply a 5-$\mu$s linearly polarized
(parallel to y) laser pulse resonant to the hyperfine transition $^2S_{1/2},
F=1$ $\rightarrow$ $^2P_{3/2}, F'=1$. Because this simple light field
configuration pumps the atom also to a dark state of the hyperfine ground
level $F=2$ we apply simultaneously two additional laser fields, depopulating
$F=2$ (see Fig. \ref{pict:exp_zeeman_pumpen}).

\subsection{Detection of Zeeman superposition states}

Now, the prepared atomic superposition state $\ket{S_{prep}}$ can be analyzed
by a state-selective Stimulated-Raman-Adiabatic-Passage (STIRAP) technique. So
far, the experimental setup as shown in chapter \ref{chapter:diptrap} was only
slightly modified by pump beams which have been derived from cooling and
repump lasers of the MOT. But for the adiabatic population transfer additional
lasers are necessarily operating at 795 nm. Before focussing on the
experimental results I will first give a short overview about the extended
experimental setup.

\subsubsection{Experimental setup}

The STIRAP-pulses are generated by two independent laser diodes locked via
Doppler-free saturation spectroscopy \cite{Vrana04} to the hyperfine
transitions $^2S_{1/2}, F=1 \rightarrow$ $^2P_{1/2}, F'=1$ and $^2S_{1/2}, F=2
\rightarrow$ $^2P_{1/2}, F'=1$ of the D1-line in Rubidium 87. The shape of each
pulse (amplitude and duration) is adjusted by an AOM in double-pass
configuration down to a minimum pulse length of approximatly 15 ns. The delay
of the Stokes pulse with respect to the pump pulse is controlled by a tuneable
electric delay line and a programmable pattern generator \cite{Kurtsiefer02},
which allows to switch on and off the AOMs with a time resolution of 20
ns. The pump and Stokes beams are overlapped on a polarizing beam-splitter
(PBS) and coupled into a single mode optical fiber. Because the STIRAP-pulses
have to be applied to the atom in the observation direction of atomic
fluorescence light, a good on/off-switching ratio of the AOMs is necessary to
eliminate unwanted background light from the STIRAP-beams. An overall optical
isolation of approximately 120 dB is realized by an additional acousto-optical
modulator AOM$_3$ switching both pulses after overlapping.

\begin{figure}[t]
\centerline{\scalebox{1}{\includegraphics[width=12cm]{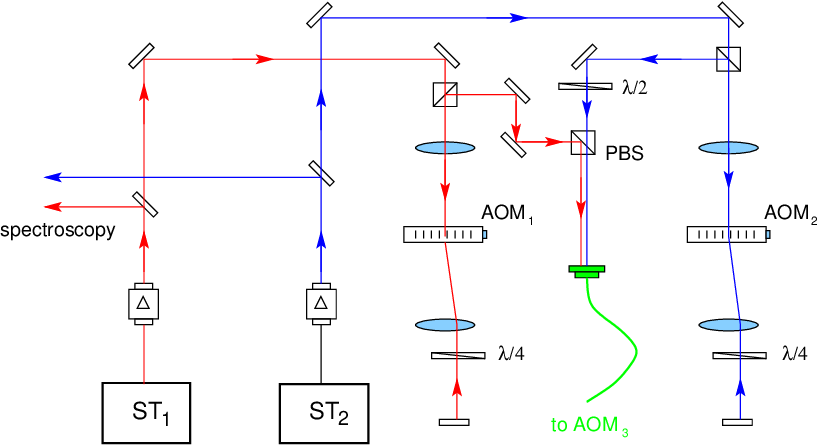}}}
\caption{Experimentel setup of the STIRAP-lasers. (ST$_1$ ... STIRAP-laser 1,
  ST$_1$ ... STIRAP-laser 2, PBS ... polarizing beam splitter, AOM
  ... acousto-optical modulator, $\lambda/2$ ... half wave plate, $\lambda/4$
  ... quarter wave plate)}
\label{pict:STIRAP_lasersetup}
\end{figure}

\begin{figure}[t]
\centerline{\scalebox{1}{\includegraphics[width=12cm]{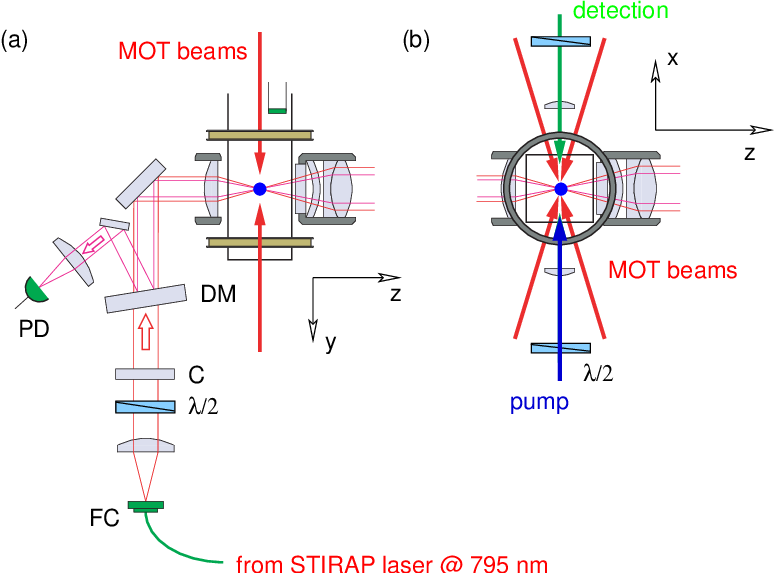}}}
\caption{Schematic view of the experimental setup to investigate a coherent
  superposition of the Zeeman states $\ket{F=1,m_F=-1}$ and
  $\ket{F=1,m_F=+1}$. (a) horizontal view; (b) upright view.}
\label{pict:STIRAPsetup_detail}
\end{figure}

To connect the preparation part of the STIRAP pulses (see
Fig. \ref{pict:STIRAP_lasersetup}) with the trap setup (see
Fig. \ref{pict:STIRAPsetup_detail}) the STIRAP beams are coupled again into a
single mode optical fiber. At the exit port of the fiber a rotable half-wave
plate ($\lambda/2$) is used to adjust the linear polarization angle $\alpha$
of the STIRAP pulses $\Omega_1$ and $\Omega_2$. In addition a tilted
birefringent crystall (C) compensates for relative phase shifts between s- and
p- polarized components that occur at the reflection on dielectric
mirrors. Finally the STIRAP-pulses are focussed with an objective onto the
atom to a waist of 5..10 $\mu$m.

For the preparation and analysis of atomic superposition states additional
pump and detection beams resonant to specific hyperfine transitions within the
D2-line of Rubidium 87 are necessary. Therefore different beams are extracted
from the cooling and repump lasers and switched independently by AOMs in
double-pass configuration. Then each beam is coupled into a single-mode
optical fiber and focussed onto the atom.

\subsubsection{Experimental process}

To verify the selective detection of Zeeman-superposition states via STIRAP,
we first load a single $^{87}$Rb atom into the dipole trap. Then a 5-$\mu$s
optical laser pulse (see preceding section) pumps the atomic population into
the dark superposition state
\begin{equation}
\ket{S_{prep}}=\frac{1}{\sqrt{2}}(\ket{1,-1}-\ket{1,+1})
\end{equation}
of the $^2S_{1/2}, F=1$ ground level. To analyze this state we apply
immediatly after the preparation pulse a 70 ns STIRAP-pulse, transferring a
superposition state
\begin{equation}
\ket{S_{trans}}=\frac{1}{\sqrt{2}}(\ket{1,-1}-e^{2i\alpha}\ket{1,+1})
\end{equation}
adiabatically to the hyperfine ground state $F=2$. Due to destructive
interference of the excitation amplitudes the orthogonal quantum state
$\ket{S_{dark}}=\frac{1}{\sqrt{2}}(\ket{1,-1}+e^{2i\alpha}\ket{1,+1})$ does
not couple to the STIRAP laser field $\Omega_1$ and remains in $F=1$. The
relative phase of these states can be controlled by the linear polarisation
angle $\alpha$ of the STIRAP laser $\Omega_1$ with respect to the
x-axis. After the transfer we apply a hyperfine state selective 40 $\mu$s
push-out laser pulse to discriminate the hyperfine ground states. Finally the
cooling and repump laser of the MOT are switched on and the atomic
fluorescence light is integrated for 60 ms to decide whether the atom is still
in the trap or not. The probability to redetect the atom in the dipole trap is
ideally given by
\begin{equation}
P=|\bra{S_{prep}}S_{dark}\rangle|^2=\sin^2{\alpha},
\end{equation}
which is the overlap of the prepared atomic state $\ket{S_{prep}}$ with the
dark state $\ket{S_{dark}}$.
 
\begin{figure}[t]
\centerline{\scalebox{1}{\includegraphics[]{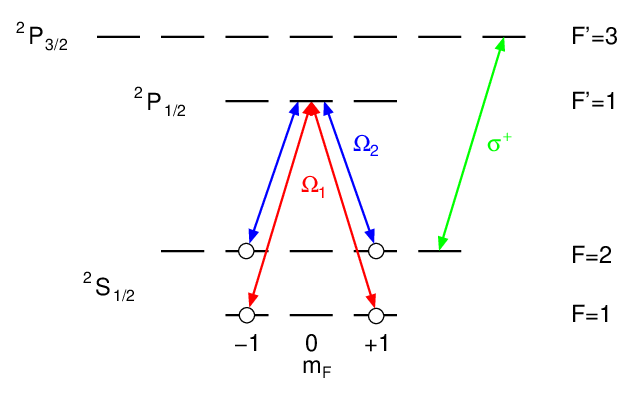}}}
\caption{Coupling scheme for the analysis of coherent superposition
  states. The STIRAP process transfers the superposition
  $\frac{1}{\sqrt{2}}(\ket{1,-1}+e^{i\phi}\ket{1,+1})$ to the $F=2$ ground
  state whereas the orthogonal state
  $\frac{1}{\sqrt{2}}(\ket{1,-1}-e^{i\phi}\ket{1,+1})$ remains in $F=1$. To
  distinguish the hyperfine ground states, a $\sigma^+$ polarized detection
  laser pulse resonant to the cycling transition $^2S_{1/2},F=2 \rightarrow$
  $^2P_{3/2},F'=3$ is used.}
\label{pict:STIRAP_termschema2}
\end{figure}

\begin{figure}[h]
\centerline{\scalebox{1}{\includegraphics[width=10cm]{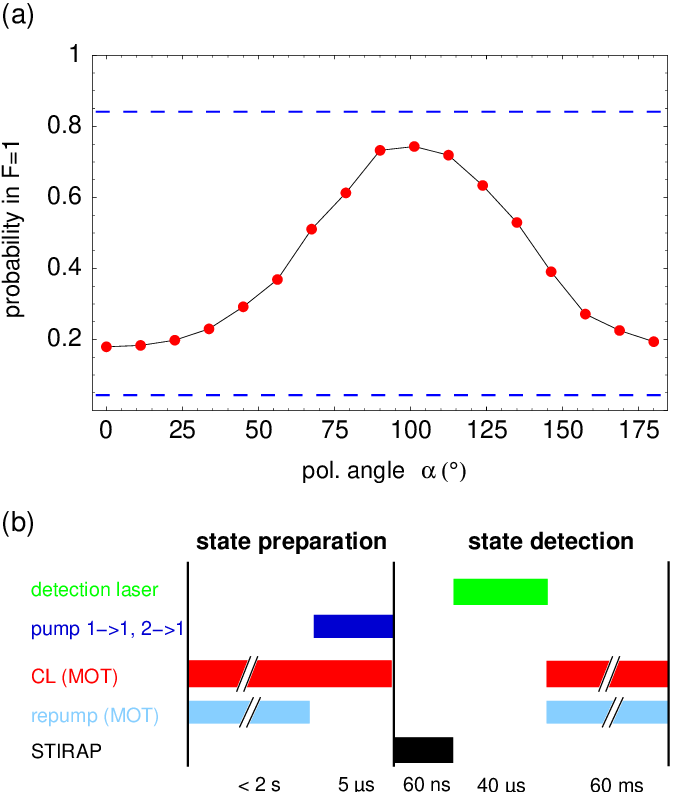}}}
\caption{{\bf (a)} Probability to detect the atom in the hyperfine ground state
  $F=1$ after application of a state-selective STIRAP and detection laser pulse
  as a function of the linear polarization angle $\alpha$ of the STIRAP
  laser. Initially the atom was prepared by a 5-$\mu$s pump pulse in the
  superposition state $\frac{1}{\sqrt{2}}(\ket{1,-1}+\ket{1,+1})$. For
  $\alpha=0$ the prepared state is adiabatically transferred to $F=2$, whereas
  for $\alpha=90^{\circ}$ it is a dark state with respect to
  the polarization of the STIRAP laser and remains in $F=1$. Atoms in $F=1$
  survive the detection pulse and are redetected, while atoms in $F=2$ are
  removed from the trap. {\bf (b)} Timing sequence for the preparation and the
  analysis of a coherent Zeeman superposition state.}
\label{pict:STIRAP_polscan}
\end{figure}

In Fig. \ref{pict:STIRAP_polscan} we show the measured probability $P$ to
detect the atom in the hyperfine ground state $F=1$ after application of the
STIRAP and push-out pulses as a function of the linear polarization angle
$\alpha$ of the STIRAP laser. For $\alpha=0$ the prepared state
$\frac{1}{\sqrt{2}}(\ket{1,-1}-\ket{1,+1})$ is transferred to $F=2$ because it
is a bright state of the vertically polarized STIRAP-laser field
$\Omega_1$. In this case the atom does not survive the detection pulse and is
not redetected. For $\alpha=\pi/2$ the state $\ket{S_{prep}}$ is dark and
therefore not transferred to $F=2$. Consequently it is not influenced by the
detection laser and redetected with a probability of 0.75.

The measured experimental data points in Fig. \ref{pict:STIRAP_polscan} are
fitted with a modified function $a+V/2 \sin^2{(\alpha+\alpha_0)}$ yielding a
visibility $V$ (defined as peak to peak amplitude) of $0.57\pm0.01$. The
reduction of the observed visibility can be explained by two effects. First,
the state preparation via optical pumping is not perfect. 20 percent of the
atomic population are pumped to a Zeeman dark state of the $F=2$ hyperfine
ground state. This situation effects the Zeeman state analysis in the
following way. Suppose that 20 percent of the atomic population are in $F=2$
and 80 percent are ideally in the prepared dark state
$\frac{1}{\sqrt{2}}(\ket{1,-1}-\ket{1,+1})$. Then, if the polarization of the
STIRAP laser is chosen such that it couples maximally to the prepared
superposition in $F=1$ and maximally to the population in $F=2$, the fraction
in $F=1$ will be transferred to $F=2$, whereas population initially in $F=2$
is partially transferred to $F=1$. Suppose the STIRAP laser does not couple
to the superposition in $F=1$ because the prepared Zeeman superposition state
is a dark state with respect to the chosen polarisation, then population in
$F=2$ will be partially transferred to $F=1$. To get an estimation about the
impact of this process on the measured data we numerically solved a master
equation of the STIRAP-process on the basis of a simplified three-level model
including spontaneous emission from the intermediate level. We get the result
that after the STIRAP-pulse maximally 84 percent of the atomic population will
be in $F=1$ provided the STIRAP polarization is chosen such that the STIRAP
does not couple to the prepared Zeeman state in $F=1$, whereas if the STIRAP
couples initially to population in $F=1$ and $F=2$, after the pulse 5 percent
will remain in $F=1$. On the basis of this estimation we calculate a corrected
visibility of $0.71\pm0.01$.

A second important source of errors is the imperfect state transfer via the
STIRAP pulses. For further optimization of the process, it is useful to
decouple the inefficient state preparation from the state detection. After
all, the preparation of a ``clean'' initial atomic state can be best realized
by the polarization measurement of a spontaneously emitted photon, provided
the photon was initially entangled with the Zeeman state $m_F=\pm1$ of the
atom.

\section{Conclusion and discussion}

We have set up a laser system which allows to read out a Zeeman superposition
state of a single atom. Depending on the polarisation of the STIRAP pulses a
superposition $\frac{1}{\sqrt{2}}(\ket{F=1,m_F=-1}+e^{i\phi}\ket{F=1,m_F=+1})$
of the $^2S_{1/2}, F=1, m_F=\pm1$ hyperfine ground state is adiabatically
transferred to the $F=2$ state. Due to destructive interference of the
excitation amplitudes the orthogonal quantum state
$\frac{1}{\sqrt{2}}(\ket{F=1,m_F=-1}-e^{i\phi}\ket{F=1,m_F=+1})$ does not
couple to the STIRAP pulse and remains in $F=1$. To discriminate these states
we apply a detection laser pulse resonant to the cycling transition
$^2S_{1/2}, F=2 \rightarrow$ $^2P_{3/2}, F'=3$. Provided the atom is in $F=2$
it scatters photons from this laser beam, and with each scattering event the
atom acquires an additional photon momentum $\hbar \vec{k}$. After
approximately 40 to 50 scattering events the atom is pushed out of the dipole
trap. Finally, to read out the atomic state the cooling and repump beams of
the MOT are switched on and the atomic fluorescence light is integrated for 60
ms to decide if the atom is still in the trap or not.
\clearpage

%------------------------------------------------------------------------------
%------------------------------------------------------------------------------

\chapter{Observation of atom-photon entanglement}
\label{chapt:ObservationAtomPhoton}

\section{Introduction}

Atom-photon entanglement has been implicit in many previous experimental
systems, from early measurements of Bell's inequality violations in atomic
cascade systems \cite{Freedman72,Aspect82} to fluorescence studies in trapped
atomic ions \cite{Eichmann93,DeVoe96} and atomic beam experiments
\cite{Kurtsiefer97}. However, it has not been directly observed until quite
recently \cite{Blinov04}, as the individual atoms and photons have not been
under sufficient control.

In our experiment a single photon is spontaneously emitted from a single
optically trapped $^{87}$Rb atom, which is initially excited to a state which
has multiple decay channels. Along a certain emission direction two decay
channels are selected and the photon polarization is maximally entangled with
two particular Zeeman sublevels of the hyperfine ground states of the atom.

To verify entanglement of the generated atom-photon state one has to disprove
the possibility that the two-particle quantum system can be a statistical
mixture of seperable states. This task is closly connected to a violation of
Bell's inequality and requiress correlated local state measurements of the
atom and the photon in complementary bases.

In this chapter I will report in detail on the generation and analysis of
atom-photon entanglement. The experimental process is described and first
experimental results are discussed confirming spin-entanglement between the
atom and the photon.

\section{Experimental process}
 
First, we load a single $^{87}$Rb atom from a magneto-optical trap - operated
in a pulsed mode - into the optical dipole trap. Then a 5.5-$\mu$s linearly
polarized optical pulse pumps the atom into the $^2S_{1/2},\ket{1,0}$ dark
state (for details see preceding chapter), from where the atom is excited by a
30-ns $\pi$-polarized optical $\pi$-pulse to the $^2P_{3/2}, \ket{0,0}$
state. Here $\ket{F,m_F}$ denotes the internal atomic quantum numbers of the
total angular momentum and its projection along the quantization axis $z$.
\begin{figure}[h]
\centerline{\scalebox{1}{\includegraphics[width=11cm]{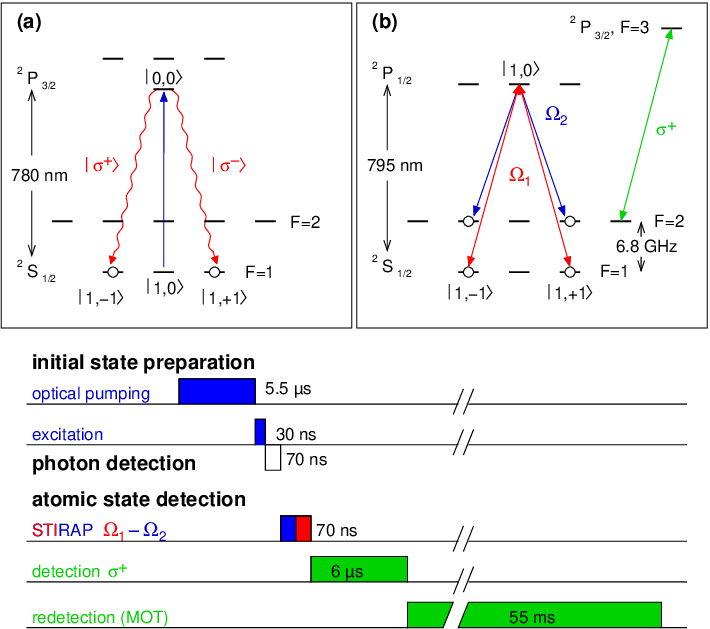}}}
\caption{The experimental process (time not to scale). {\bf (a)} The atom is
  initialized to the $\ket{1,0}$ hyperfine ground state by optical pumping (not
  shown) and excited by a 30-ns $\pi$-polarized $\pi$-pulse to the $\ket{0,0}$
  excited state. In the following spontaneous decay the polarization state of
  the photon is entangled with magnetic hyperfine state of the atom. {\bf (b)}
  After the polarization measurement of the photon the internal state of the
  atom is measured by a 70-ns STIRAP and a 6-$\mu$s detection pulse, before
  atomic fluorescence light is accumulated for 55 ms.}
\label{pict:EntVerification_term}
\end{figure}

In the following spontaneous emission the atom decays either to the
$\ket{1,-1}$ ground state while emitting a photon in the polarization state
$\ket{\sigma^+}$, or to the $\ket{1,0}$ state while emitting a
$\ket{\pi}$-polarized photon or it decays to the $\ket{1,+1}$ ground state and
emits a $\ket{\sigma^-}$-polarized photon. Because the residual magnetic field
is smaller than 100 mGauss, the Zeeman-splitting of these states is two orders
of magnitude smaller than the natural linewidth of the transition. Therefore
these decay channels are spectrally indistiguishable and a coherent
superposition of separable atom-photon states is formed, entangling the
magnetic quantum number $m_F$ of the atom with the polarization state of the
emitted photon. Along the observation direction, defined by the aperture of
the microscope objective, $\pi$-polarized photons are not emitted. Thus, the
resulting atom-photon state is maximally entangled:
\begin{equation}
\ket{\Psi^+}=\frac{1}{\sqrt{2}}(\ket{1,-1}\ket{\sigma^+} +
                                \ket{1,+1}\ket{\sigma^-}). 
\end{equation}

\clearpage

\begin{figure}[h]
\centerline{\scalebox{1}{\includegraphics[width=10cm]{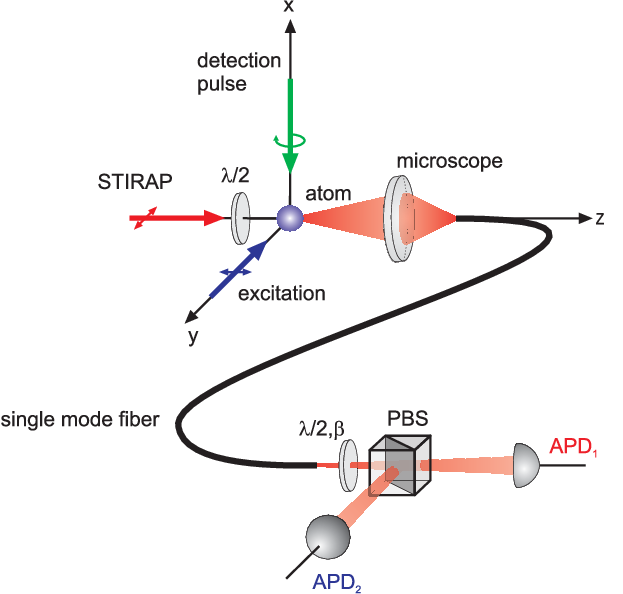}}}
\caption{Simplified setup of the experimental apparatus. The $\pi$-polarized
  excitation beam propagates perpendicular to the quantization axis $z$
  defined by the optical axis of the detection optics. The scattered photons
  are collected by a microscope objective, coupled to a single mode optical
  fiber and directed to a polarizing beam splitter (PBS). Two single photon
  detectors APD$_1$ and APD$_2$ register the H- and V-polarized photons,
  respectively. The $\lambda/2$-waveplate is used to rotate the photon
  polarization for photonic qubit measurements in different bases. The atomic
  state is analysed by a STIRAP-pulse driving coherent two-photon transitions
  between the hyperfine levels of the atomic ground state. The atomic
  measurement basis is defined by the angle $\alpha$ of the linearly polarized
  STIRAP laser field $\Omega_1$ with respect to the y-axis.}
\label{pict:entanglement_setup}
\end{figure}

To investigate the nonclassical correlation properties of this state, the
single photon from the spontaneous decay is collected with a microscope
objective (see Fig. \ref{pict:entanglement_setup}) and coupled into a
single-mode optical fiber guiding it to a polarization analyzer consisting of
a rotable $\lambda/2$-halfwave plate, a polarizing beam splitter (PBS) and two
avalanche photo-diodes APD$_1$ and APD$_2$ for single photon
detection. Triggered by the detection of a photon, the internal quantum state
of the atom is analyzed by a state-selective
Stimulated-Raman-Adiabatic-Passage (STIRAP) technique transferring the
superposition $\frac{1}{\sqrt{2}}(\ket{1,-1}+e^{2i\alpha}\ket{1,+1})$ to the
hyperfine ground state $F=2$ (see chapter
\ref{chapter:AtomicStateDetection}). Due to destructive interference of the
excitation amplitudes the orthogonal quantum state
$\frac{1}{\sqrt{2}}(\ket{1,-1}-e^{2i\alpha}\ket{1,+1})$ does not couple to the
STIRAP laser field $\Omega_1$ (see Fig. \ref{pict:EntVerification_term}(a))
and remains in $F=1$. Here the relative phase $2\alpha$ is defined by the
linear polarization angle $\alpha$ of the STIRAP laser $\Omega_1$ with respect
to the y-axis. The phase settings $\alpha=0,\pi/2$ and $\alpha=\pi/4,3\pi/4$
define two sets of complementary basis-states
$\{\frac{1}{\sqrt{2}}(\ket{1,-1}+\ket{1,+1}),\frac{1}{\sqrt{2}}(\ket{1,-1}-\ket{1,+1})\}$
and
$\{\frac{1}{\sqrt{2}}(\ket{1,-1}+i\ket{1,+1}),\frac{1}{\sqrt{2}}(\ket{1,-1}-i\ket{1,+1})\}$,
which allow to measure the atomic qubit in the $\sigma_x$- and
$\sigma_y$-basis, whereby $\ket{1,\pm1}$ are eigenstates of $\sigma_z$.

\begin{figure}[h]
\centerline{\scalebox{1}{\includegraphics[width=10cm]{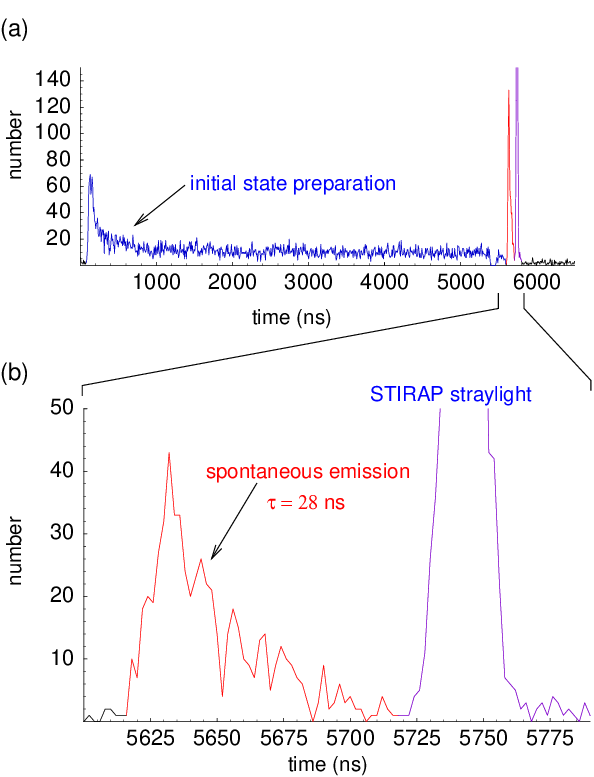}}}
\caption{Number of scattered photons as a function of time during {\bf (a)}
  preparation of the initial state via optical pumping into the dark state
  $^2S_{1/2}, \ket{1,0}$ (integration time bin $\Delta t=8$ns), {\bf (b)}
  excitation to $^2P_{3/2}, \ket{0,0}$ and subsequent spontaneous decay to
  $^2S_{1/2}, \ket{1,\pm1}$ (integration time bin $\Delta t=2$ns).}
\label{pict:PiPulse}
\end{figure} 
\clearpage

After the state-selective population transfer the atom is in a superposition
of the hyperfine ground states $F=1$ and $F=2$. To discriminate these states
we apply a 6-$\mu$s detection laser pulse resonant to the cycling transition
5$^2S_{1/2}, F=2 \rightarrow$ 5$^2P_{3/2}, F=3$. If the atom is in the
hyperfine ground state $F=2$, it is removed from the dipole trap due to photon
recoil heating. If the atom was in state $F=1$, fluorescence is observed after
the cooling and repump lasers of the MOT are switched on. Counts are
integrated for 55 ms to decide whether the atom is still in the trap or not.

In Fig. \ref{pict:PiPulse} the number of scattered photons during the
preparation of the initial state $^2P_{3/2}, \ket{0,0}$ is shown as a function
of time. When the pump pulse is switched on, the number of scattered photons
decreases after few 100 ns as the atomic population is pumped to the Zeeman
dark state $^2S_{1/2}, \ket{1,0}$. The intensity of the following 30-ns
optical $\pi$-pulse is chosen such, that at the end of the pulse the maximum
atomic population is in the $^2P_{3/2}, \ket{0,0}$ state. In the following
spontaneous decay - indicated by the red data trace in
Fig. \ref{pict:PiPulse}(b) - the number of scattered photons drops
exponentially with a measured time constant of 28 ns. To correlate the atomic
state detection with the internal state of a spontaneously emitted photon we
restrict the detection of photons to a well defined time window of 70 ns
following the optical excitation pulse (see
Fig. \ref{pict:EntVerification_term}).

The generation of entangled atom-photon pairs is probabilistic because a
spontaneously emitted photon is detected with a probability of $\eta=5 \times
10^{-4}$. This means, that the atom has to be excited approximately 2000 times
until a photon is detected. To achieve the best rate of detected atom-photon
pairs, the excitation cycle - consisting of optical pumping to 5$^2S_{1/2},
\ket{1,0}$ and following excitation to 5$^2P_{3/2}, \ket{0,0}$ - is repeated
as long until a photon is detected. Once a photon is detected the atom is
measured with almost perfect efficiency, limited only by loss from the dipole
trap during the fluorescence detection.

\section{Experimental results}

To verify atom-photon entanglement we measure the conditional probability of
detecting the atomic qubit in the complementary bases $\sigma_x$ and
$\sigma_y$ as a function of the polarization state of the detected photon.

\begin{figure}[h]
\centerline{\scalebox{1}{\includegraphics[]{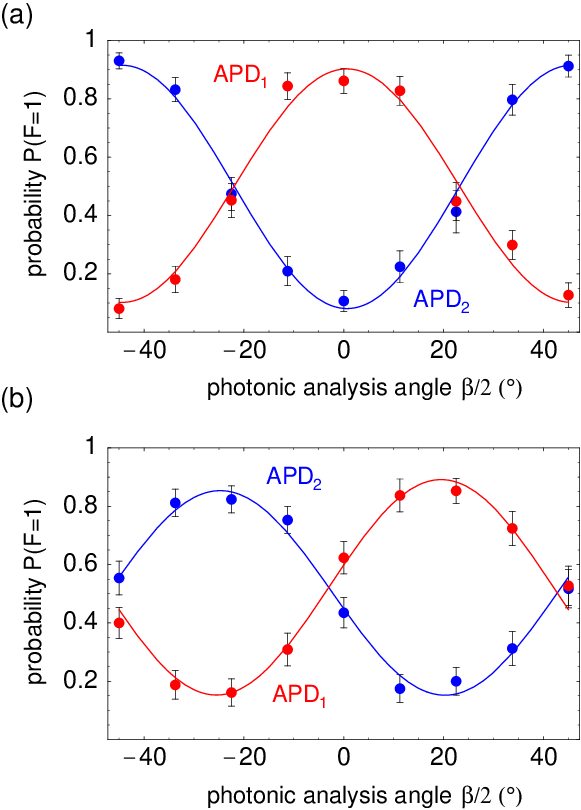}}}
\caption{Probability of detecting the atom in the ground state $F=1$ (after
  the STIRAP pulse) conditioned on the detection of the photon in detector
  APD$_1$ (red data points) or detector APD$_2$ (blue data points) as the
  phase $\phi=2\beta$ of the photonic rotation is varied. The atomic qubit is
  projected {\bf (a)} to the states $1/\sqrt{2}(\ket{1,-1}\pm\ket{1,+1})$,
  {\bf (b)} to the states $1/\sqrt{2}(\ket{1,-1}\pm i\ket{1,+1})$
  corresponding to a spin measurement in $\sigma_x$ and $\sigma_y$,
  respectively, whereas the photonic qubit is projected onto the states
  $\frac{1}{\sqrt{2}}(\ket{\sigma^+}\pm e^{i\phi}\ket{\sigma^-})$.}
\label{pict:Entanglement_result}
\end{figure}

In Fig. \ref{pict:Entanglement_result}(a) the polarisation of the
STIRAP-lasers is set to $\alpha=0$ defining the $\sigma_x$ measurement basis
for the atom and the polarization of the photon is rotated on the
Poincare-sphere by a variable angle $2\beta$. We observe strong correlations
between the polarization state of the detected photon and the internal quantum
state of the atom. As expected, if the photon is detected in the polarization
state $\frac{1}{\sqrt{2}}(\ket{\sigma^+}+\ket{\sigma^-})$ (detection in
APD$_1$) the atom is projected to the corresponding superposition state
$\frac{1}{\sqrt{2}}(\ket{1,-1}+\ket{1,+1})$. This atomic state is a ``dark''
state for the chosen linear polarization of the STIRAP laser field
$\Omega_1$. Hence the atomic population will remain in $F=1$ after the
STIRAP-pulse. If the photonic qubit is rotated by a phase of $\phi=\pi$ before
state reduction the single photon detectors exchange their role and a
detection at APD$_1$ projects the atom to
$\frac{1}{\sqrt{2}}(\ket{1,-1}-\ket{1,+1})$. Now, this atomic state can be
transferred to the hyperfine ground state $F=2$, i.e. $P(F=1)$ goes to a
minimum. From a fit of the measured correlations we obtain a visibility
(defined as peak to peak amplitude) of $0.81\pm0.04$. To verify entanglement
this measurement has to be repeated for an atomic basis conjugate to the first
one. Thus the atomic qubit was analyzed in the complementary $\sigma_y$ basis
(see Fig. \ref{pict:Entanglement_result}(b)). Therefore the STIRAP
polarization angle $\alpha$ is set to $\pi/4$. Again, strong correlations are
observed with a visibility of $0.70\pm0.04$. But now the measured atom-photon
correlations are shifted by $\beta=\pi/4$, as we expect from an entangled
state.

To quantify the amount of entanglement we determine the entanglement fidelity
$F=\langle\Psi^+|\rho|\Psi^+\rangle$, which is defined as the overlap of the
measured state - characterized by the density matrix $\rho$ - with the
maximally entangled state $|\Psi^+\rangle$ we expect to generate. Because the
atomic state detection is imperfect, we model these errors by a quantum
channel which is subjected to white noise. Therefore, the density matrix of
the detected state can be expressed as
$\rho=V|\Psi^+\rangle\langle\Psi^+|+\frac{1-V}{4}\hat 1$, where $V$ is the
mean visibility of the observed correlations in the two complementary
bases. From this we derive an entanglement fidelity of $0.82\pm0.04$.

\section{Conclusion and discussion}

In the current experiment we excite a single optically trapped $^{87}$Rb atom
by a short optical pulse to a state which has multiple decay channels and
detect the subsequent spontaneously emitted single photon. Along a certain
emission direction - defined by the optical axis of the detection optics - two
decay channels are selected and the photon polarization is maximally entangled
with two particular Zeeman sublevels of the hyperfine ground states of the
atom. The entanglement is directly verified by appropriate polarization
analysis of the photon and Zeeman state detection of the trapped atom. We
observe strong atom-photon correlations in complementary measurement bases
yielding an entanglement fildelity of $0.82\pm0.04$. 

\clearpage

%-----------------------------------------------------------------------------
%-----------------------------------------------------------------------------

\chapter{Conclusion and Outlook}

The goal of this thesis was the experimental generation and analysis of
entanglement between the spin state of a single neutral $^{87}$Rb atom and the
polarization state of a single spontaneously emitted photon suitable for
long-distance transport in optical fibers and air. This task required in a
first step the efficient detection and manipulation of a single $^{87}$Rb
atom. For this purpose a far-off-resonance optical dipole trap was set
up. Because of the small trap volume and due to light-induced two-body
collisions, which are present during the loading stage within the light fields
of a magneto-optical trap, only single atoms are captured. This blockade
effect was confirmed by the observation of photon antibunching in the
detected fluorescence light. To generate the entangled atom-photon state the
atom is excited to a state which has two decay channels. Due to conservation
of angular momentum in the following spontaneous emission the polarization
state of the photon is maximally entangled with two particular Zeeman
sub-levels of the atomic ground state. To detect entanglement of this
atom-photon state we performed correlated local measurements of the
polarization state of the photon and the internal quantum state of the
atom. The atomic spin state was analyzed by means of a
Stimulated-Raman-Adiabatic-Passage (STIRAP) process, where the polarization of
the analyzing laser light defined the atomic measurement basis. Strong
nonclassical atom-photon correlations in complementary measurement bases were
observed yielding an entanglement fidelity of $0.82\pm0.04$.

The realization of entanglement between a single optically trapped $^{87}$Rb
atom and a spontaneously emitted single photon described in this work marks a
first successful step towards the remote state preparation of a single atom
over large distances and a first loophole-free test of Bell's inequality
\cite{Saucke02,Simon03}.

\subsubsection{Remote state preparation of a single atom}
\markright{7 Conclusion and Outlook}

\begin{figure}[t]
\centerline{\scalebox{1}{\includegraphics[width=7cm]{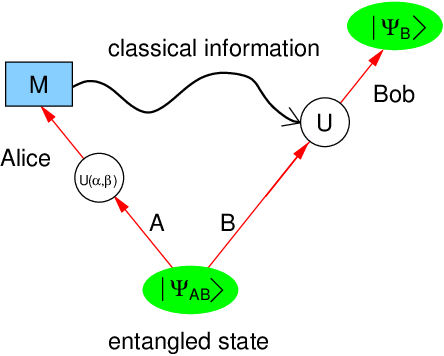}}}
\caption{Principle of Remote State Preparation.}
\label{pict:atomatom}
\end{figure}

A first step towards entanglement swapping or the closely related quantum
teleportation is the so-called remote state preparation
\cite{Boschi98,Pati02}. In this protocoll generic, state independent unitary
transformations $U$ on the atom are sufficient to prepare it in the arbitrary
quantum state $\ket{\Psi_B}=\cos\alpha\ket{\uparrow} - \sin\alpha e^{i\beta}
\ket{\downarrow}$. This process works as follows: Suppose a single atom is
entangled with a single photon. Then the photon propagates to Alice who
performs a unitary operation $U(\alpha,\beta)$ with two free parameters
$\alpha,\beta$ in an extended four-dimensional Hilbert space, before she
projects the photon onto a generalized Bell-basis \cite{Michler96}. Depending
on her measurement result, Bob now has to perform a unitary spin rotation $U$
in order to prepare the required quantum state of the atom.

This scheme generalizes to teleportation, if one introduces yet another
photon, allowing a direct transfer of a photonic qubit to an atom which can be
used for robust long-term storage of quantum information. To reconvert the
stored information onto a single photon one can reexcite the atom on a
so-called V-type transition \cite{Lim03}. Furthermore, if one puts the atom
into a high Q cavity, it is possible to force the atom to emit the photon into
one selected spatial mode. These properties fulfill all requirements for
efficient quantum networking. But teleportation or entanglement swapping is
the key for a loophole free Bell experiment.

\subsubsection{Loophole-free test of Bell's inequality}

For a loophole-free test of Bell's inequality one needs two entangled quantum
systems which can be detected with high efficiency and which are separated far
enough that local observations of each system are space-like separated. This
task can be accomplished by two entangled atoms. First, atoms have the
important property that they can be detected with almost perfect
efficiency. Second, entanglement between two distant atoms can be generated in
a robust way by an entanglement swapping \cite{Zukowski93} process between two
independently created entangled atom-photon pairs
\cite{Saucke02,Simon03}. Suppose one starts with two atoms in two distant
traps, where in each trap only one atom is stored. Then each atom is entangled
with one spontaneously emitted photon. The independently generated photons
then propagate to an intermediate location where a partial Bell-state analysis
is performed, projecting the atoms onto an entangled state.

\begin{figure}[t]
\centerline{\scalebox{1}{\includegraphics[width=5.5cm]{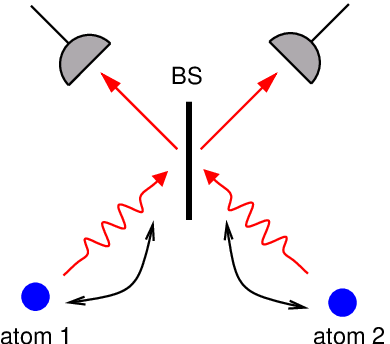}}}
\caption{Atom-atom entanglement by two-photon interference.}
\label{pict:atomatom}
\end{figure}

In the present experiment verifying atom-photon entanglement the atomic state
detection is performed with an efficiency of one because every state
measurement gives a result. Hence, the measured entanglement fidelity already
includes all experimental errors of the atomic state detection and no
additional fair-sampling assumption has to be made for a Bell test with
entangled atoms. To close at the same time the locality loophole, the atoms
have to be space-like separated with respect to the duration of the atomic
state detection. That is, in our experiment the minimum distance of the atoms
is determined by the duration of the detection laser pulse. In detail the
reduction of the atomic state is performed by scattering photons from a closed
atomic transition. After approximately 10 lifetimes ($\tau=26$ ns) of the
5$^2$P$_{3/2}$ excited state, the state reduction is completed with a
probability of more then 99$\%$. Together with the STIRAP-process this yields
an overall measurement time of less than 0.5 $\mu$s corresponding to a minimum
distance between the two atoms of 150 m. This requirement can easily be
achieved since the transmission losses of single photons at a wavelength of
780 nm passing through an optical fiber or air are very low.

The observation of entangled atom-photon pairs is probabilistic with a success
pro\-bability given by the total detection efficiency $\eta=5\times10^{-4}$ of
a single photon. This argument holds also for the generation of entanglement
between two distant atoms. But, given the case of a coincident detection of
the two photons in the two output ports of the beamsplitter (BS) in
Fig. \ref{pict:atomatom}, the atoms are projected with certainty onto a
maximally entangled Bell state. The total success probability for this
entanglement swapping process is given by $\frac{1}{4}\eta^2 T^2$, where the
factor 1/4 accounts for the fact that only one of four photon Bell-states is
detected and $T=\sqrt{0.95}$ corresponds to the transmission probability of
the photonic qubit through 75 m (halfway distance) optical fiber. The maximum
repetition rate of this experiment is determined by the time it takes for the
photon to travel from the locations where the atoms are trapped to the
intermediate location where the photons are detected. For a halfway distance
of 75 m, a minimum cycle time of 1 $\mu$s for the experimental generation of
entangled atoms is possible. This corresponds to a maximum repetition rate of
$1\times 10^6$ s$^{-1}$. From this estimation one can calculate an expected
generation rate of entangled atom-atom pairs of $0.8$ per minute, including
the limited lifetime of the atoms in the dipole traps.

For a violation of Bell's inequality a minimum entanglement fidelity of 0.78
is necessary. This threshold yields a minimum entanglement fidelity of the
generated atom-photon pairs of 0.88. Recently we performed new measurements
verifying atom-photon entanglement with an overall fidelity of
$(89.4\pm0.7)\%$ \cite{Volz05}. Hence we expect an ent\-anglement fidelity of
80 $\%$ for the entangled atom-atom state. The loophole-free violation of a
CHSH-type Bell's inequality \cite{CHSH69} by three standard deviations would
require approximately 7000 atom pairs. On the basis of the above estimations a
total measurement time of 12 days should be feasible, starting with the
performance of the setup developed in this work.

\markright{7 Conclusion and Outlook}

%---------------------------------------------------------------------------
%---------------------------------------------------------------------------

\begin{appendix}
%\includegraphics[width=5cm]{empty.eps}
%\thispagestyle{empty}
%\pagebreak

\chapter{Appendix}

\begin{table}[h]{ \bf \caption{\label{tab:Rb87} Physical properties of
      $^{87}$Rb \cite{Steck87}. }}
\begin{center}
\begin{tabular}{|l|c|c|}
\hline
Atomic Number & $Z$ & 37 \\\hline
Total Nucleons & $Z+N$ & 87 \\\hline
Relative Natural Abundance & $\eta$ & 27.83(2) \% \\\hline
Nuclear Spin & $I$ & 3/2 \\\hline
Atomic Mass & $m$ & 86.9092 $u$ \\\hline
Vacuum Wavelength D$_1$-Transition & $\lambda_{D1}$ & 794.979 nm \\\hline
Vacuum Wavelength D$_2$-Transition & $\lambda_{D2}$ & 780.246 nm \\\hline
Lifetime 5 $^2$P$_{1/2}$ & & 27.70 ns \\\hline
Lifetime 5 $^2$P$_{3/2}$ & & 26.24 ns \\\hline
Natural Line Width D$_1$-Transition & $\Gamma_{D1}$ & $2\pi \times
5.746(8)$ MHz \\\hline
Natural Line Width D$_2$-Transition & $\Gamma_{D2}$ & $2\pi \times
6.065(9)$ MHz \\\hline
Ground State Hyperfine Splitting & $\nu_{HFS}$ & 6834.68 MHz \\\hline
Recoil Velocity D$_2$-Transition & $v_r$ & 5.8845 mm/s \\\hline
Recoil Temperature D$_2$-Transition & $T_r$ & 361.95 nK \\\hline
Doppler Temperature D$_2$-Transition & $T_D$ & 146 $\mu$K \\\hline
Dipole Matrix Element D$_2$-Transition& $\bra{J}|er|\ket{J'}$ &
$3.584(4)\times 10^{-29}$ C m \\\hline
Saturation Intensity & & \\
5$^2$S$_{1/2}, F=2, m_F=\pm2 \rightarrow$ & $I_S$ & 1.67 mW/cm$^2$ \\
5$^2$P$_{3/2}, F=3, m_F=\pm3$ Transition & & \\\hline
\end{tabular}
\end{center}
\end{table}

\begin{figure}[h]
\centerline{\scalebox{1}{\includegraphics[width=12cm]{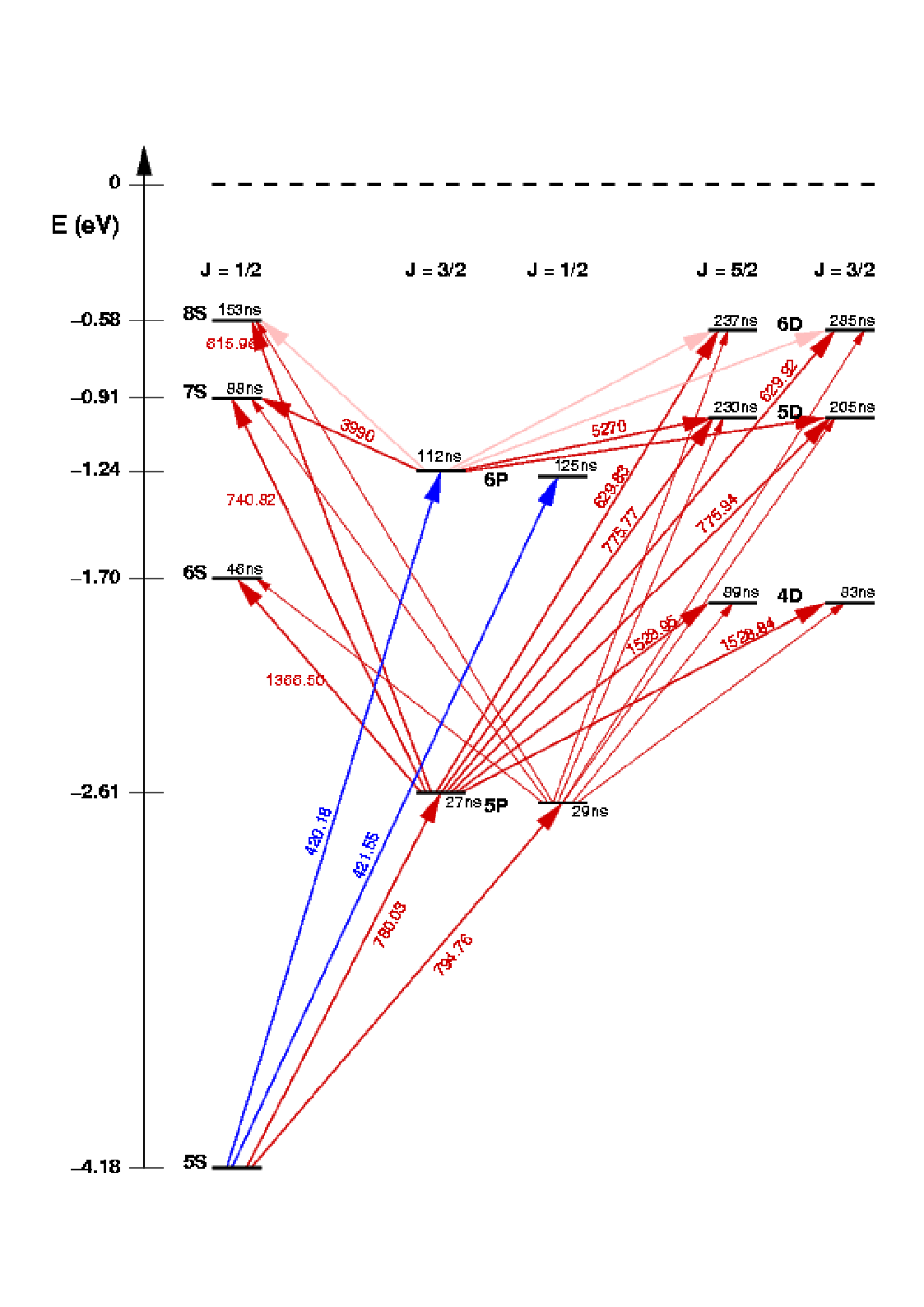}}}
\caption{Combined partial energy-level - Grotrian diagramm for
$^{87}$Rb with lifetimes of atomic levels and transition wavelengths
in air \cite{Schulz02}.}\label{pict:Rb87termschema}
\end{figure}

\begin{figure}[h]
\centerline{\scalebox{1}{\includegraphics[]{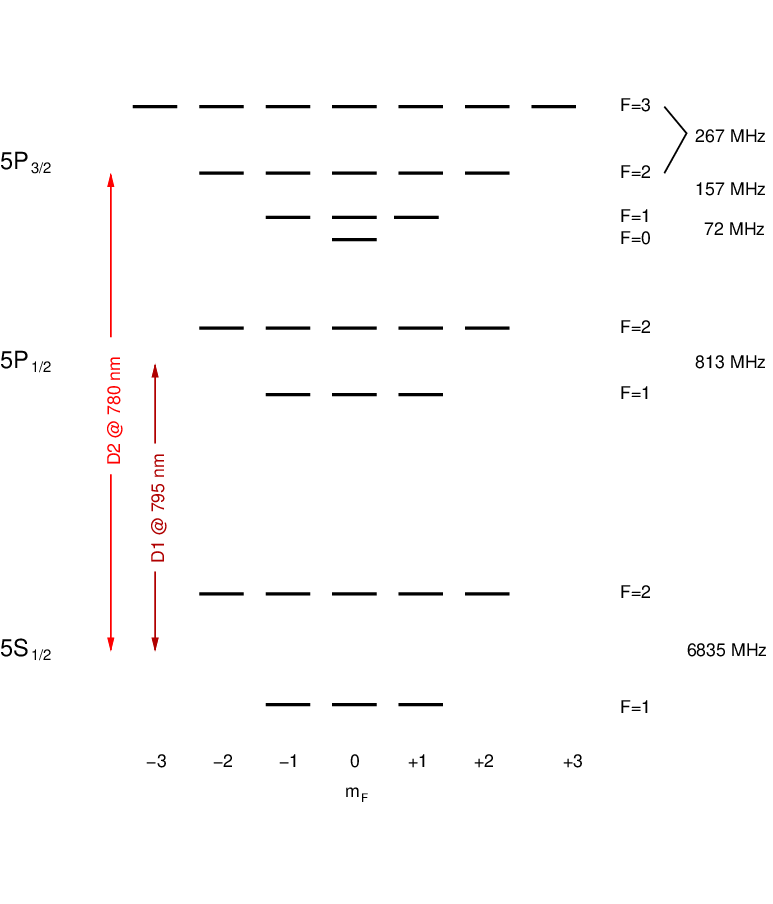}}}
\caption{Level scheme of $^{87}$Rb with nuclear spin $I=3/2$.} 
\label{pict:Rb87D1D2line}
\end{figure}
\markright{A Appendix}

\end{appendix}

%----------------------------------------------------------------------------

\bibliographystyle{unsrt}
\bibliography{literatureDATAbase}

\begin{thebibliography}{100}
\bibitem{footnote1}
This procedure is justified because the statistical error on the data points of
  the reference laser spectrum is much smaller than the error on the
  fluorescence data.
\end{thebibliography}
\end{document}